\documentclass[10pt, a4paper, onecolumn, twoside]{article}
\usepackage[top=1.45in, bottom=1.45in, left=1in, right=1in]{geometry}

\usepackage[T1]{fontenc}     

\usepackage{lipsum}

\usepackage{authblk}
\usepackage{blindtext}

\usepackage{mathtools}  

\usepackage{empheq}
\newcommand*\widefbox[1]{\fbox{\hspace{2em}#1\hspace{2em}}}

\usepackage{fixltx2e}           

\usepackage{color}

\usepackage{url}       

\usepackage[notref,notcite,final]{showkeys} 

\usepackage{bm}

\newcommand{\sample}[1]{\overline #1}
\newcommand{\exact}[1]{#1^*}
\newcommand{\estimated}[1]{\hat #1}

\newcommand{\samplem}[1]{\overline{\bm{#1}}}
\newcommand{\exactm}[1]{\bm{#1}^*}

\newcommand{\params}{\bm{\theta}}
\newcommand{\param}{\theta}

\newcommand{\ordo}{\mathcal{O}}

\newcommand{\figProblem}{Figure S1 in Supplementary Information}  
\newcommand{\figHistogramFewTrajectories}{Figure S2 in Supplementary Information}  
\newcommand{\figBias}{Figure S3 in Supplementary Information}  
\newcommand{\figHighOrderJack}{Figure S4 in Supplementary Information}  
\newcommand{\figFitGoodness}{Figure S5 in Supplementary Information}  
\newcommand{\figExample}{Figure S6 in Supplementary Information}  
\newcommand{\figResample}{Figure S7 in Supplementary Information}  
\newcommand{\figBma}{Figure S8 in Supplementary Information}  

\newcommand{\tabRealWorldDataNoJack}{Table S1 in Supplementary Information}  

\newcommand{\secNewMethod}{Section A in Supplementary Information} 
\newcommand{\secReviewFittingProcedures}{Section B in Supplementary Information} 
\newcommand{\secPrototypicalSystems}{Section C in Supplementary Information} 
\newcommand{\secSimulation}{Section D in Supplementary Information} 
\newcommand{\secBiasEffect}{Section E in Supplementary Information} 
\newcommand{\secDistributionForParameterEstimates}{Section F in Supplementary Information} 
\newcommand{\secJackknife}{Section G in Supplementary Information} 
\newcommand{\secResampling}{Section H in Supplementary Information} 
\newcommand{\secGoodnessOfFit}{Section I in Supplementary Information} 
\newcommand{\secParticleTracker}{Section J in Supplementary Information} 
\newcommand{\secSimulationParameters}{Section D.5 in Supplementary Information} 

\newcommand{\figHistogram}{Figure 1 in the main text}  
\newcommand{\figSigma}{Figure 2 in the main text} 

\newcommand{\tabRealWorldData}{Table 1 in the main text} 

\newcommand{\eqChiTwo}{eq. (2) in the main text} 
\newcommand{\eqC}{eq. (3) in the main text} 
\newcommand{\eqGA}{eq. (4) in the main text} 
\newcommand{\eqPhiAb}{eq. (4b) in the main text} 

\begin{document}

\title{Fitting a function to time-dependent ensemble averaged data}

\author[1]{Karl Fogelmark}
\author[2]{Michael A. Lomholt}
\author[1]{Anders Irb\"ack}
\author[1,*]{Tobias Ambj\"ornsson}
\affil[1]{Computational Biology and Biological Physics, Department of Astronomy and Theoretical Physics, Lund University, 223 62 Lund, Sweden}
\affil[2]{Department of Physics, Chemistry and Pharmacy, University of Southern Denmark, Campusvej 55, 5230 Odense M, Denmark}
\affil[*]{tobias.ambjornsson@thep.lu.se}

\renewcommand*{\thefootnote}{\fnsymbol{footnote}}

\maketitle

\flushbottom

\thispagestyle{empty}

\subsection*{Abstract}
Time-dependent ensemble averages, i.e., trajectory-based averages of some
observable, are of importance in  many fields of science. A
crucial objective when interpreting such data is to fit these averages (for
instance, squared displacements) with a function and extract parameters (such
as diffusion constants). A commonly overlooked challenge in such function
fitting procedures is that fluctuations around mean values, by construction,
exhibit temporal correlations. We show that the only available general purpose
function fitting methods, correlated chi-square method and the weighted least
squares method (which neglects correlation), fail at either robust parameter
estimation or accurate error estimation. We remedy this by deriving a new
closed-form error estimation formula for weighted least square fitting. The
new formula uses the full covariance matrix, i.e., rigorously includes
temporal correlations, but is free of the robustness issues, inherent to the
correlated chi-square method. We demonstrate its accuracy in four examples of
importance in many fields: Brownian motion, damped harmonic oscillation,
fractional Brownian motion and continuous time random walks. We also
successfully apply our method, weighted least squares including correlation in
error estimation (WLS-ICE), to particle tracking data. The WLS-ICE method is
applicable to arbitrary fit functions, and we provide a publically available
WLS-ICE software.

 \section*{Introduction}
\noindent


Time-dependent ensemble averages appear in several scientific fields. Examples include: particle tracking experiments where mean square
displacements (MSD) are measured at different sampling times~\cite{saxton2008},
human travel dynamics where dispersal distance as a function of time are
measured\cite{brockmann2006}, single-molecule pulling
experiments\cite{desouza2012}, applications of fluctuation theorems
\cite{seifert2012} such as the Jarzynski equality \cite{jarzynski1997},
measurements of the time-dependence of donor-acceptor distance
dynamics\cite{kou2004}, tracer particle dynamics in complex
systems\cite{szymanski2009} and correlation functions in spin systems and
lattice gauge theories\cite{rothe2012}. The final step when interpreting ensemble averages is often to fit a function to these averages in order to extract parameters.

Fitting a function to data is done so readily in science
that one seldom considers the correctness of the standard go-to solution of
the (linear and non-linear) weighted least squares (WLS) method~\cite{press2007,bos2007,sivia2006}. One of the
crucial implicit assumptions of the ``standard'' version of this
method is that the fluctuations around mean values are independent. However, since for time-dependent ensemble averages the data is sampled along
trajectories, this independence assumption is in general \emph{not} satisfied when
analyzing ensemble averages; heuristically, if
in one trajectory an observable, such as the square displacement, was smaller
than its ensemble averaged value at some time, it is typically still so at the
next time step. For an illustrative example, see \figProblem{}, which shows the time-evolution in simulations of fractional Brownian motion (FBM).  Thus, the fluctuations around an ensemble averaged
(time-dependent) observable will in general
exhibit temporal correlations.  Herein, the term trajectory is used in its
widest sense: an observable (such as squared displacement) is chosen, and
a trajectory is then measurements of this observable at different consecutive
sampling times.

The question now arises of how severe the consequences of neglecting the
temporal correlations in least squares fitting are. We demonstrate that such neglect leads to unreliable error estimation for parameters and can in some cases lead
to underestimated errors for fitted parameters (such as diffusion constants) by
more than one order of magnitude for our prototype systems (see below). The unreliability of the estimated errors
can have detrimental effects when statistically interpreting the data: The $1\sigma$ ($2\sigma$) rule for Gaussian statistics states that 68 \% (95 \%) of the observed data should (on average) fall within $\pm 1$ ($\pm 2$) $\sigma$ from the estimated mean. For this rule to be meaningful one must have a correct estimator for the variance in estimated parameters, $\sigma^2$.

 To our knowledge,
the only previous method for dealing fully with correlation in data for
function fitting to ensemble-averages is the
correlated chi-square method (CCM) ~\cite{gottlieb1988,michael1994}. This method is known to
the lattice quantum chromodynamics community, but does not seem to have found wide spread use.
This could partly be due to that, while mathematically sound, numerical robustness issues have been identified~\cite{seibert1994,yoon2013}.
We here carefully examine the CCM method and demonstrate that it in general
only provides correct parameter estimation in a small region of the "phase space" $(N,M)$,
where $N$~is the number of sampling times and $M$~is the number of trajectories.
Thus, it appears that the CCM is of limited general purpose use for fitting of time-dependent ensemble averages to a model function.

Although the least squares and WLS methods are common techniques for parameter
  estimation from ensemble averages,  alternative methods exist, e.g., for inferring parameters from trajectories for biological systems.\cite{meroz2015,hofling2013,norregaard2017} In particular,
  for Brownian motion (BM) an optimal estimator for the diffusion constant
   has recently been derived\cite{berglund2010,michalet2012,vestergaard2014}.  Bayesian methods
  \cite{sivia2006,persson2013,monnier2015,el2015,robson2013,krog2017} have also
   been used for parameter estimation for certain classes of systems.
In general, when they apply, these methods give more precise parameter estimates than the WLS method. However, these newer approaches require as input a full stochastic
  model of the process, and we refer to this type of approach as {\em model
  matching} methods. By a full stochastic model we here refer to
  a model from which (in principle) any probability or average of a measured
  observable can be calculated. A simple example is BM, where the time-evolution is described by a Langevin equation with a noise term for which the statistics is fully specified. In contrast, the WLS and CCM
  methods are parametric {\em function fitting}\cite{gershenfeld1999} type
  methods, which can be used even if a full stochastic model is not available
  to describe the data at hand. An example from single-particle tracking, where function fitting is useful, is if one wants to determine a power-law exponent for the scaling of the mean-square displacement with time. In this situation, a function fitting procedure such as WLS can be used, without making any assumption about the underlying dynamics.  Also,
  even if a full stochastic model is indeed available, it might be impractical to  carry out a full model matching procedure.

In this article, we derive a mathematically rigorous expression for the variance and covariance of estimated parameters in WLS fitting. Our new error estimation formula for fitted WLS parameters takes into account the temporal
correlations, which are intrinsic to ensemble averages based on trajectories. To avoid
confusion we term the ``standard'' WLS method\cite{press2007,bos2007,sivia2006} (i.e., weighted least squares neglecting correlation) as WLS-ECE
(Weighted Least Squares Excluding Correlation in Error estimation), whereas our new
approach is referred to as WLS-ICE (Weighted Least Squares Including Correlation in Error
estimation). In figures and discussion where we only consider parameter values and not the associated errors, we only use the term WLS. In contrast to the previous
two methods (WLS-ECE and CCM), our new method has the
desirable unique features of providing both (1) robust parameter estimates in
the full phase space $(N,M)$ with mean parameter values in
agreement with theory for our prototype systems; (2) error estimates that
reproduce the observed spreads in our fitted parameters.

As prototype models we use BM, damped harmonic oscillation (DHO) in a heat
bath, FBM and continuous time random walks (CTRW). These have been
identified as important model systems  in a wide range of systems. BM is of
interest to many fields of science
\cite{metzler2000,pigeon2017,mehrer2009}. Variants of DHO appear in physics, engineering and chemistry.\cite{bloch2013} FBM has been
applied, for instance, to protein dynamics\cite{kou2004}, in financial
modeling\cite{bouchaud1994}, for analyzing climate time series\cite{yuan2014},
to describe tracer particle diffusion\cite{szymanski2009,barkai2012} and for
modeling earth quake phenomena\cite{tsai1997}. Recent
applications of CTRW\cite{metzler2000,metzler2004} include modeling of human
travel patterns\cite{brockmann2006} and of molecular motions in cells and cell
membrane\cite{weigel2011,barkai2012}. However, we point out that our model
systems are merely convenient examples for illustrating our WLS-ICE function  fitting procedure, which can be applied to arbitrary fit functions.
Our four model systems provide ideal test beds for our method, because the
functions to be fitted, the mean position and MSD, are known analytically for
these systems. Moreover, trajectories are fast to generate for these systems,
which, which facilitates stringent testing of the fitting methods based on a
relatively large number of trajectories.

We finally point out two restrictions on the scope of our study: First, we  do not concern ourselves with the
model selection problem~\cite{machta2013,sivia2006}, i.e., how to choose the ``best''
model or ``best'' form for the fit function. Second, in single particle tracking (one of the application fields of our results),
it is common to separate between time-averaged observables (such as the
time-averaged MSD) and ensemble averaged
observables.\cite{kepten2013,metzler2014} In certain cases, these averages are described by the same functional form, but this is not always
so.\cite{metzler2014}   In this study our sole focus is on ensemble averaged
observables.

\section*{Methods}

\label{sec:method}In what follows, we provide a ready-to-use method, which is further motivated
and detailed in \secNewMethod{}.

\subsection*{ The WLS-ICE procedure}

In experiments or simulations one records a set of trajectories, here
 denoted by~$m$. The
task at hand is to fit some functional form $f(t_i; \params)=f_i(\params)$,
with $K$ free fitting parameters $\params = \param_1, \ldots, \param_{K}$ to
some ensemble averaged observable $\sample{y}(t_i) = \sample{y}_i$ over the trajectories,
i.e., to a sample mean of the form
\begin{equation}
  \label{eq:msd}
  \sample{y}_i = \frac{1}{M}\sum_{m=1}^{M} y_i^{(m)}
\end{equation}
where the index $i$ is over the $N$ sampling times
$\bm{T}=T_1,\ldots, T_{N}$ (with $N\ge K$). Herein, we use bold
  symbols to denote vectors or matrices. For BM, FBM and CTRW (see Results), which are all zero mean processes, the observable used is the squared
displacements, i.e., $y_i^{(m)} = |\bm{x}^{(m)}(T_i)-\bm{x}^{(m)}(0)|^2$,
where $\bm{x}^{(m)}(t)$ is the
  position (a vector with $d$ components, where $d$ is the number of spatial
  dimensions) at process time~$t$ for trajectory~$m$, and the start time for the
simulation/experiment is $t=0$. For DHO, our
  non-zero-mean prototype process, we instead use the position directly
as relevant observable, $y_i^{(m)}={x}^{(m)}(T_i)$. It is important to point
out, however, that in the fitting procedure
the quantity
$y_i^{(m)}$ can be any observable for trajectory $m$ at sampling time~$T_i$.  We shall consistently use a 'bar'
to denote a sample estimator (we only make use of sample means and sample covariances). The
challenge in function fitting procedures~\cite{bos2007} is to
fit some function $f_i(\params)$ to the data $\sample{y}_i$ and thereby extract the
model parameters,~$\params$. This problem has previously been
tackled using the WLS-ECE or CCM methods
(reviewed in \secReviewFittingProcedures{}).

Our approach, the WLS-ICE method, extends the WLS-ECE procedure with a correct error
estimation formula which takes correlations in
  fluctuations around ensemble averages into account (see Introduction). For completeness and ease of application, we here provide
the full details of the proposed WLS-ICE fitting procedure.  We start by
introducing a cost function, $\chi^2$, based on the the difference between the
sample average and the fit function $\Lambda_i = f_i(\params) - \sample{y}_i $ for all time points, according to
\begin{equation}
  \label{eq:chi2}
  \chi^2 = \bm{\Lambda}^T \bm{R}\ \bm{\Lambda},
\end{equation}
 where~$\bm{R}$ is a symmetric positive
definite matrix. This cost function is to be minimized with respect
to $\params$ in order to determine the best parameter
values,~$\estimated{\param}_a$
($a=1,\ldots,K$)~\cite{transtrum2010}. We  use a 'hat' to denote
  parameters which have been estimated through minimization of the $\chi^2$
  cost function above and for the estimated (co)variance of such parameters.   In the WLS method one uses weights $R_{ij}=\sample{R}_{ij} = \delta_{i,j}/\sample{C}_{ij}$, where $\delta_{i,j}$ is the Kronecker delta,
and the (unbiased) sample ``covariance matrix of the mean'' is defined as
$\sample{C}_{ij} = \sample{Q}_{ij}/M$, with $\samplem{Q}$ being the sample covariance matrix
\begin{equation}
  \label{eq:C}
   \sample{Q_{ij}} = \frac{1}{M-1} \sum_{m=1}^{M} (y_i^{(m)} - \sample{y}_i)
   (y_j^{(m)} - \sample{y}_j).
\end{equation}
While this specific choice of $\bm{R}$ is used in our applications, we note that the results in this section, including the new error formula
below, is valid for arbitrary choices of~$\bm{R}$. In \secNewMethod{} we elaborate on one "non-conventional"
choice of $\bm{R}$ particularly adapted for BM.

The parameters, $\estimated{\param}_a$, obtained by minimizing $\chi^2$ in equation~\eqref{eq:chi2}, have a
(co)variance $\Delta_{ab} = \langle (\estimated{\param}_a - \exact{\param}_a)(\estimated{\param}_b -
\exact{\param}_b)\rangle$, where $\langle \ldots \rangle $ denotes
  ensemble average. Throughout this study we use a 'star' to denote exact parameter values, i.e., estimated values as $M\rightarrow\infty$. The variances of the fitted parameter
are~$\sigma_a^2=\Delta_{aa}$.  As noted in the Introduction, this covariance
depends on the temporal correlations. For a stationary process, it is
  well-known how to estimate the variance of a mean in the presence of
  temporal correlations, typically by expressing the variance in terms
  of the sum or integral of the auto-correlation function~\cite{flyvbjerg1989,berg2008}. In the
  present context, such an estimation corresponds to fitting to a constant, $f_i(t) = \param_1$, and assuming all correlation functions only
  depend on time differences.

We here extend the above-mentioned results to
  non-stationary processes and arbitrary fit functions by deriving the
  analogous expression for $\estimated{\Delta}_{ab}$ by using the full multivariate  probability density for the fluctuations around mean
  values. Briefly, the covariance for the estimated parameters is defined $\estimated{\Delta}_{ab} =
  \langle  (\estimated{\param}_a - \exact{\param}_a)
  (\estimated{\param}_b - \exact{\param}_b)\rangle$
where $\langle F(\samplem{y}) \rangle = \int  F(\samplem{y})
\rho(\samplem{y};\exact{\params} )d\sample{y}_1d\sample{y}_2\cdots d\sample{y_N}  $ denotes an average
over the multivariate probability density,
$\rho(\samplem{y};\exact{\params})$. We note that the dependence of the estimated parameters $\estimated{\params}$ on
$\samplem{y}$ is implicitly determined by the minimization condition $\partial
\chi^2/\partial \param_a$ = 0. Now, because all $\sample{y}_i$ are averages
over $M$ identically distributed random numbers, for large $M$, it immediately
follows from the multivariate central limit theorem that the function $\rho$
takes the Gaussian form: 
$\rho(\samplem{y};\exact{\params}) = Z^{-1}  \exp( -(\samplem{y} - \exactm{y})^T {\exactm{C}}^{-1} (\samplem{y} - \exactm{y}) / 2)$
with normalization constant
$Z=(2\pi)^{N/2}\sqrt{\det(\exactm{C})}$~\cite{kampen1992}. Two
complications that occur in evaluating  $\estimated{\Delta}_{ab}$ in
closed-form are that the $\samplem{y}$-dependence of $\estimated{\params}$ is
implicit, and, in general, non-linear. Both of these challenges are solved by making a Taylor series expansion of $\estimated{\param}_a -
\exact{\param}_a$ in terms of $\samplem{y} -\exactm{y}$ and 
implicitly using the minimization condition.  The full derivation is given in \secNewMethod{}. The final result is the following estimator:
  \begin{subequations}
    \label{eq:GA}
    \begin{equation}
      \estimated{\Delta}_{ab} = \frac{\estimated{\phi}_{ab}}{M},
    \end{equation}
    \begin{equation}
      \label{eq:phi_ab}
      \estimated{\phi}_{ab} = 4 \sum_{c,d}\sum_{i,j} (\bm{\estimated{h}}^{-1})_{ac}
      \left.\frac{\partial f_i(\params)}{\partial \param_c}\right|_{\params=\estimated{\params}}
      (\bm{R}^T \samplem{Q} \ \bm{R})_{ij}
      \left.\frac{\partial f_j(\params)}{\partial \param_d}\right|_{\params=\estimated{\params}}
      (\bm{\estimated{h}}^{-1})_{db},
    \end{equation}
    and
    \begin{equation}
      \label{eq:H_ab}
      \estimated{h}_{ab} =  2 \sum_{i,j}
      \left.\frac{\partial^2f_i(\params)}{\partial\param_a \partial\param_b}
      \right|_{\params=\estimated{\params}}
      R_{ij} \Lambda_j  + 
      2 \sum_{i,j}
      \left.\frac{\partial f_i(\params)}{\partial \param_a}\right|_{\params=\estimated{\params}}
      R_{ij}
      \left.\frac{\partial f_j(\params)}{\partial \param_b}\right|_{\params=\estimated{\params}},
    \end{equation}
  \end{subequations}
where the indices $a,b=1,\ldots, K$. Equation~\eqref{eq:GA} gives a mathematically
rigorous expression (to lowest order in $1/M$) for the covariance of the estimated parameters, and is our
key result. It allows us to accurately estimate the covariance of any
parameter fitted by minimizing the cost function in
equation~\eqref{eq:chi2}.  Notice that the correlations in
fluctuations around mean values enter through the quantity $\samplem{Q}$, which is
estimated using the usual sample estimate above. In practice, our general formula, equation~\eqref{eq:GA} is simple to implement and computationally fast.

The new error estimation formula, equation~\eqref{eq:GA}, reduces to previously known results in specific limits.
(i) Neglecting the off-diagonal
elements of $\samplem{Q}$ above we recover the WLS-ECE error estimation
formula~\cite{press2007}.
(ii) By setting $\samplem{R} = \samplem{C}^{-1}$ above we
recover the covariance estimation formula for CCM~\cite{gottlieb1988,bos2007}.
(iii) For a stationary process one seeks to fit a constant, $f_i(\param_1)
=\param_1$, to  data. For such a case, the minimization procedure (solving
$\partial \chi^2/\partial \param_1 = 0$ with $R_{ij}=(1/\sigma^2)\delta_{i,j}$, where $\sigma$ is the time-independent variance) yields
$\estimated{\param}_1 = (1/N) \sum_i \sample{y}_i$, i.e., the parameter estimate is the
mean of the data. The error estimation equation~\eqref{eq:GA}, then
reduces to the usual result~\cite{flyvbjerg1989,berg2008} $\estimated{\Delta} = (1/M) \sum_{i,j} \sample{Q}_{ij}/N^2 $ used,
for instance, in analyzing Monte Carlo and molecular dynamics
simulations. (iv) For linear fit functions, $f_i(\params)=\param_1 t_i$, equation~\eqref{eq:GA} reduces
to previously known expressions (equation~5.253 in van den
Bos~\cite{bos2007}).

\subsection*{Validation procedure}

We tested the different fitting procedures on simulation data for
our four prototype systems (generated as described in \secSimulation{}).
Estimated parameters, $\estimated{\param_a}$, were compared to their known
exact values $\exact{\theta}_a$
(see \secPrototypicalSystems{}). For BM, the MSD behaves as $\langle [\bm{x}(t)-\bm{x}(0)]^2 \rangle =
f(\param, t) = \param_1 t$. The corresponding expression for FBM and
CTRW is $\langle [\bm{x}(t)-\bm{x}(0)]^2 \rangle =  f(\params, t) = \param_1 t^{\param_2}$.
For DHO (at critical damping and with the initial conditions $x(0) = x_0$ and $v(0) = 0)$,
the mean position has the form  $\langle x(t) \rangle =  f(\param, t) = x_0(1+\param_1 t)\exp(-\param_1 t)$.

For validating the WLS-ICE estimator for $\Delta_{ab}$, we generated $S$ simulation sets (with $S=500$) each consisting of $M$ trajectories.
Using these $S\times M$  trajectories, we obtained $S$ number of
 parameter estimates $\estimated{\param}_a$. From these $S$ estimates we calculate the
covariance $\Delta_{ab}$ (using sample estimators), which then
  serves as true $\Delta_{ab}$ (``ground truth''). This true $\Delta_{ab}$ is then
compared to estimates based on the WLS-ICE error formula,
equation~\eqref{eq:GA} (which requires only one set of simulations), and the
corresponding error estimates for WLS and CCM.

\subsection*{Code availability}
\label{sec:code-availability}
Computer codes (Python, Octave/\textsc{matlab}, and Lisp) which performs the
associated fitting (determining $\estimated{\param}_a$) and error estimation
(calculating $\estimated{\Delta}_{ab}$), using
a set of measured observables for different trajectories and at different
times as input, is freely available under the \textsc{gnu} General Public License
(\textsc{gpl})~\cite{gplv3} at \texttt{http://cbbp.thep.lu.se/activities/wlsice/}.

\section*{Results}
\label{sec:results}

\begin{figure}[!htp]
  \centering
  \includegraphics{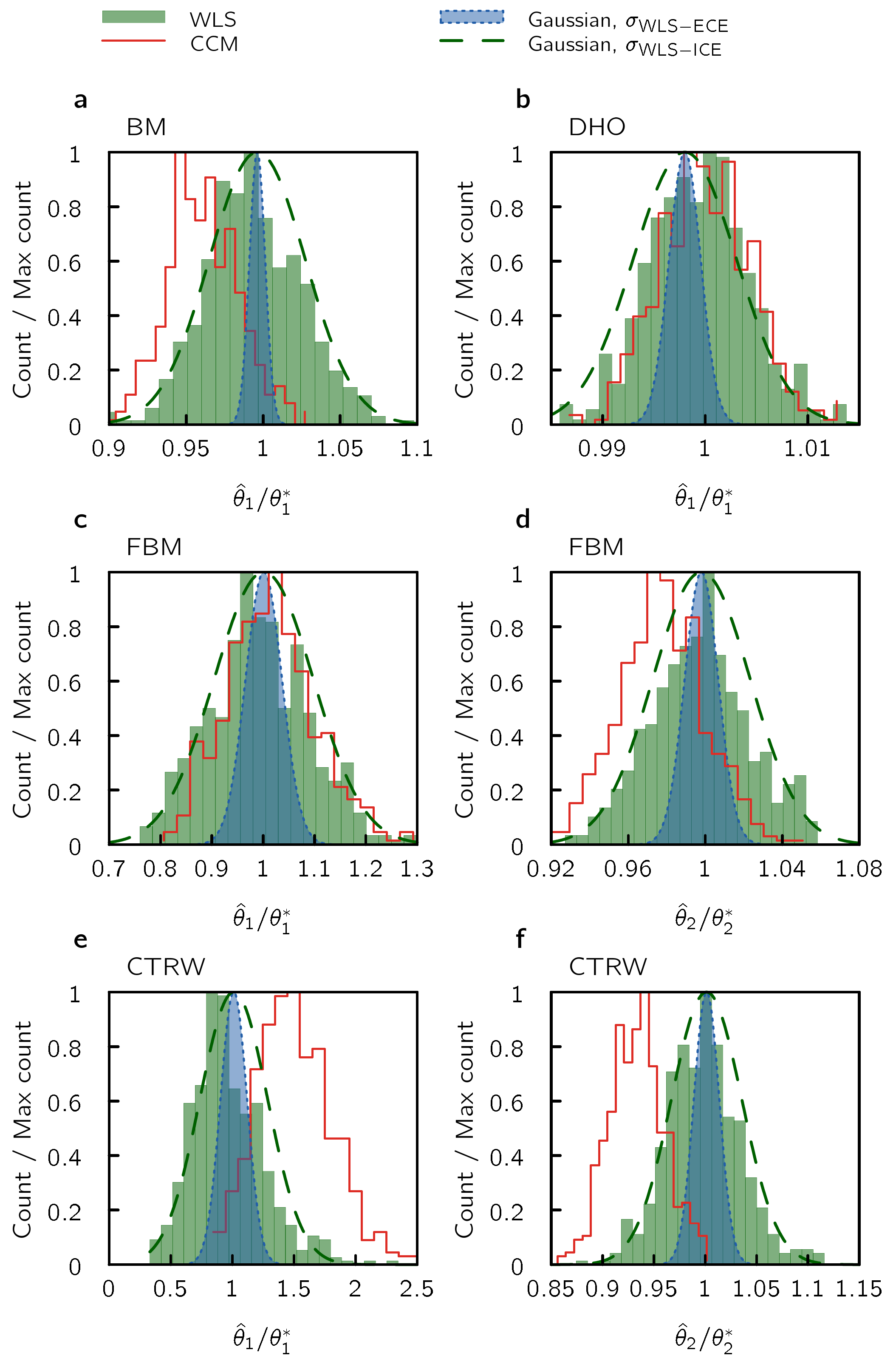}
  \caption{ \textbf{Histograms of fitted parameters for two WLS methods and CCM compared to theoretical predictions.} Each method is tested on:
    (\textbf{a}) Brownian motion (BM),
    (\textbf{b}) damped harmonic oscillation (DHO)
    (\textbf{c}--\textbf{d}) fractional Brownian motion (FBM), and
    (\textbf{e}--\textbf{f}) continuous time random walk (CTRW).
    In each test, we generate~$S=500$ data sets, each consisting of $M=1000$
    trajectories sampled at $N=75$ time points (histograms).
    Panel (\textbf{a}) shows the MSD prefactor (proportional to the diffusion constant) for BM, panel (\textbf{b}) shows DHO fitting parameter $\param_1$, while panels (\textbf{c-f}) left and right panels show the MSD prefactor $\param_1$, and the exponent $\param_2$, respectively.
For comparing WLS-ICE and WLS-ECE to the  histograms based on the $S$ data sets, we place Gaussian functions with their center positions at the
mean of the WLS-fitted parameters. The  widths of the Gaussians correspond to the parameter uncertainty estimated by the fit method (averaged over the $S$  number of fits).
The CCM fits for BM and CTRW exhibits a strong bias in the parameter
value (not centered on the analytical prediction), and the WLS-ECE fit gives
an error estimation, see \secReviewFittingProcedures{}, that is much too
small. The new WLS-ICE procedure (Methods) works well, i.e., exhibits
negligible bias for all model systems and yields correct error estimation,
equation~\eqref{eq:GA}.
The rather large number of trajectories ($M=1000$) was
used in order to avoid ill-conditioness and major bias issues for the CCM fitting, compare to
Fig.~\ref{fig:phasespace}. Results for a smaller ensemble size are found in
\figHistogramFewTrajectories{}, where we see that also for FBM there can be
pronounced bias effects for CCM fitting. For simulation parameters, see \secSimulationParameters{}. }
  \label{fig:histogram}
\end{figure}

\begin{figure}[!htp]
  \centering
  \includegraphics{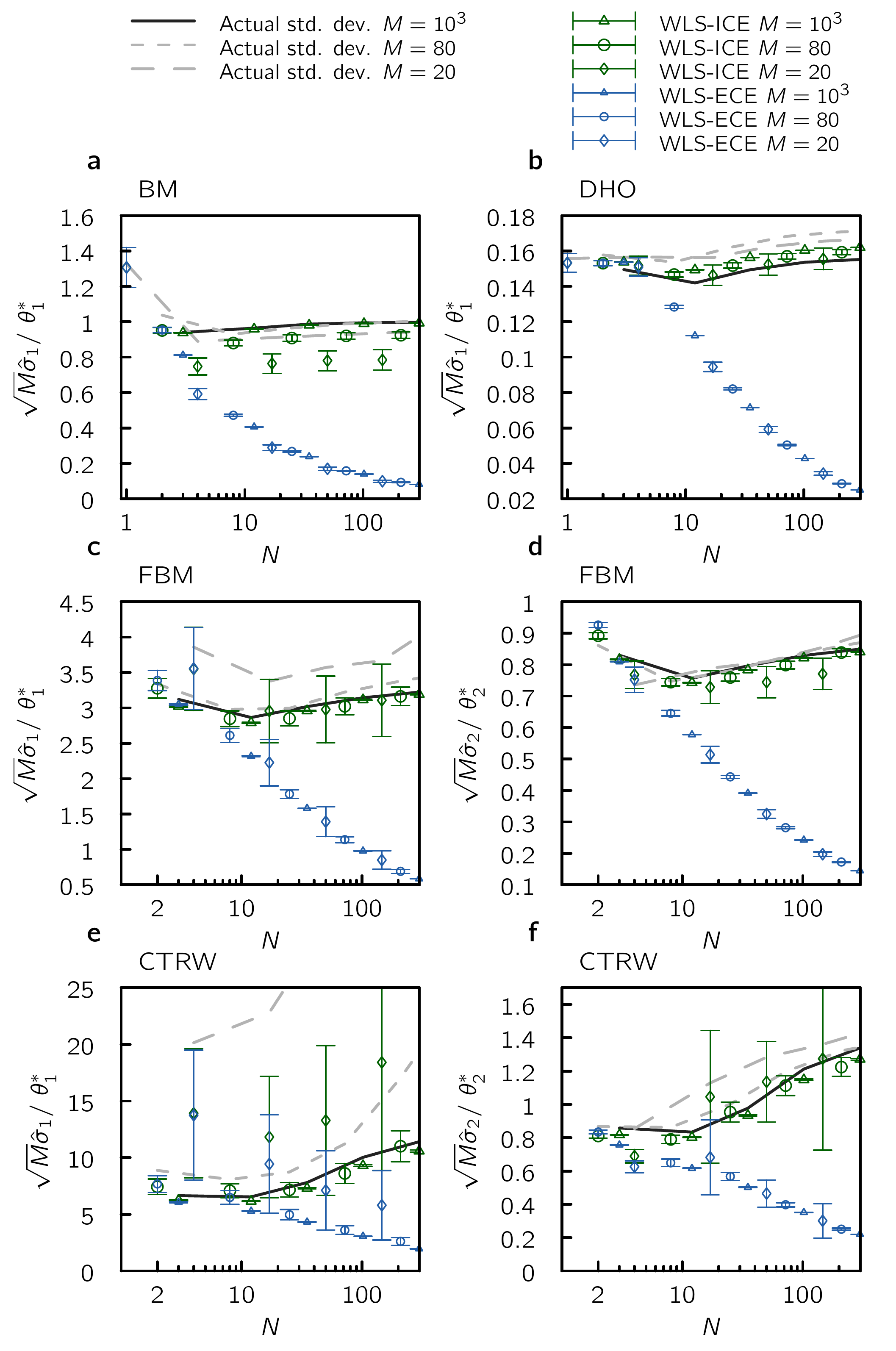}
  \caption{\textbf{Error estimation.} Standard
    deviation from the WLS-ECE and WLS-ICE parameter fits as a function of
    the number of sampling points,~$N$, used in the
    fitting procedure (log-scale on the horizontal axis for visibility).
    Each method is applied to $S=500$ realizations of data from
    (\textbf{a}) Brownian motion (BM),
     (\textbf{b}) damped harmonic oscillation (DHO),
    (\textbf{c}--\textbf{d}) fractional Brownian motion (FBM), and
    (\textbf{e}--\textbf{f}) continuous time random walk (CTRW).
    In conjunction we show the true standard deviation of each of these
    methods computed from the parameters from the fit (lines),
    i.e., the width seen in Fig.~\ref{fig:histogram}, but for an extended range
    of~$N$.
    It is evident that the standard deviation from the WLS-ECE fit is far too small
    for almost all~$N$. Error bars show standard error of the
    mean. For panels
    a-d there are small biases for $M=20$ and $M=80$ in the observable $\estimated{\sigma}$, as compared to actual standard deviation.
    These biases can be removed using the jackknife procedure applied to
    equation~\eqref{eq:phi_ab}, see \secJackknife{}.  For panel e,
      $M=20$, there is discrepancy between the WLS-ICE estimate $\estimated{\sigma}$, and the
      actual standard deviation; we assign this to slow convergence
      towards the asymptotic form for the multivariate distribution $\rho$
      (see Methods) for CTRW (see also \figHistogramFewTrajectories). 
    For simulation parameters, see \secSimulationParameters{}. }
  \label{fig:sigma}
\end{figure}

Our first test of the fitting methods involve comparing histograms of fitted parameters for our four prototype systems (the number of trajectories, $M$, and number of sampling times, $N$, were kept fixed). For
both CCM and WLS the $S$ fitted values of a given parameter were
binned to a histogram, see Fig.~\ref{fig:histogram}, and compared to a
Gaussian centered on the mean of the estimated parameters with a variance from the average
of the error estimates, using either the WLS-ECE or WLS-ICE method. For WLS, the histogram of fitted parameters is centered close to the true value
(see also \figBias{}). However, only the
WLS-ICE method gives a correct error estimation, equation~\eqref{eq:GA}, as the
predicted width of the WLS-ECE method, see \secReviewFittingProcedures{}, is much too narrow. Clearly, the new error
estimation of the WLS-ICE method performs extremely well. By contrast, the
WLS-ECE method does not provide correct errors of the estimated
parameters; this result extends beyond the chosen parameters for ($N$,$M$) in Fig.~\ref{fig:histogram}, and
holds true under rather general conditions, see
Fig.~\ref{fig:sigma} (the exception is the prefactor for CTRW for very small
$M$). Notice that while the parameters from the WLS-ICE and WLS-ECE methods are centered on the analytical
prediction, this is not true for parameters from the CCM method, which
show a strong bias (Fig.~\ref{fig:histogram}) for BM, FBM and CTRW (not for
DHO). Thus, the WLS-ICE is the only method which yields an acceptable bias and
correct error estimation for all model systems. Note that for the ensemble
size used in \figHistogramFewTrajectories{}, the distribution of fitted
parameters is well described by a Gaussian, see
\secDistributionForParameterEstimates{} for a discussion on this topic. For a smaller ensemble size there are deviation from a Gaussian distribution, see \figHistogramFewTrajectories{}, in particular for the prefactor for CTRW. From Fig.~\ref{fig:sigma} we notice that the variance in the estimated parameter does not approach zero as $N\rightarrow \infty$.  Hence, the only way to decrease the variance in estimated parameters further is to increase $M$ (the WLS estimator is consistent with respect to $M$).

As we have seen (Fig.~\ref{fig:histogram}), the CCM method gives a pronounced bias in the parameter
estimate for a specific choice of the number of sampling times~$N$ and
trajectories~$M$ for BM, FBM and CTRW
systems, but not for DHO. In order to understand the generality of these
findings,  we numerically quantified the bias for an extended range of $(N,M)$
values, and find that the pronounced bias for BM, FBM and CTRW (and lack of
bias for DHO) is rather general, see \figBias{}. In \secBiasEffect{} we investigate the expected bias for the CCM
method further by analytical means. Indeed, we find that the parameter
estimate from CCM fitting is unbiased for DHO. Mathematically, this result
follows from the fact that the observable (mean position) used for the fitting
is a linear function of the noise (this is in contrast to BM, FBM and CTRW,
where the squared displacements are used as relevant observables). For BM, our
analytical calculation in \secBiasEffect{}  shows that for large $N$ the bias
for CCM fitting becomes $\langle \estimated{\param} \rangle = \exact{\param} +
DG(N)/M$,  where $G(N) \approx - 8N/(\ln N + \gamma + 2 \ln 2)$ and
$\gamma \approx 0.5772 $ is the Euler-Mascheroni constant.
Thus, with increasing number of sampling points~$N$, the bias increases as $N/\ln N$ (see \figBias{}). The bias for CCM
appears also in the FBM and CTRW systems, as seen in Fig.~\ref{fig:histogram} \figBias{}. A similar calculation for the WLS parameter estimate,
see \secBiasEffect{}, yields only a
minor, essentially $N$-independent,
bias with $G(N) = -4 (1-1/N)$ for BM.

In order to further investigate practical implications of the pronounced bias for CCM fitting, as well as other known issues with the CCM
method~\cite{seibert1994,yoon2013}, we quantified in what parts of phase
space $(N,M)$ the CCM fitting and WLS-ICE provides ``acceptable'' (see below)
parameter estimation, see Fig.~\ref{fig:phasespace}. First, we find that for
large $N$ and moderate to small $M$, the sample estimate for the covariance
matrix $\bm{C}$ is ill-conditioned (the condition number is larger than the
machine precision). In practice this means that it cannot be
numerically inverted, as required in the CCM parameter estimation procedure,
without uncontrollable numerical errors. Second, for parts of phase space where
ill-conditioness is not an issue, we, rather generously, defined an acceptable fit as one where
the bias is smaller than 10\% (compared to the analytic value,
$\exact{\param}_a$). We find that for BM, FBM and CTRW there is indeed a thin
region of the $(N,M)$-phase space (large $M$ and small $N$) where CCM works. For DHO, the bias effect is negligible, as previously noted. However,
the ill-conditioness issue is as pronounced for DHO as for BM, FBM and
CTRW. In contrast, for WLS ill-conditioness is not a problem (no matrix
inversion is required in this procedure), and the bias in the parameter
estimation is acceptable for most
parts of the phase space. The bias inherent in the CCM method (for observables which are not linear functions of the noise (MSD for BM, FBM and CTRW))
can be reduced by applying the common jackknife procedure~\cite{miller1974},
which removes bias terms proportional to $1/M$, see \secJackknife{}. By applying the (first-order) jackknife procedures  to BM, FBM and CTRW
(Fig.~\ref{fig:phasespace}), we find that the bias is reduced which expands somewhat the
region of the phase space where the CCM method may be used reliably.  Note that the computational time is a
factor $g$ (i.e., the number of groups into which the trajectories are
pooled) larger for the first-order jackknife procedure compared to the
non-jackknife case. Finally, the jackknifing procedure can be extended to
remove higher order bias terms (proportional to $1/M^n$, with
$n=2,3,\ldots$)~\cite{miller1974}. However, for the present case there is no
guarantee that these higher order terms have this functional form with respect
to $M$, see \secBiasEffect{}.  Also,
our results show that the second-order jackknife increased, rather than
decreased, the bias in the parameter estimations for most parts of the phase
spaces (Fig.~\ref{fig:phasespace}). For BM, \figHighOrderJack{} indicates that
the reason for this is that the third order term (term proportional to
$1/M^3$) is generally larger in amplitude (but of opposite sign) than the second order one.  Higher order bias reduction comes at a computational
price, since the number of numerical evaluations required for second order jackknife
is $g(g+1)/2$ times that of non-jackknifed parameter estimation.  Due to these
findings and the lack of a formal functional form for the bias, beyond the
$1/M$ term (see above), we do not recommend applying the jackknife procedure
beyond first order. Finally, we point out that the new error estimation
formula, equation~\eqref{eq:GA}, remains valid also for jackknifed parameters,
see \secJackknife{}.

\begin{figure}[!htp]
  \centering
  \includegraphics{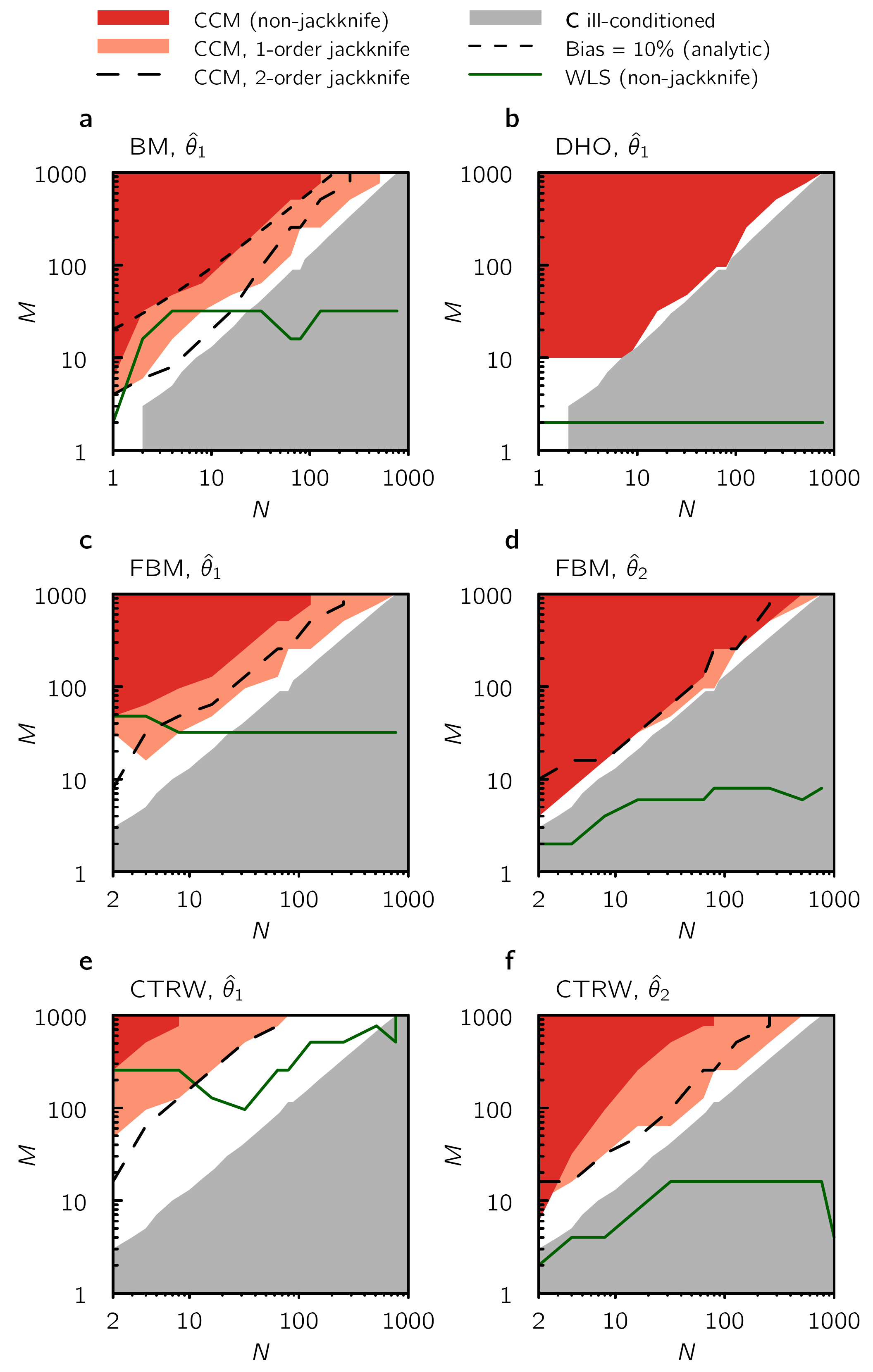}
  \caption{ \textbf{Phase space of reliable parameter estimation for CCM and WLS.} For each of our example
    systems,
    (\textbf{a}) Brownian motion (BM),
    (\textbf{b}) damped harmonic oscillation (DHO)
    (\textbf{c}--\textbf{d}) fractional Brownian motion (FBM), and
    (\textbf{e}--\textbf{f}) continuous time random walk (CTRW),
    we investigate for which number of sampling times~$N$, and number of
    trajectory realizations~$M$, the fitting is more than 10\% off from its
    analytical value, averaged over $S=500$ simulations. As indicated, CCM is only reliable
    in a limited region (large $M$, small $N$), which can be extended by a
    first order jackknife correction.
    For BM we also include when the analytically predicted first order bias term
    for CCM, $G(N)$, see \secBiasEffect{}, gives a bias that is 10\% of the
    true parameter value.  We also show the boundary for when more
    than half of the~$S$ generated covariance matrices become ill-conditioned.
    Interestingly, for the CCM a second order jackknife generally does more harm than good compared
    to the first order, which we elaborate more on in \figHighOrderJack{}. In contrast to CCM (non-jackknifed), the
    parameter estimations for the WLS method are acceptable for most $N,M$ (region above the green curve), and can be extended
    even further using a jackknife approach (data not shown). For simulation parameters, see \secSimulationParameters{}.
  }
  \label{fig:phasespace}
\end{figure}

In \figFitGoodness{} we investigated the "goodness of
fit" for the WLS and CCM procedures using a standard $R^2$ metric (see \secGoodnessOfFit). Examples
of fitted curves are found in \figExample{}. A good fit is characterized
by $R^2 \approx 1$.  We find that, in this sense, the new method provides
"good" fits. In contrast, the CCM method provides "bad" fits  for BM, FBM and CTRW with
$R^2 \ll 1$ for large~$N$. We point out that for the present type of data, $R^2$ is only a heuristic goodness-of-fit metric --- its distributional properties are not known for general fit functions and correlated data.

When computational times are not a concern, error estimation using bootstrap
resampling (or the related jackknife error estimation procedure) are common
method (see \secResampling).\cite{efron1994}
 We here find that bootstrap resampling performs as well as WLS-ICE in general
 for our four models (jackknife error estimation is slightly worse), see \figResample{}. Thus, our numerical
 results indicate that for the type of observables and fit functions used in
 our model systems, the
 bootstrap can be used for calculating the variance for parameters
 estimated through $\chi^2$ minimization. However, we point out that such resampling techniques require us to repeat the $\chi^2$ minimization several
 (herein, 100) times (the WLS-ICE method requires only one $\chi^2$
 minimization). Such minimization can be computationally costly, especially
 for the case when the number of unknown parameters is large. Moreover, one
 must bear in mind that the bootstrap method is in general a heuristic method
 (there are cases when it does not apply\cite{efron1994}).

As a final alternative to the WLS-ICE method, we now briefly turn to error estimation using subsampling~\cite{berg2008}. Subsampling refers to the method of choosing
sampling times sufficiently sparsely in order to make the data points
essentially uncorrelated (the ``brute force'' method in \figProblem{} is an extreme case of subsampling where only one data point
per trajectory is kept). After subsampling, error analysis is performed using
standard error analysis for independent data. In order to properly choose $N$
within this method, $N$ is systematically decreased until the variance
saturates to a constant, which is assumed to be the true
variance~\cite{berg2008}. Notice for stationary time series, rather than reducing the number of sampling times, one can make full use of the data through the blocking method.\cite{flyvbjerg1989} For non-stationary processes the blocking method cannot be used, however. Fig.~\ref{fig:sigma} shows how
estimated errors from our WLS-ECE and WLS-ICE analyses depend on the number of data
points used,~$N$. We find that temporal correlations are so strong that the
WLS-ECE method underestimates the errors down to very small~$N$. Moreover,
finding a sufficiently small $N$ is difficult, since the error does not in
general saturate to a constant level as $N$ is reduced. These
  problems are circumvented by instead using the error estimation from the
WLS-ICE method (i.e., using equation~\eqref{eq:GA} instead of the WLS-ECE equations in
\secReviewFittingProcedures{}).

As a final test of our method, we now turn to "real world" data. To that end,
we use particle tracking data used in
a competition for testing particle tracking software where 14 teams
world-wide participated.\cite{chenouard2014} We choose to analyze this data
set for two reasons. First, it served as  standard benchmark data within the particle
tracking community. Second, since these movies are based on noisified and
pixelated simulations (aiming to mimic actual experimental data), we know the values of the underlying model parameters. We
used their Supplementary Videos~1 (medium particle density), 5 (low particle
density) and 6 (high particle density). All these movies correspond to
BM of vesicles for which the expected MSD for the data sets are $\langle [\bm{x}(t)-\bm{x}(0)]^2 \rangle =
f_{\rm BM}(\param, t) = \param_1 t$, with
$\param_1 = 2dD = 8$. For particle detection in the movies and linking of
particle positions into trajectories  we used Method 1\cite{chenouard2014},
i.e., the tracking method described by
Sbalzarini et al.\cite{sbalzarini2005}, and
implemented as the ImageJ plugin "Particle Tracker" by the MOSAIC
group~\cite{particletracker}. Parameter settings for the
plug-in are listed in \secParticleTracker{}. For each video
we extracted trajectories which were subsequently cut into trajectories
consisting of $7$ discrete process times (there is no memory in BM, so the start time
is inessential).  Notice that for the higher particle density, fewer sufficiently long trajectories were produced
as compared to the low density scenario (values for $M$ are listed in Table
\ref{tab:real_world_data}). We subsequently divided the trajectories for each movie into two data sets
each with $M$ trajectories. For the fitting procedures the first process time point,
$t_0=0$, was discarded (since at $t_0$ the position is precisely known, the
variance = 0 and can not be used as a weight in equation~\eqref{eq:chi2}), thus leaving us with $N=6$ sampling times. Results for the estimated parameters, $\estimated{\param}_1$ and
associated standard deviation, $\estimated{\sigma}$ are found in  Table \ref{tab:real_world_data}.  We notice that the CCM
method fails at predicting the correct parameter for high and medium particle
densities. This finding is simply due to the smaller ensemble size for these cases which, in turn, is a result of the tracking software's inability to track and link particles in high and medium density settings. Comparing the WLS-ECE and WLS-ICE method, we see that the WLS-ECE underestimates
the error by factors $\approx$ 2 for all movies. While, this underestimation may seem minor it will affect
conclusions drawn from particle tracking data (see discussion in Introduction), in particular it is noteworthy
that for the WLS-ECE method only 2 out of 6 estimates fall within
$2\sigma$ (confidence level
95 \%) of the expected result ($=8$). In contrast, for the WLS-ICE all six
observed parameter estimations for $\param_1$  fall within $2\sigma$ of the expected value.

\begin{table}[h]
\centering
\begin{tabular}{l|l|l|l|l|l|l|l}
\textbf{Description} & & \multicolumn{2}{c|}{Low density} & \multicolumn{2}{c|}{Medium density} & \multicolumn{2}{c}{High density} \\
\textbf{Video} & & \multicolumn{2}{c|}{S5} &  \multicolumn{2}{c|}{S1}  & \multicolumn{2}{c}{S6} \\
\textbf{Number of trajectories} & & \multicolumn{2}{c|}{$M=310$} & \multicolumn{2}{c|}{$M=16$} & \multicolumn{2}{c}{$M=5$} \\
\hline
\textbf{Method} & Observable & \multicolumn{2}{c|}{} & \multicolumn{2}{c|}{} & \multicolumn{2}{c}{} \\
\hline
  WLS-ICE & $\estimated{\param}_1$ & 8.49 &  8.63 & 11.41 &
  8.14 &  7.60 & 5.45 \\
       & $\estimated{\sigma}$ & 0.38  &  0.38 & 2.17 &
       1.81 &   2.53 & 1.93 \\
\hline
WLS-ECE & $\estimated{\param}_1$ &  8.49 &  8.63 & 11.41 &
  8.14 &  7.60 & 5.45 \\
 & $\estimated{\sigma}$ & 0.20 & 0.19 & 1.25 & 0.93 & 1.56 & 1.00 \\
\hline
CCM & $\estimated{\param}_1$ &  8.63 & 8.33 & 10.83 & 3.79 & ill-cond. & ill-cond. \\
 & $\estimated{\sigma}$ & 0.37 &  0.35 & 1.14 & 1.60  & ill-cond. & ill-cond. \\
\end{tabular}
\caption{ \textbf{Results of the three fitting methods for ``real world''
      particle tracking data}. Particle trajectories where extracted from the
    ``Vesicle'' Supplementary videos from the article by Chenouard et al
    \cite{chenouard2014} using the ``Particle Tracker'' software (MOSAIC
    group). The trajectories where cut into shorter trajectories, all of
    length 7 discrete process times. The short trajectories were then divided into
    two independent sets of
    size $M$. We then performed fitting using the WLS-ICE, WLS-ECE and CCM
    methods for BM, discarding the first process time point, resulting in $N=6$ sampling times.
  Expected parameter value is $\param_1=8$ (data are noisified and pixelized
  simulations with known properties). Since $M$ was very small for video S6, we applied the jackknife
  procedure both in parameter and error estimation (all videos). Results before
  jackknifing are found in \tabRealWorldDataNoJack{}. We notice that the CCM method gives
  ill-conditioness issues for the high density movie, where few trajectories could
  be extracted. The WLS-ECE method underestimates the error as compared to WLS-ICE
  method. }\label{tab:real_world_data}
\end{table}

Let us finally briefly discuss how well one is expected to be able to
  estimate a parameter based on experimental/simulation data. For
   model
  matching procedures (see
  Introduction), the Cramer-Rao bound is useful by providing an
  expression for the smallest possible variance in the estimated
  parameter.\cite{bos2007} For the case of BM, optimal estimators
  (i.e., estimators which reach the Cramer-Rao bound) based on the measured
  displacements have been derived for model matching type
  fitting\cite{berglund2010,michalet2012,vestergaard2014}. For function fitting, the
  question is rather whether an optimal cost function, i.e., an optimal weight
  matrix $\bm{R}$, can be found (see equation~\eqref{eq:chi2}). If the
  covariance matrix for the process is independent of the inferred parameters
  (up to a proportionality constant), and for linear fit functions, then the
  generalized least squares method can be shown to be optimal among
  unbiased WLS methods.\cite{kariya2004}. Since the generalized
    least squares method requires as
  input the inverse of the true covariance matrix, it can be viewed as a
  hybrid method in between model matching and function
    fitting. In \figBma{} we
  show results for the  generalized least squares for BM (we use the term BMALS -- Brownian motion
  adapted least squares) where we see that, indeed, the variance in estimated
  parameter value is smaller for BMALS as compared to WLS-ICE, although the
  difference is not dramatic. Also notice that for $M$ and $N$ values where the
CCM ``works'' (acceptable bias, see Fig.~\ref{fig:phasespace}) the variance in
estimated parameters for CCM and BMALS agree, as it should.

\section*{Discussion, conclusion and outlook}
\label{sec:summary-outlook}

A common task in many fields of science is that of fitting a
model to the time-evolving mean of some observable.  Since fluctuations around observed mean values, calculated
based on trajectories, are in general correlated in time, the error estimates
provided by a ``standard'' weighted least squares (WLS-ECE) fit  can be more than one order of magnitude too small, see Fig.~\ref{fig:sigma}. Further, the correlated chi-square method (CCM), involving numerical inversion of a noisy
covariance matrix, often show numerical instabilities (ill-conditioning) or a
strong bias in the fitted
parameters, see Fig.~\ref{fig:phasespace}.  To overcome these problems, we derived a new error estimation
formula, see equation~\eqref{eq:GA}, for weighted least squares
fitting, which does not require inversion of a noisy covariance matrix. With
this formula at hand, a simple, yet accurate, function fitting procedure,
WLS-ICE, can be followed:  (A) perform a
weighted least squares fit to the data, (B) use the new formula to estimate
the errors. We demonstrated on four simulated prototype systems that
the WLS-ICE
method provides robust results, with a negligible bias in the fitted
parameters and accurate error estimates.
 Our method's estimated errors are comparable to errors estimated
  using bootstrap and jack-knife resampling for the four model
  systems. A strength of our method is that the
  fitting procedure does not have to be repeated multiple times. 

We separated between two types of parameter
estimation procedures: model matching  where a full stochastic model is
matched to the data, and function fitting  in which a full stochastic model is
not known and one rather seeks to fit a function to the chosen
ensemble-averaged observables. The weighted least-squares method is a
procedure of function fitting type.

We have in this study not discussed methods for dealing with
  experimental errors, such as missing data etc. Such errors depend on the experimental setup and
  typically have to be dealt with in different ways depending on setup. For the
  single-particle tracking field (one of the application fields  of our
  results), two major sources of experimental errors are: effects due to the
  finite size of pixels in cameras used to record the trajectory and motional
  blur effects (in a single time frame, a fluorescent molecule moves while
  being imaged). Methods for correcting these types of errors are discussed by Savin et al.\cite{savin2005}, Martin et
  al.\cite{martin2002}, Berglund\cite{berglund2010} and Calderon.\cite{calderon2016}

Parameter estimation through $\chi^2$ minimization is
ubiquitous throughout many fields of science, and we hope that our
method and publically available software will be found useful in these
fields.


\section*{Acknowledgments}
We are grateful to Bo S\"oderberg and Bj\"orn Linse for fruitful discussions. T.A. was supported
by the Swedish Research Council (grant nos 2009-2924 and 2014-4305).
K.F. was supported by the Swedish Research Council (grant no 2010-5219). M.A.L.
acknowledges funding from the Danish council for Independent Research-Natural Sciences (FNU), grant
number 4002-00428B.

\section*{Author contributions statement}

M.A.L. and T.A. conceived the idea of the project. All authors contributed to
the conceptual design of the WLS-ICE method. K.F. performed the simulations and
wrote the analysis software supervised by T.A. K.F. prepared all figures. T.A. and
K.F. wrote the manuscript with help from A.I. and M.A.L. T.A. derived the new error
estimation formula (with and without jackknife). M.A.L. derived the bias
correction prediction for BM with input from K.F. and T.A. A.I.
suggested the use of jackknife for CCM fitting. T.A.
coordinated the project.

\section*{Competing interests}

The authors declare no competing interests.

\newpage


\renewcommand*{\thefootnote}{\arabic{footnote}}
\setcounter{footnote}{0}

\makeatletter
\renewcommand{\thefigure}{S\@arabic\c@figure}
\renewcommand{\thetable}{S\@arabic\c@table}
\renewcommand{\theequation}{S.\arabic{equation}}
\renewcommand{\thesection}{\ifstrequal{\arabic{section}}{3}{S.M}{\ifstrequal{\arabic{section}}{2}{S.F}{S.T}}}
\makeatother


\thispagestyle{empty}

\vspace*{2cm}

\section*{\Large Supplementary Figures}

\setcounter{figure}{0}
\renewcommand{\figurename}{Supplementary Figure}

\begin{figure}[!htp]
  \centering
  \includegraphics{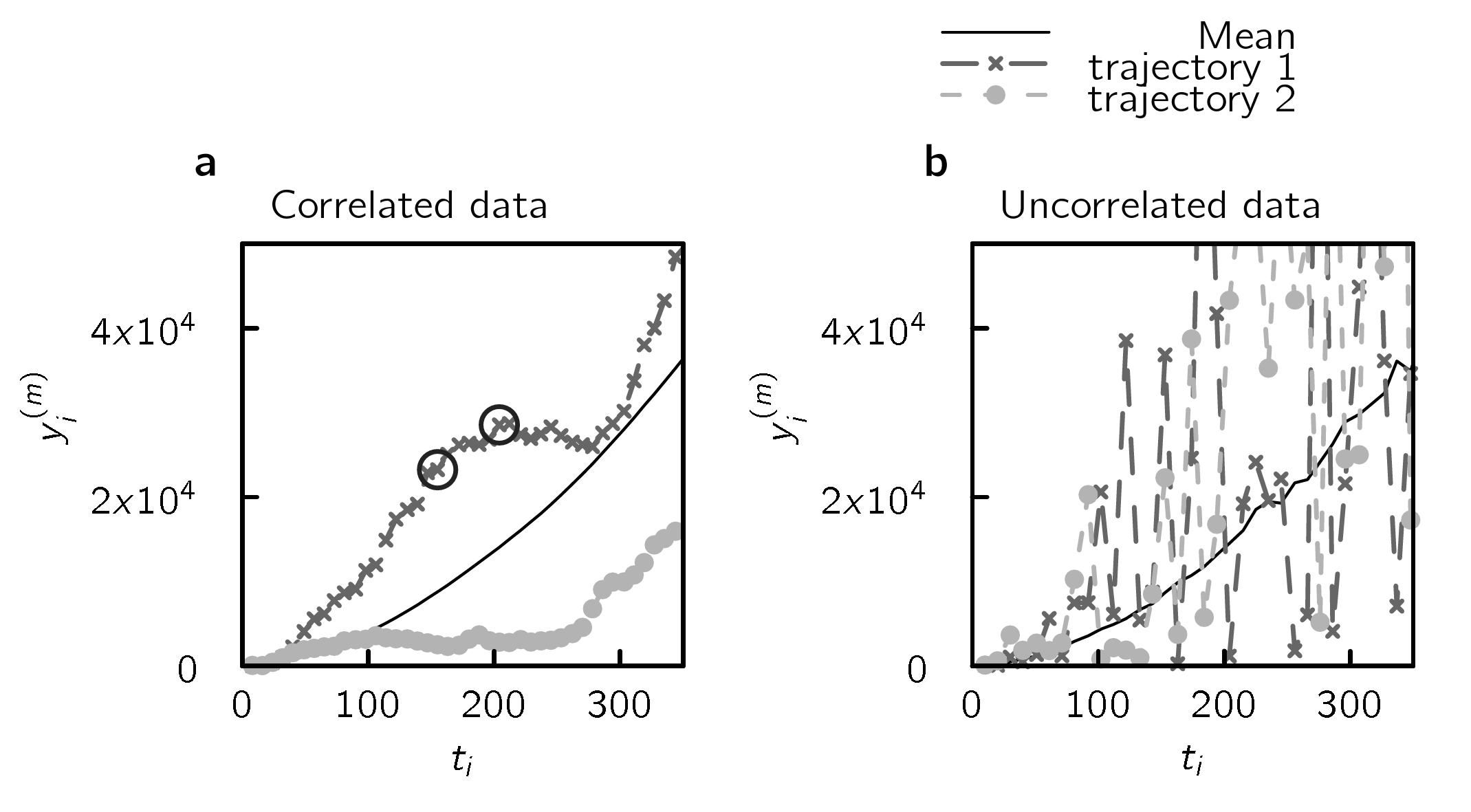}
  \caption{\textbf{Correlations in fluctuations around ensemble
        averages for real trajectories compared to
      uncorrelated fluctuations (synthetic data).} The displacement squared $y^{(m)}_i =
    [\bm{x}^{(m)}(t_i)-\bm{x}^{(m)}(0)]^2$ for fractional Brownian motion (FBM) as a
    function of time,~$t$, for two trajectories, labeled by $m$, and the mean
    of a large ensemble ($M=10^3$) of trajectories. Panel (\textbf{a}) shows
    actual trajectories which exhibit strong temporal correlation, meaning: if we are
    above the mean for some time point on a trajectory, we are likely to still
    be above the mean for time points close to it (circled). In panel (\textbf{b}) we have
    constructed ``synthetic'' trajectories for comparison by only using one data
    point from each real trajectory, and "throw away" the rest, resulting in
    (computationally expensive) uncorrelated data.  That is, within this
    ``brute force'' method, to generate a single uncorrelated trajectory of
    $N$ sampling points, we need to use the same amount of real trajectories,
    and throw away all data points save one. Data was generated from a
    one-dimensional FBM simulation with Hurst
    parameter~$H=0.9$, see Supplementary Methods section~\ref{sec:fbm}.}
  \label{fig:problem}
\end{figure}

\clearpage

\begin{figure}[!htp]
  \centering
  \includegraphics{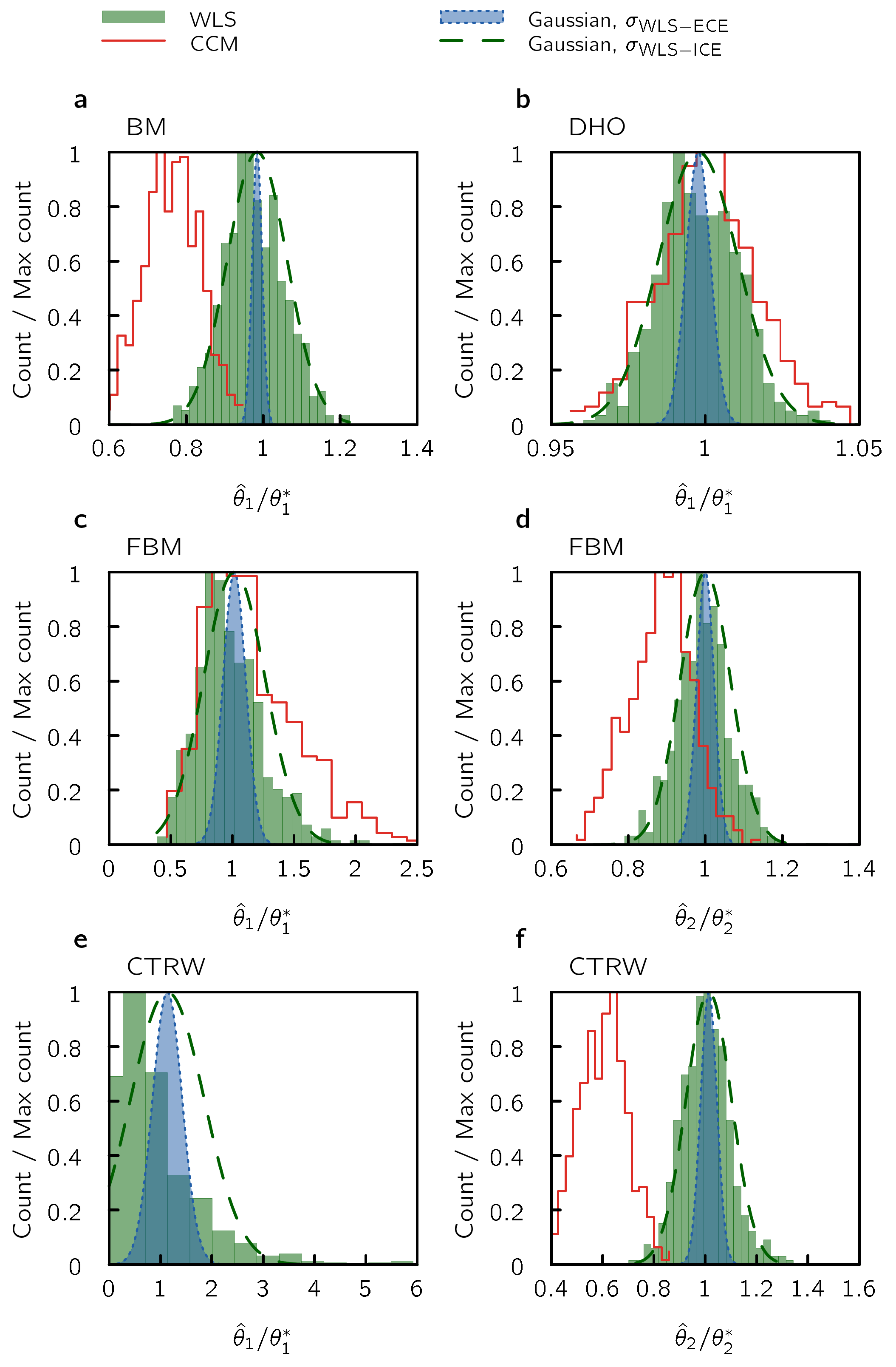}
  \caption{ \textbf{Histograms of fitted parameters for two WLS methods and
      CCM compared to theoretical predictions for a small ensemble size.} All
    panels are identical to those in \figHistogram{} except that we
    here only used $M=150$ trajectories (instead of $M=1000$). In panel
    \textbf{e} (the CTRW prefactor), the CCM fitting procedure gave a vastly
    incorrect parameter estimate ($\langle \sample{\param}_1 \rangle
    /\exact{\param}_1= 13.2$) and the associated histogram is therefore not
    displayed. Due to the smaller $M$ value used here as compared to
     \figHistogram{} the histogram of fitted parameters are
    non-Gaussian for panel \textbf{e}, see Supplementary Methods
    Sec. \ref{sec:distribution_for_parameter_estimates} for a discussion on
    this topic. The other panels converged to normal distributions for smaller
    $M$ values. Examples of parameter fits to the MSD data are shown in Fig. \ref{fig:example}.}
  \label{fig:histogram_few_trajectories}
\end{figure}

\clearpage

\begin{figure}[!htp]
  \centering
  \includegraphics{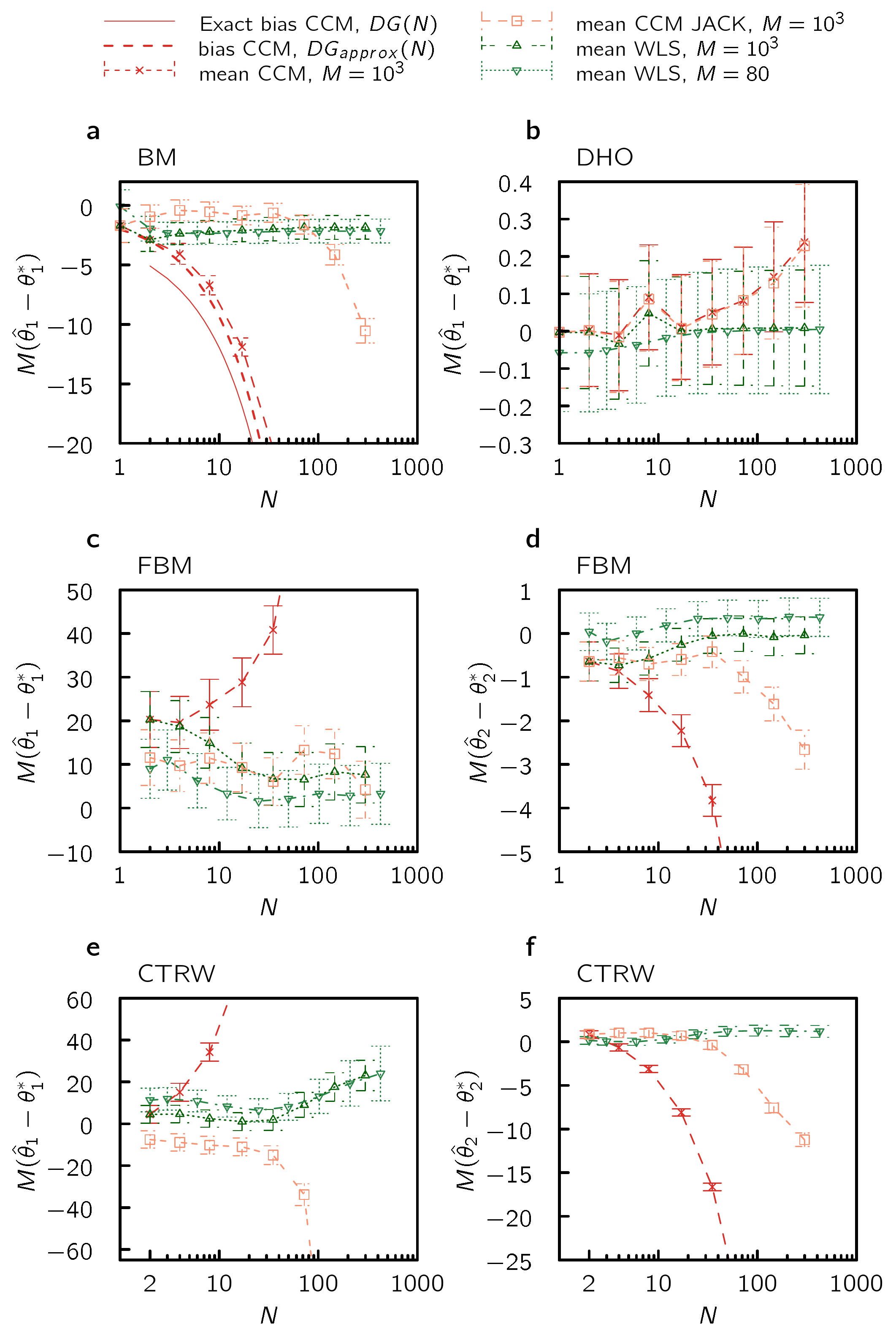}
  \caption{\textbf{Bias in the parameter fit.} The residual bias in the fit
    (multiplied by the number of trajectories~$M$) as a function of sampling
    points,~$N$ (log-scale for the horizontal axis for visibility), averaged
    over parameters from fitting to $S=500$ realizations of the mean.
    (\textbf{a}) For the Brownian motion (BM) CCM fit, the analytical
    prediction, $G(N)$, (full line) for the first order bias follows the
    observed bias for $M=10^3$, data (Supplementary Methods
    section~\ref{sec:bias_corr}). For (\textbf{b}) damped harmonic
      oscillation (DHO) the bias in CCM and WLS are both small, but for
    (\textbf{c}--\textbf{d}) fractional Brownian motion (FBM), and
    (\textbf{e}--\textbf{f}) continuous time random walk (CTRW), the bias term
    in CCM is much larger than the WLS bias. The bias can be alleviated
    to some degree by a Jackknife procedure. Error bars show standard error of
    the mean. For simulation parameters, see Supplementary Methods
    section~\ref{sec:simulation_parameters}.}
  \label{fig:bias}
\end{figure}

\clearpage

\begin{figure}[!htp]
  \centering
  \includegraphics{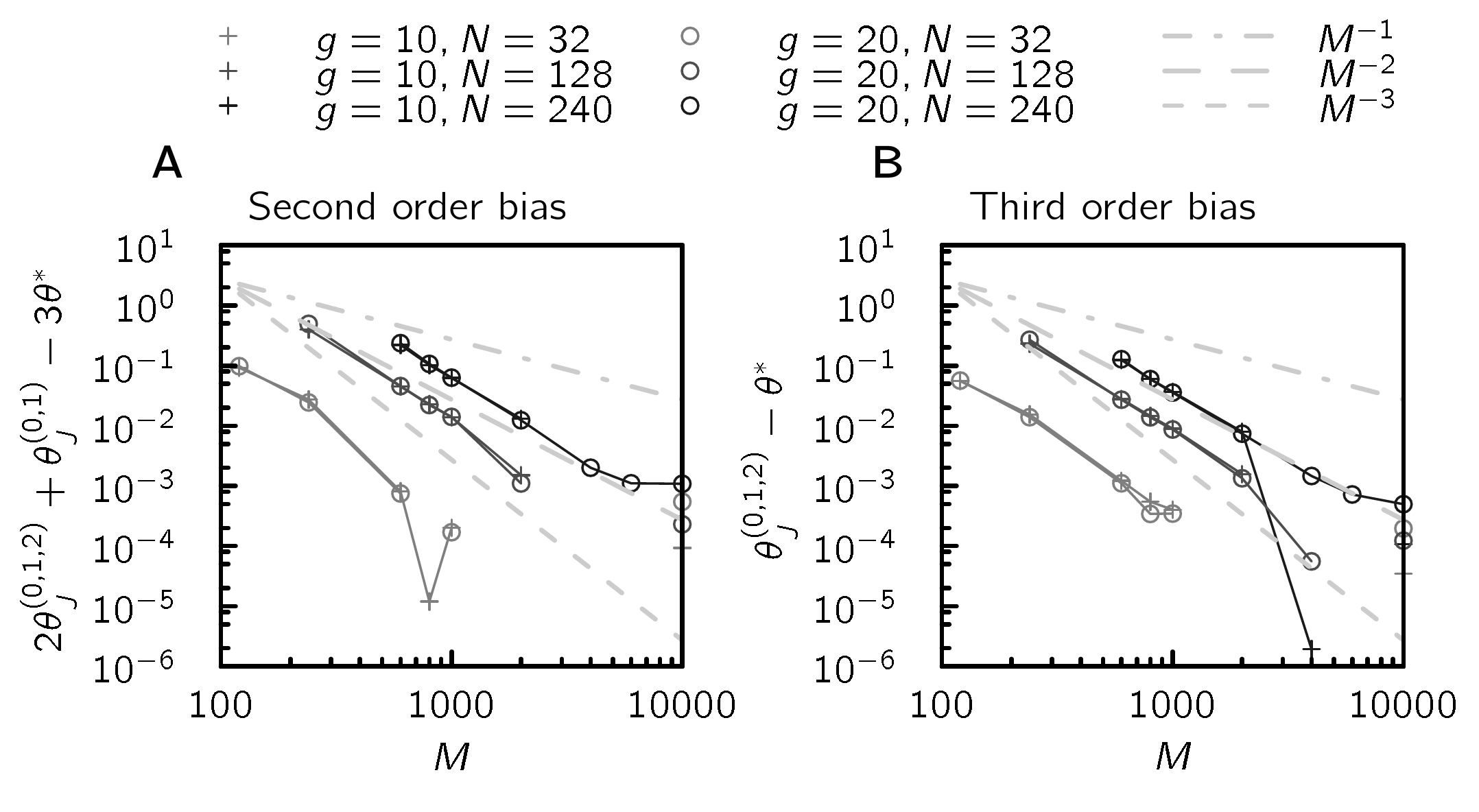}
  \caption{\textbf{High order bias contribution in CCM fitting for BM.} The bias in the parameter
    estimation is commonly assumed
    to be of the form $\estimated{\param} = \exact{\param} + a/M + b/M^2 +
    c/M^3+\ordo (M^{-4})$, see Supplementary Methods section~\ref{sec:bias_effect}.
    In panel (\textbf{a}) the vertical axis shows the (negative) second order
    bias term $-b/M^2$, and in (\textbf{b}) the (positive) third order
    term, $c/M^3$, for three different number of sampling times~$N$.
    Note that these are of comparable magnitude, but
    opposite sign. Thus a second order jackknife, which removes terms
    proportional to $a/M$ and $b/M^2$, may yield more unfavorable results
    than a first order jackknife, which only removes the $a/M$ term.
    We note that the slope of the second order bias term approximately
    corresponds to $M^{-2}$, and the third order is slightly more. For panel (a)
    the second order bias was extracted combining equation~\eqref{eq:jack1_bias}
    and equation~\eqref{eq:jack2_bias}, to give $-b/M^2 = 2\param_J^{(0,1,2)}
    + \param_J^{(0,1)} - 3\exact{\param}$, and for panel (b) we have
    $(\param_J^{(0,1,2)} - \exact{\param})=c/M^3$, which follows immediately
    from equation~\eqref{eq:jack2_bias}. For simulation parameters, see
    Supplementary Methods
    section~\ref{sec:simulation_parameters}.
  }
  \label{fig:high_order_jack}
\end{figure}

\begin{figure}[!htp]
  \centering
  \includegraphics{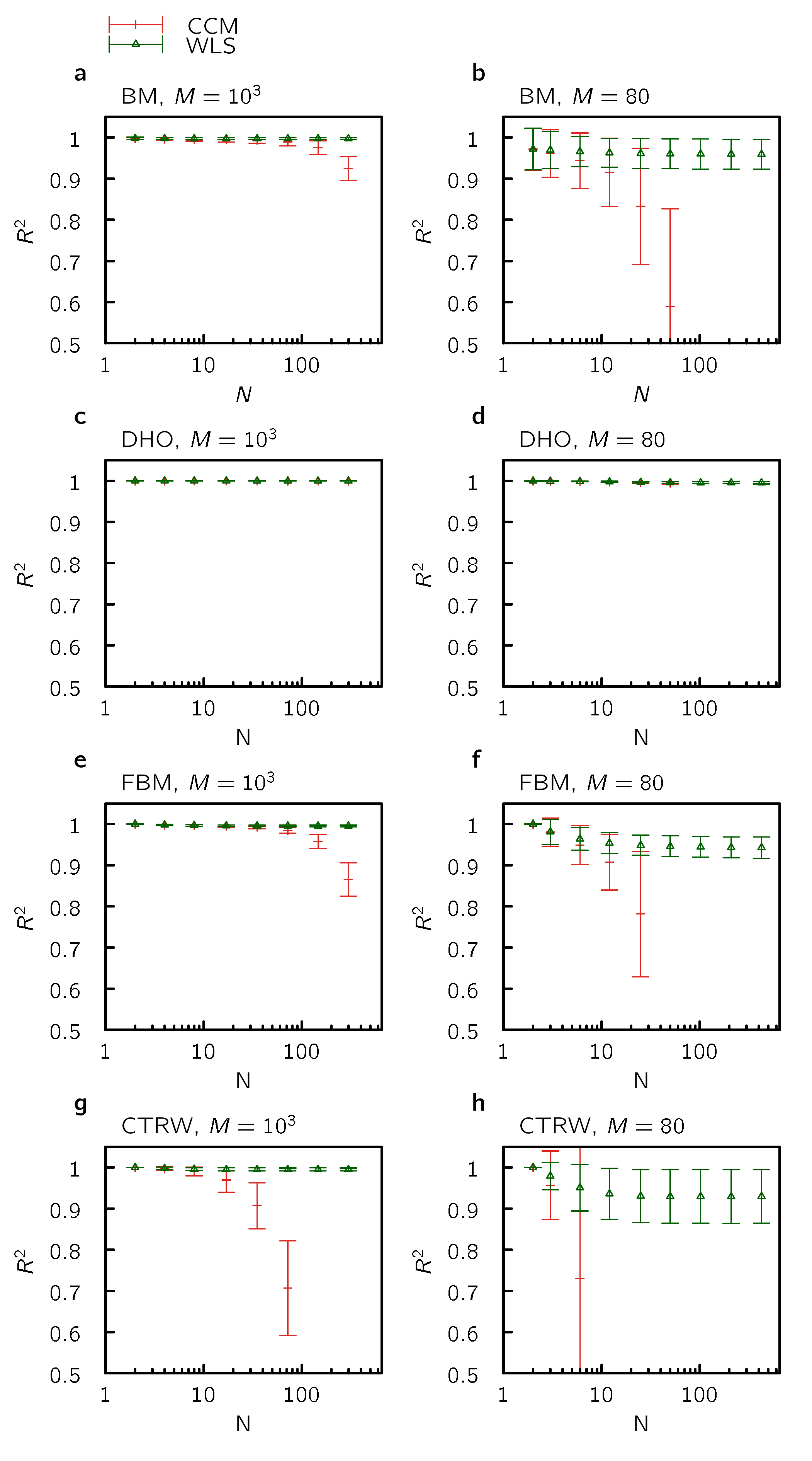}
  \caption{\textbf{Heuristic goodness-of-fit using the the coefficient of
      determination, $R^2$.} The quality of the CCM and WLS fits are heuristically quantified by
    the coefficient of determination, $R^2$, as a function of sampling
    points,~$N$ (horizontal axis on log-scale for visibility), for our four
    prototype systems: (\textbf{a}--\textbf{b}) Brownian motion (BM),
    (\textbf{c}--\textbf{d}) damped harmonic oscillation (DHO),
    (\textbf{e}--\textbf{f}) fractional Brownian motion (FBM), and
    (\textbf{g}--\textbf{h}) continuous time random walk (CTRW). A perfect fit
    yields unit value, while a bad fit results in~$R^2 \ll 1$ (see Supplementary Methods, Sec. \ref{sec:goodness_of_fit}). The number of
    trajectories used in the ensemble average was either $M=10^3$ (left), or $M=80$
    (right). All data was averaged over $S=500$ realizations, with standard
    deviation given by the error bars. For panels (\textbf{b},\textbf{f}) only a few
    data points could be obtained, due to numerical instability of CCM, and for
    panels (\textbf{g},\textbf{h}) $R^2 < 0$ for larger $N$. For simulation parameters, see Supplementary Methods section~\ref{sec:simulation_parameters}. }
  \label{fig:fitgoodness}
\end{figure}

\clearpage

\begin{figure}[!htp]
  \centering
  \includegraphics{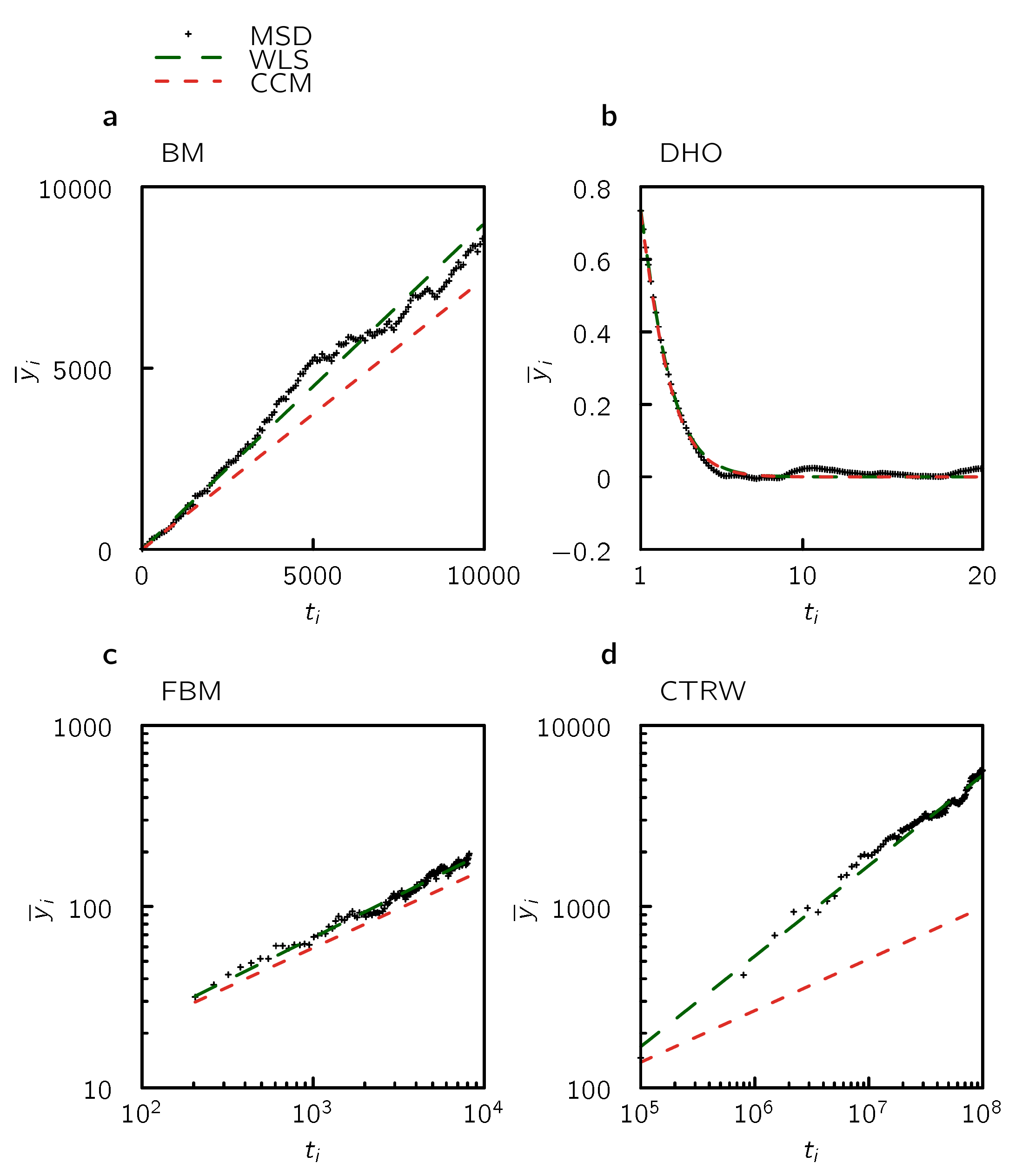}
  \caption{\textbf{Example of a fit to the mean of ensemble data for the
      WLS and CCM methods.} An illustrative example of
    a typical fit to average ensemble trajectory data, based on $M=150$ trajectories, for
    (\textbf{a}) Brownian motion (BM),
    (\textbf{b}) damped harmonic oscillation (DHO),
    (\textbf{c}) fractional Brownian motion (FBM), and
    (\textbf{d}) continuous time random walk (CTRW).
    The model parameters were fitted to the data using either WLS
    or CCM fitting procedure, for $N=75,\ M=150$. For CCM fitting to the FBM data, we
    see that although the exponent is almost the same, the pre-factor is
    inaccurate. For CCM fitting to CTRW data, both exponent and pre-factor
      is poor. For simulation parameters, see Supplementary Methods
    section~\ref{sec:simulation_parameters}.}
  \label{fig:example}
\end{figure}

\clearpage

\begin{figure}[!htp]
  \centering
  \includegraphics{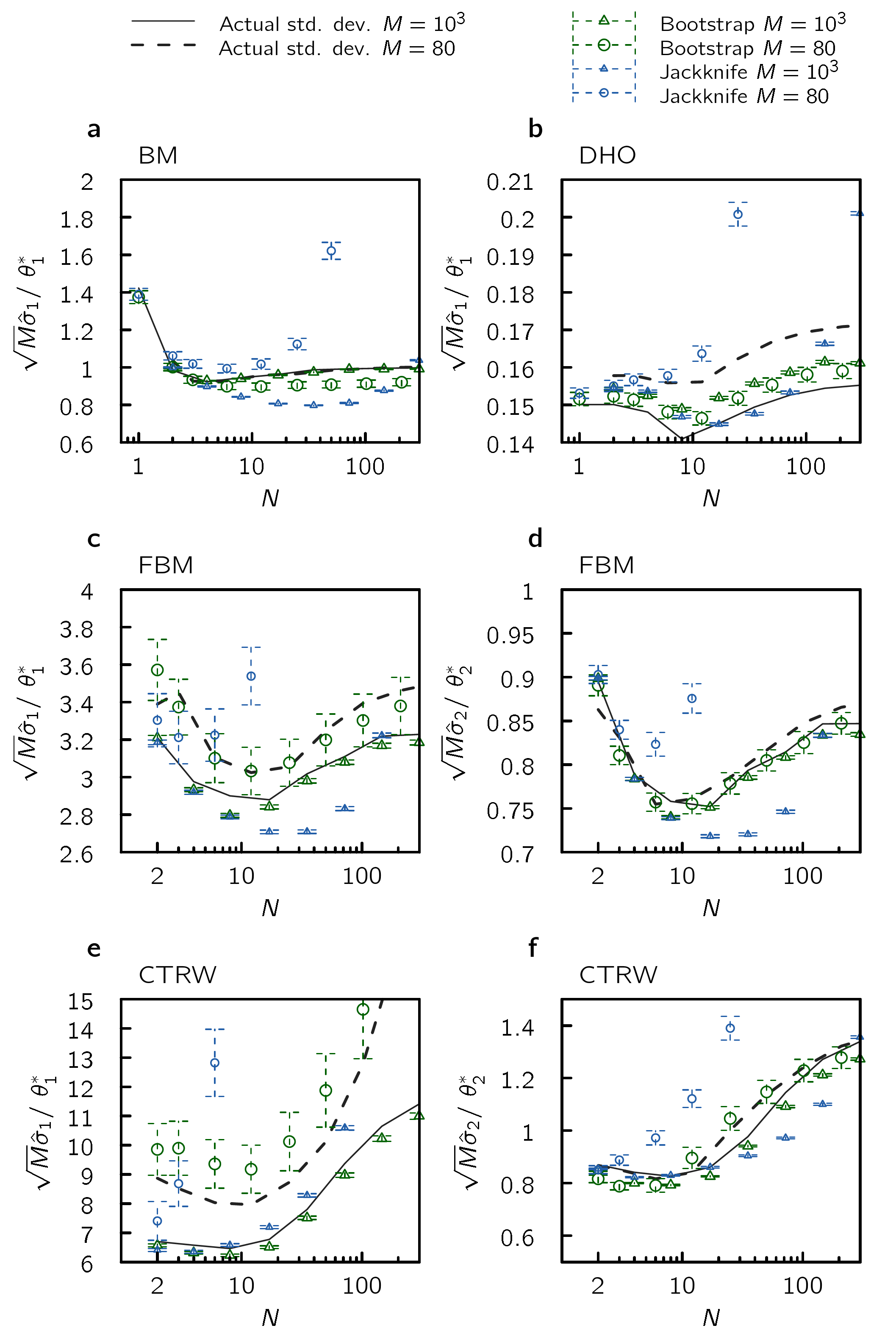}
  \caption{\textbf{Error estimation using bootstrap resampling and jackknife error estimation.} Standard
    deviation for the parameter fits as a function of
    the number of sampling points,~$N$, used in the
    fitting procedure.
    Each method is applied to $S=500$ realizations of data from
    (\textbf{a}) Brownian motion (BM),
     (\textbf{b}) damped harmonic oscillation (DHO),
    (\textbf{c}--\textbf{d}) fractional Brownian motion (FBM), and
    (\textbf{e}--\textbf{f}) continuous time random walk (CTRW).
    The associated standard deviation in parameter estimates serve as "actual"
    standard deviation. These actual values are compared to estimates using
    bootstrap resampling and jackknife error estimation procedures, see
    Supplementary Methods, Sec. \ref{sec:resampling}. We see that the
    bootstrap method gives rather reliable error estimates which are similar
    to that of the WLS-ICE procedure, compare to \figSigma{}.
    However, note that the bootstrap method is associated with a substantially
    larger computational time compared to the WLS-ICE. The jackknife error
    estimation performs worse than bootstrap resampling in general. For the
    jackknife error estimation, we used $100$ groups. For the
    bootstrap results, trajectories were resampled with replacement and the
    $\chi^2$ minimization performed 100 times. For simulation parameters, see
    Supplementary Methods section~\ref{sec:simulation_parameters}.}
  \label{fig:resample}
\end{figure}

\begin{figure}[!htp]
  \centering
  \includegraphics{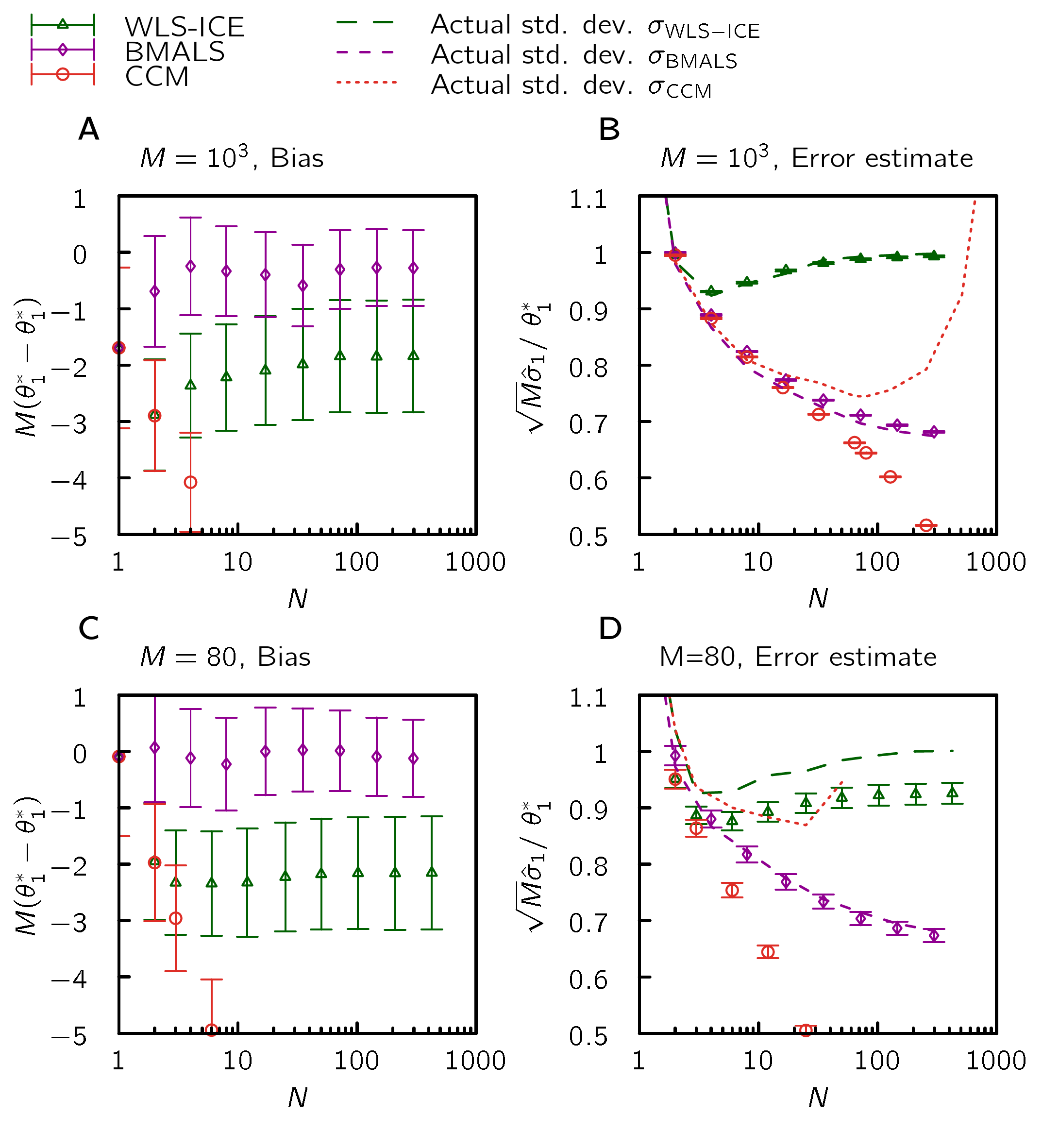}
  \caption{\textbf{Bias and variance of Brownian motion adapted least squares
      (BMALS) compared to the WLS-ICE and CCM methods.} We
    show the bias in parameter fit (left panels) and their variance compared to
    estimates from fitting procedure (right panels), as a function of the number
    of sampling times,~$N$. The MSD based on two different data sizes,~$M$
    (number of trajectories)
    was considered:
    (\textbf{a},\textbf{b})~$M=10^3$ and
    (\textbf{c},\textbf{d})~$M=80$; averaged over $S=500$ realizations.
    Notice the lower variance in BMALS as compared to WLS-ICE, and that as $M$
    is increased the CCM variance approach the variance for BMALS. The BMALS
    is a hybrid between  model matching and function fitting procedures as it requires the true covariance matrix as input. Error bars
    show standard errors of the mean. For simulation parameters, see Supplementary Methods
    section~\ref{sec:simulation_parameters}.
  }
  \label{fig:bma}
\end{figure}

\clearpage

\newpage


\appendix

\newpage

\section*{\Large Supplementary Tables}

\setcounter{table}{0}
\renewcommand{\tablename}{Supplementary Table}

\begin{table}[ht]
\centering
\begin{tabular}{l|l|l|l|l|l|l|l}
\textbf{Description} & & \multicolumn{2}{c|}{Low density} & \multicolumn{2}{c|}{Medium density} & \multicolumn{2}{c}{High density} \\
\textbf{Video} & & \multicolumn{2}{c|}{S5} &  \multicolumn{2}{c|}{S1}  & \multicolumn{2}{c}{S6} \\
\textbf{Number of trajectories} & & \multicolumn{2}{c|}{$M=310$} & \multicolumn{2}{c|}{$M=16$} & \multicolumn{2}{c}{$M=5$} \\
\hline
\textbf{Method} & Observable & \multicolumn{2}{c|}{} & \multicolumn{2}{c|}{} & \multicolumn{2}{c}{} \\
\hline
  WLS-ICE & $\estimated{\param}_1$ & 8.46 & 8.62 & 10.30 & 7.70 & 6.75 & 5.22 \\
       & $\estimated{\sigma}$ &  0.38 & 0.38 & 1.88 & 1.64 & 2.56 & 1.93\\
\hline
WLS-ECE & $\estimated{\param}_1$  & 8.46 & 8.62 & 10.30 & 7.70 & 6.75 & 5.22 \\
 & $\estimated{\sigma}$ & 0.20 & 0.19 & 1.14 & 0.86 & 1.61 & 1.04 \\
\hline
CCM & $\estimated{\param}_1$ &  8.53 &  8.27 &  9.25 &  3.84  & ill-cond. & ill-cond. \\
 & $\estimated{\sigma}$ & 0.36 & 0.35 &   1.01 & 1.17  & ill-cond. & ill-cond. \\
\end{tabular}
\caption{ \textbf{Results of the three fitting method for ``real world''
    particle tracking data, without jackknife}. Results
  shown are for the same data as in \tabRealWorldData{}, but
  here before the jackknife procedures were applied. Comparing to
  \tabRealWorldData{} we see that biases is rather large for video
  S6 (few trajectories) but minor for video S5 (large number of
  trajectories). }
\label{tab:real_world_data_no_jack}
\end{table}

\begin{table}[h]
\centering
\begin{tabular}{|l|l|l|}
\hline
\textbf{Abbreviation} & & \textbf{Comment} \\
\hline
WLS-ICE &  weighted least squares including &  \\
&  correlations in error estimation& new method \\
WLS-ECE & weighted least squares excluding & \\
&  correlations in error estimation & old method \\
CCM & correlated chi-square method & old method \\
\hline
BM & Brownian motion & zero-mean process without memory \\
DHO & damped harmonic oscillation & process with a time-dependent mean \\
FBM & fractional Brownian motion & zero-mean process with memory\\
CTRW & continuous time random walk & zero-mean, ageing process \\
\hline
\end{tabular}
\caption{ \textbf{List of abbreviations}.  }\label{tab:abbreviations}
\end{table}

\newpage

\section*{\Large Supplementary Methods}

\tableofcontents

\renewcommand\theequation{S\arabic{equation}}
\setcounter{equation}{0}

\vspace*{1cm}
 In this Supplementary Methods, details of the derivations,
  simulations and methods are provided. For convenience, Table
  \ref{tab:abbreviations} lists all abbreviations used herein.

\section{Weighted Least Squares Including Correlations in Error estimation (WLS-ICE)}
\label{sec:newmethod}

We here describe our new fitting procedure, the WLS-ICE method, in detail.
As demonstrated in the main text, the  previous standard methods for fitting of ensemble averages, the WLS-ECE or CCM procedures
(section~\ref{sec:review_fitting_procedures}), are of limited general applicability for fitting of correlated data:
 the WLS-ECE method assumes data points are independent resulting in flawed error estimation, whereas the CCM method
 (involving inversion of a noisy sample covariance matrix) provides ill-conditioned results or
 strong bias in the parameter estimation.
We here formulate the problem at hand as a
minimization of a ``cost function'', $\chi^2$, which can be chosen rather
general. Minimizing this  cost function provides an estimate, $\estimated{\params}$, for the model parameters of interest. However, unlike the WLS-ECE fitting procedure, where fluctuations around mean values are assumed
to be independent, we use the full multivariate probability density function
for the mean values, eq.~\eqref{eq:rho_wlsice} (which is Gaussian due to the
multivariate central limit theorem), when estimating the standard error and
covariance in the fitted parameters. This provides a mathematically rigorous
way of avoiding the problems with previous fitting methods.

\subsection{Parameter estimation}
\label{sec:new_param_estimation}

The cost function used herein is a $\chi^2$ functional
(\eqChiTwo{}) on the form:
\begin{equation}
  \label{eq:chi2_vectorform}
  \chi^2 = (\bm{f} - \samplem{y} )^T \bm{R}\ (\bm{f} - \samplem{y}),
\end{equation}
where $\samplem{y} = (\sample{y}_1,\ldots, \sample{y}_{N})$,
$\bm{f} = (f_1,\ldots,f_N)$, $f_i=f(T_i;\params)$ with sampling
times $T_i$ ($i=1,\ldots,N$) and where $(\ldots)^T$ denote transpose.  We find
the best parameters $\estimated{\params}$ by minimizing $\chi^2$, i.e., by solving:
\begin{equation}
  \label{eq:chi2firstderiv}
  \left.\frac{\partial \chi^2}{\partial \param_a}\right|_{\params =
    \estimated{\params}}  = 0 = 2 \sum_{i,j} \left.\frac{\partial
    f_i(\params)}{\partial \param_a}\right|_{\params = \estimated{\params}}  R_{ij}(f_j(\estimated{\params})-\sample{y}_j),
\end{equation}
where $a=1,\ldots,K$.
As in the main text, a 'bar' denotes a sample estimator, a 'hat' denotes
parameters obtained through $\chi^2$ minimization, and a 'star' is used to
denote the true value of a parameter. For a linear fit function,
$f_i(\param_1)=\param_1 T_i $, eq. \eqref{eq:chi2firstderiv} can be
solved analytically:
\begin{equation}
    \label{eq:wls_linear}
    \estimated{\param}_1 = \frac{\samplem{y}^T \bm{R}\bm{T}}{\bm{T}^T\bm{R}\bm{T}}.\\
 \end{equation}
 Note that the positive definite symmetric matrix $\bm{R}$ in eq. \eqref{eq:chi2_vectorform}
could potentially be custom made for particular applications. In the main text the observables $\sample{y}_i$ are mean positions or
mean square displacements at different sampling times,~$T_i$. We note, however,
that our WLS-ICE procedure is valid for any type of ensemble averaged observables
(the matrices $\samplem{C}$ and $\samplem{Q}$ below are then the covariance
matrix for those particular observables).

For the matrix~$\bm{R}$, we consider three main choices:
\begin{description}
\item[1. Correlated Chi-Square Method (CCM):] Here we make use of the full covariance
  matrix, (see section~\ref{sec:corr-chi2-fitt}):
  \begin{equation}
    \label{eq:bias_4}
    \bm{R}=\samplem{R}^{[CCM]} = \samplem{C}^{-1},
  \end{equation}
  where $\samplem{C}$ is the covariance matrix of the mean,
  $\samplem{C}=\samplem{Q}/M$, as defined in \eqC{}.
\item[2. Weighted least squares (WLS):] Here we only
  make use of the diagonal elements,
  \begin{equation}
    \label{eq:bias_6}
    R_{ij}=\sample{R}_{ij}^{[WLS]} = \delta_{i,j} /\sample{C}_{ii},
  \end{equation}
 where $\delta_{i,j}$ is the Kronecker delta-function.
\item[3. Brownian motion adapted least squares (BMALS):] Finally we probe our fitting
  method by the following choice:
  \begin{equation}
    \label{eq:bias_8}
    \bm{R}=\samplem{R}^{[BMALS]} = \frac1M {\exactm{Q}_{\rm BM}}^{-1},
  \end{equation}
  where $\exactm{Q}_{\rm BM}$ is the exact covariance matrix
  for BM, see eq. \eqref{eq:bias_14}.  For comparison of the BMALS method to WLS-ICE, please see Supplementary Figure~\ref{fig:bma}.
\end{description}

\subsection{Error estimation}
\label{sec:new_error_estimation}
The covariance for the estimated parameters (i.e., the parameters $\estimated{\params}$ obtained
by solving eq.~\eqref{eq:chi2firstderiv}) is defined
\begin{equation}
  \label{eq:paramerror}
  \estimated{\Delta}_{ab} =
  \langle
  (\estimated{\param}_a - \exact{\param}_a)
  (\estimated{\param}_b - \exact{\param}_b)\rangle,
\end{equation}
where $\langle F(\samplem{y}) \rangle = \int  F(\samplem{y})
\rho(\samplem{y};\exact{\params})d\sample{y}_1d\sample{y}_2\cdots d\sample{y_N}  $ denotes an average
over the multivariate probability density, $\rho(\samplem{y};\exact{\params})$.
Due to the multivariate central limit theorem (note that $\samplem{y}$ is a sum of
$M$ identically distributed random numbers), for large $M$ this probability density is a multi-variate
Gaussian:
\begin{equation}
  \rho(\samplem{y};\exact{\params}) = Z^{-1} \exp\left( -\frac{1}{2}(\samplem{y} -
    \exactm{y})^T {\exactm{C}}^{-1} (\samplem{y} - \exactm{y})\right),\label{eq:rho_wlsice}
\end{equation}
with normalization constant
$Z=(2\pi)^{N/2}\sqrt{\det(\exactm{C})}$~\cite{kampen1992} and
$\exactm{C}=\exactm{Q}/M$, where $\exactm{Q}$ is the exact covariance matrix.

In order to derive an explicit expression for $\estimated{\Delta}_{ab}$ we follow
the lines of thought of Gottlieb \textit{et~al.}~\cite{gottlieb1988} and make a
first order Taylor series expansion of the
estimated parameter values in terms of deviations of the estimated
$\samplem{y}$ from their true values:
\begin{equation}
\label{eq:param_taylor}
\estimated{\param}_a - \exact{\param}_a = \left. \sum_k \frac{\partial
  \estimated{\param}_a}{\partial \sample{y}_k}\right|_{\samplem{y} = \exactm{y}}
\ (\sample{y}_k -\exact{y}_k) \ + \ \ordo [ (\sample{y}_k -\exact{y}_k) (\sample{y}_l -\exact{y}_l)].
\end{equation}
Substituting this expression
into eq.~\eqref{eq:paramerror} and using the definition of the covariance
matrix:
$\exact{C}_{kl} =  \langle (\sample{y}_k -\exact{y}_k)(\sample{y}_l
-\exact{y}_l)\rangle $ [this result follows from eq. \eqref{eq:rho_wlsice}] we find, to first order,
\begin{equation}
\label{eq:Delta_ab}
\estimated{\Delta}_{ab} = \sum_{k,l} \left.\frac{\partial \estimated{\param}_a}{\partial
  \sample{y}_k}\right|_{\samplem{y} = \exactm{y}}
 \exact{C}_{kl}
\left.\frac{\partial \estimated{\param}_b}{\partial
  \sample{y}_l}\right|_{\samplem{y} = \exactm{y}}.
\end{equation}
In order to obtain an explicit expression for $\partial
\estimated{\param}_a/\partial \sample{y}_k$ we differentiate eq.
\eqref{eq:chi2firstderiv} with respect to~$\sample{y}_k$. This yields
\begin{equation}
\label{eq:param_diff}
0  = \sum_b \estimated{h}_{ab} \frac{\partial \estimated{\param}_b}{\partial \sample{y}_k} - 2
\left.\sum_i \frac{\partial f_i(\params)}{\partial \param_a}\right|_{\params
=\estimated{\params}} R_{ik}
\end{equation}
where we introduced
\begin{equation}
    \estimated{h}_{ab} =  2 \left.\sum_{i,j} \frac{\partial^2 f_i(\params)}{\partial \param_a \partial
      \param_b }\right|_{\params = \estimated{\params}}
R_{ij} (f_j(\estimated{\params}) - \sample{y}_j) +
    2 \left. \sum_{i,j} \frac{\partial f_i(\params)}{\partial \param_a}\right|_{\params =\estimated{\params}}
    R_{ij}
    \left.\frac{\partial f_j(\params)}{\partial \param_b}\right|_{\params =\estimated{\params}}.
\end{equation}
Solving eq.~\eqref{eq:param_diff} we obtain:
\begin{equation}
 \frac{\partial \estimated{\param}_a}{\partial \sample{y}_k} =  2 \sum_i
 \sum_b (\bm{\estimated{h}}^{-1})_{ab}\left.\frac{\partial f_i(\params)}{\partial \param_b}\right|_{\params
=\estimated{\params}} R_{ik},
\end{equation}
which when substituted into eq.~\eqref{eq:Delta_ab} yields the following expression for the
covariance of the estimated parameter, $\estimated{\params}$:
\begin{equation}
\label{eq:delta_ab_final}
\estimated{\Delta}_{ab} = \left( 4 \sum_{c,d} \sum_{j,k,l,m} (\bm{\estimated{h}}^{-1})_{ac} \left.\frac{\partial
  f_j(\params)}{\partial \param_c}\right|_{\params=\estimated{\params}} R_{jk}
\exact{C}_{kl}R_{lm} \left.\frac{\partial
  f_m(\params)}{\partial \param_d}\right|_{\params=\estimated{\params}}
(\bm{\estimated{h}}^{-1})_{db} \right)_{\samplem{y} = \exactm{y}}.
\end{equation}
We finally replace all exact quantities above by the corresponding sample
estimators (and use $\bm{C}=\bm{Q}/M$), giving the key result, \eqGA{}. The replacement of exact ensemble averages by sample estimates introduces bias terms which, to first order, are
proportional to $1/M$, where $M$ is the number of trajectories, see section~\ref{sec:origin_of_bias}.
For WLS-ICE/WLS-ECE procedures, we find that the bias is in practice
often negligible (see main text). Just as the parameter estimates $\estimated{\param}_a$ are typically biased, so
will the quantity $\estimated{\phi}_{ab}$ in \eqGA{} also be, as it is a
nonlinear function of sample estimates, see section~\ref{sec:origin_of_bias}.
This bias can
be reduced using the jackknife procedure applied to $\estimated{\phi}_{ab}$ (see section~\ref{sec:jackknife}).

\section{Review of previous fitting procedures}
\label{sec:review_fitting_procedures}

In this section we investigate the two previous ubiquitous $\chi^2$ methods
for model fitting, namely WLS-ECE (uncorrelated~$\chi^2$) fitting and
CCM (correlated~$\chi^2$) fitting.

\subsection{WLS-ECE fitting}
\label{sec:uncorr-chi2-fitt}
The previous most common  method of functional fitting to data is the
``standard''  weighted least squares (WLS-ECE in the main text) method (uncorrelated $\chi^2$ fitting), which
is reviewed in this section. In this method, one assumes that all fluctuations
around mean values are uncorrelated.

\subsubsection{General fit functions}

In the WLS-ECE method one maximizes the probability for
the function $f(T_i;\params) = f_i(\params)$ to have a good fit to the data:
\begin{equation}
  \label{eq:P_uncorrelated}
  P(\samplem{y};\params) \propto \prod^{N}_{i=1} \exp\left(-\frac{1}{2} \frac{(\sample{y}_i - f_i(\params))^2}{\sample{\sigma}_i^2}\right).
\end{equation}
Note that this probability is a product over the observations, $\samplem{y}$,
hence the data is assumed to be statistically \emph{independent}. Within this
assumption, the unbiased estimator of variance of the mean is
\begin{equation}
  \label{eq:sigma}
  \sample{\sigma}^2_i = \frac1M \frac{1}{M-1} \sum_{m=1}^{M} (y_i^{(m)} - \sample{y}_i)^2.
\end{equation}
Maximizing the probability $P$ is equivalent to minimizing
\begin{equation}
  \label{eq:chi2_uncorrelated}
  \chi^2 = \sum^{N}_{i=1} \frac{(\sample{y}_i - f_i(\params))^2}{\sample{\sigma}_i^2},
\end{equation}
from which we get estimated parameters $\estimated{\params}$, by solving
\begin{equation}
  \label{eq:Dchi2_uncorrelated}
  \left.\frac{\partial \chi^2}{\partial \param_a}\right|_{\params = \estimated{\params}} = 0 = 2 \left.\sum_{i}\frac{\partial
    f_i(\params)}{\partial \param_a}\right|_{\params = \estimated{\params}} \frac{1}{\sample{\sigma}_i^2} (f_i(\estimated{\params})-\sample{y}_i).
\end{equation}
For $\chi^2$ close to the estimated parameter set $\estimated{\params}$ we have
the Taylor expansion
\begin{equation}
  \label{eq:chi2_taylor}
  \begin{split}
    \chi^2 =& \left.\chi^2\right|_{\estimated{\params}}
    + \sum_{a=1}^{K} (\param_a - \estimated{\param}_a) \left.\frac{\partial
        \chi^2}{\partial \param_a}\right|_{\params = \estimated{\params}}\\
    &+ \frac12 \sum_{a,b=1}^{K} (\param_a - \estimated{\param}_a)(\param_{b} - \estimated{\param}_b)
    \left.\frac{\partial^2
        \chi^2}{\partial \param_a \partial \param_{b}}\right|_{\params=\estimated{\params}},
  \end{split}
\end{equation}
which we can insert back into the expression for $P$, eq.~\eqref{eq:P_uncorrelated}, to yield
\begin{equation}
  \label{eq:P_correlated_optimal}
  P(\params) = W \exp\left( -\frac{1}{4} \sum_{a,b=1}^{K}
    \bm{\estimated{H}}_{ab}(\param_a - \estimated{\param}_a)(\param_{b} - \estimated{\param}_b)  \right),
\end{equation}
where $W$ is a normalization constant and
\begin{equation}
  \label{eq:H_lsm_covariance_matrix}
  \estimated{H}_{ab} = \left.\frac{\partial^2 \chi^2}{\partial \param_a \partial \param_{b}}\right|_{\params=\estimated{\params}}
\end{equation}
is the Hessian matrix, and we used $\partial
\chi^2/ \partial \param_a|_{\params = \estimated{\params}} =0$. From
eq.~\eqref{eq:P_correlated_optimal} we find that
\begin{equation}
  \label{eq:covWLS}
  \estimated{\Delta}_{ab}\equiv  \langle (\estimated{\param}_a - \exact{\param}_a)(\estimated{\param}_b -
  \exact{\param}_b)\rangle = 2(\bm{\estimated{H}})^{-1}_{ab},
\end{equation}
i.e., the inverse of the Hessian
matrix determines the covariances of the estimated parameters.

\subsubsection{Linear fit functions}
For the case that the fit function is linear, i.e., $f_i(\param_1) = \param_1 T_i $,
eq.~\eqref{eq:Dchi2_uncorrelated} can be solved analytically (Press
\textit{et~al.}~\cite{press2007}). The same can be done for the variance, $\sigma^2$,
in the estimated parameter. We have
\begin{subequations}
  \begin{align}
    \label{eq:lsm_mu}
    \estimated{\param}_1&= \frac{\sum_i \sample{y_i}T_i/\sample{\sigma}_i^2}{\sum_i T_i^2/\sample{\sigma}_i^2}\\
    \label{eq:lsm_sigma}
    \estimated{\sigma}^2 &= \estimated{\Delta}_{11} = \frac{1}{\sum_i T_i^2/\sample{\sigma}_i^2}.
  \end{align}
\end{subequations}

\subsection{CCM fitting}
\label{sec:corr-chi2-fitt}
In this section we review  CCM (correlated chi-square method)
fitting procedure~\cite{bos2007,gottlieb1988,seibert1994,michael1994}.

\subsubsection{General fit functions}
Where a WLS-ECE fit only makes use of the diagonal (variance) of the
covariance matrix, CCM makes use of the full matrix, defined as in
eq.~\eqref{eq:C_sample}, where the diagonal will be the square of the standard
error of the mean, $s_i^2=\sigma_i^2/M$. The task of fitting a function
$f(t_i;\params)$, reduces to
maximizing the probability which is taken as the multi-variate Gaussian:
\begin{equation}
  \label{eq:P_correlated}
  P(\samplem{y};\params) = Z^{-1} \exp\left( -\frac{1}{2}(\samplem{y} -
    \bm{f}(\params))^T \samplem{C}^{-1} (\samplem{y} - \bm{f}(\params))\right),
\end{equation}
where (for a good fit: $\exactm{y}\approx \bm{f}$)  $\samplem{C}=\samplem{Q}/M$ can be estimated through \eqC{},
and the normalization constant
$Z=(2\pi)^{N/2}\sqrt{\det(\samplem{C})}$~\cite{kampen1992},
$\samplem{y} = (\sample{y}_1,\ldots, \sample{y}_{N})$,
$\bm{f} = (f_1,\ldots,f_N)$, with $f_i=f(T_i;\params)$, and $(\ldots)^T$ denotes transpose.
For uncorrelated data the covariance matrix estimator, $\samplem{C}$, will be diagonal and
eq.~\eqref{eq:P_correlated} reduces to eq.~\eqref{eq:P_uncorrelated}, and the
WLS-ECE method is attained.

As for WLS-ECE, maximizing $P$ is equivalent to minimizing the cost function
\begin{equation}
  \label{eq:chi2_correlated_vectorform}
  \chi^2 = (\samplem{y} - \bm{f}(\params))^T \samplem{C}^{-1}(\samplem{y} - \bm{f}(\params)).
\end{equation}
Thus, we get our estimated parameters $\estimated{\param}_a$ ($a=1,\ldots, K$) by solving:
\begin{align}
\label{eq:Dchi-Dp}
\begin{split}
  \left.\frac{1}{2}\frac{\partial \chi^2}{\partial \param_a}\right|_{\params = \estimated{\params}} = 0 =& -
  \left.\frac{1}{2}\frac{\partial \bm{f}}{\partial \param_a}\right|_{\params = \estimated{\params}} \samplem{C}^{-1} (\samplem{y} -
  \bm{f}(\estimated{\params})) + (\samplem{y} - \bm{f}(\estimated{\params})) \samplem{C}^{-1}
  \left(\left.-\frac{1}{2}\frac{\partial
      \bm{f}}{\partial \param_a}\right|_{\params = \estimated{\params}}\right) \\
  =& \left.\frac{\partial \bm{f}}{\partial \param_a}\right|_{\params = \estimated{\params}} \samplem{C}^{-1} (\bm{f}(\estimated{\params}) - \samplem{y}),
\end{split}
\end{align}
where in the last step we used the symmetry property of~$\samplem{C}$, i.e., that
$\sample{C}_{ij} = \sample{C}_{ji}$.

The derivation of the covariance, $\Delta_{ab}$, of the CCM estimated
parameters, $\estimated{\param}_a$ follows along identical lines as for WLS-ECE (previous
section). Hence, $\Delta_{ab}$ is given by eq.~\eqref{eq:covWLS} where $\estimated{\param}_a$ is now obtained by
solving eq.~\eqref{eq:Dchi-Dp} (instead of solving
eq.~\eqref{eq:Dchi2_uncorrelated} as for WLS).

We finally note that the CCM is a maximum likelihood estimation procedure "asymptotically". More precisely, if $M$ is large enough so that $\samplem{y}$s are  Gaussian by  the multi-variate central limit theorem, if the fit is "good" in the sense that $\exactm{y}\approx \bm{f}$, and if the errors on the estimated elements of the covariance matrix are negligible, then the CCM is a maximum likelihood estimation method.

\subsubsection{Linear fit functions}
\label{sec:analytical_lomholt_linear}
For fitting a \textit{linear} function, $f_i(\params)=\param_1 T_i$, to data one can determine the minimum
of the CCM $\chi^2$ function, eq.~\eqref{eq:chi2_correlated_vectorform}, analytically.
In particular, such a fit function is of relevance for BM
 (section~\ref{sec:brownian-motion}). Eq.~\eqref{eq:Dchi-Dp} becomes
\begin{equation}
  0 = \frac{1}{2}\left.\frac{\partial \chi^2}{\partial \param_1}\right|_{\param_1 = \param_1^*}
  = (\samplem{y} - \param_1^*\ \bm{T})^T \samplem{C}^{-1}\bm{T}.
\end{equation}
Taking the second derivative we get
\begin{equation}
  \left.\frac{\partial^2 \chi^2}{\partial \param_1^2}\right|_{\param_1=\param_1^*} = - \bm{T}^T\samplem{C}^{-1}\bm{T}.
\end{equation}
From these results, as well as using eq.~\eqref{eq:H_lsm_covariance_matrix} and
eq.~\eqref{eq:covWLS}, we get the estimated value for the parameter $\param_1$ and its variance $\sigma^2$ as
\begin{subequations}
  \begin{align}
    \label{eq:lomholt_mu}
    \estimated{\param}_1 &= \frac{\samplem{y}^T \samplem{C}^{-1}\bm{T}}{\bm{T}^T\samplem{C}^{-1}\bm{T}}\\
    \label{eq:lomholt_sigma}
    \estimated{\sigma}^2 &= \estimated{\Delta}_{11} = \frac{1}{\bm{T}^T \samplem{C}^{-1}\bm{T}}.
  \end{align}
\end{subequations}

\section{Prototypical model systems}
\label{sec:prototypical_systems}

In the main text we provide results for different parameter estimation
procedures. As prototype systems we use four processes where the true
parameter values are known, namely: (i) Brownian motion (BM), (ii) damped
harmonic oscillation (DHO), (iii) fractional Brownian motion (FBM), and (iv)
continuous time random walks (CTRW). For BM and CTRW in $d$ spatial
dimensions, steps in different directions are independent. Therefore, without
loss of generality, all simulations are here performed in one dimension,
$d=1$, for these systems. Also, for consistency, we use $d=1$ in our FBM
simulations.

\subsection{Brownian motion}
\label{sec:brownian-motion}
Our first example is a simple BM, which can be used to describe, e.g., single particle diffusion
in one dimension. The mean square displacement~(MSD) at time~$t$, for
dimension~$d$, and diffusion constant~$D$, is
\begin{equation}
  \label{eq:brown_diffusion_equation}
  \langle (\bm{x}(t)-\bm{x}(0))^2 \rangle = \langle y(t) \rangle = \param t,
\end{equation}
where
\begin{equation}
\param = 2dD
\end{equation}
and
\begin{equation}
y(t) = [\bm{x}(t)-\bm{x}(0)]^2.
\end{equation}
In all simulations in the main text we use one-dimensional simulations, i.e., $d=1$.

In one-dimensional BM, the full covariance matrix for the
displacement is known~\cite{chaichian2001}. Choosing $x(0)=0$ and
discretizing time into process times $t_i = i\epsilon$ ($i=1, \ldots, N$), with time
step $\epsilon$, we have
\begin{equation}
  \label{eq:bias_11}
  \exact{V}_{ij} = \langle (x_i -\langle x_i\rangle) (x_j -\langle x_j\rangle)\rangle =  2D\min(t_i,t_j),
\end{equation}
where $x_i = x(t_i)$ and $D$ is the diffusion constant. On matrix form:
\begin{equation}
  \exactm{V} = 2 D \epsilon
  \begin{pmatrix}
    1      &  1     & 1      &  \ldots & 1      \\
    1      &  2     & 2      &         & 2      \\
    1      &  2     & 3      &         & 3      \\
    \vdots &        &        &  \ddots &        \\
    1      &  2     & 3      &         &  N     \\
  \end{pmatrix}.
\end{equation}
Of interest here is also the covariance matrix for the square displacements:
\begin{equation}
  \exact{Q}_{ij} = \langle (y_i -\langle y_i\rangle) (y_j -\langle y_j\rangle)\rangle.
\end{equation}
Using Wick's (Isserlis') theorem for zero-mean processes, we can
calculate any moment of a multivariate Gaussian according to
\begin{equation}
  \label{eq:bias_30}
  \langle x_1x_2 \cdots x_{2n}\rangle  = \sum \prod \langle x_ix_j\rangle ,
\end{equation}
where the sum is over all distinct ways of partitioning~$x_1 \ldots, x_{2n}$
into pairs~$x_ix_j$. Using eq.~\eqref{eq:bias_30} we have the following relation
between~$\exactm{Q}$ and~$\exactm{V}$:
\begin{equation}
  \label{eq:bias_13}
  \exact{Q}_{ij} = 2 (\exact{V}_{ij})^2.
\end{equation}
On matrix form:
\begin{equation}
  \label{eq:bias_14}
  \exactm{Q} = 8 (D \epsilon)^2
  \begin{pmatrix}
    1      &  1     & 1      & \ldots & 1      \\
    1      &  4     & 4      &        & 4      \\
    1      &  4     & 9      &        & 9      \\
    \vdots &        &        & \ddots &        \\
    1      &  4     & 9      &        &  N^2   \\
  \end{pmatrix}.
\end{equation}
The standard unbiased sample estimator of $\bm{Q}$ is
\begin{equation}
  \label{eq:C_sample}
  \sample{Q}_{ij} = \frac{1}{M-1}
    \sum_m (y_i^{(m)} -\sample{y}_i) (y_j^{(m)} - \sample{y}_j).
\end{equation}
where $m$ labels trajectories, see main text.

For BM, the inverse of the $\exactm{Q}$ matrix is a tridiagonal
matrix with column sum of zero, except the first. Explicitly
\begin{equation}
  \label{eq:bias_15}
  {\exactm{Q}}^{-1} = \frac{1}{8(D\epsilon)^2}
  \begin{pmatrix}
    1+\frac13 & -\frac13          &     0              &   \ldots &          & \\
    -\frac13  & \frac13 + \frac15 & -\frac15           &   0      &          & \\
    0         & -\frac15          &  \frac15 + \frac17 & -\frac17 &   0      & \\
    \vdots    &  0                & -\frac17           &   \ddots &   \ddots & \\
              &                   &  0                 &          &          & -\frac{1}{2N-1} \\
              &                   &                    &          &          &  \frac{1}{2N-1} \\
  \end{pmatrix},
\end{equation}
which can be written as
\begin{equation}
  \label{eq:bias_16}
  ({\exactm{Q}}^{-1})_{ij} = \frac{1}{8(D\epsilon)^2}
  \left[
    \left( \frac{1}{2i-1} + \frac{(1-\delta_{i,N})}{2i+1} \right)\delta_{i,j} -
    \left( \frac{1}{2i+1} \right)\delta_{i,j-1} -
    \left( \frac{1}{2i-1} \right)\delta_{i,j+1}
  \right].
\end{equation}
where $\delta_{i,j}$ is the Kronecker delta-function ($\delta_{i,j}=1$, if~$i=j$;
 $\delta_{i,j}=0$, if~$i\neq j$).
It is straightforward to show that indeed the matrix above satisfies $
({\exactm{Q}}^{-1})\cdot \exactm{Q} = \bm{I}$, where $\bm{I}$ is the identity
matrix. Note that the results above for ${\exactm{Q}}^{-1}$ assumes
  that the time of the first sampling time is equal to the distance between
  subsequent sampling times. In general, this choice of
  sampling times may not be optimal. In such situations one can
  evaluate ${\exactm{Q}}^{-1}$ using numerical inversion of $\exactm{Q}$
given in  Eqs. \eqref{eq:bias_13} and \eqref{eq:bias_11}.

\subsection{Damped Harmonic Oscillation in a heat bath (DHO)}
\label{sec:harmonic-oscillation}

Following N{\o}rrelykke and Flyvbjerg\cite{norrelykke2011} we consider the
dynamics of a damped harmonic oscillation in a heat bath (DHO). Physically, this process corresponds to the motion of a particle in a harmonic potential (i.e., the particle experiences a restoring force proportional to the displacement from the botttom of the potential) in a viscous liquid. Besides exerting friction on the particle, the molecules in the viscous liquid act as a noise source by providing thermal kicks on the particle. The equation of motion is:
\begin{equation}\label{eq:DHO}
 m \frac{d^2x(t)}{dt^2} + \gamma \frac{dx(t)}{dt} + \kappa x(t) = F_{\rm therm}(t),
\end{equation}
where $x(t)$ is the particle position at time $t$, $m$ is the mass, $\gamma$ is the friction constant, $\kappa$ is the spring constant and $F_{\rm therm} = (2k_BT \gamma)^{1/2} \eta(t)$ is the thermal noise, which is assumed to be zero mean Gaussian and delta-correlated, i.e.,
 \begin{equation}
 \langle \eta(t) \rangle = 0
 \end{equation}
 and
 \begin{equation}
 \langle \eta(t) \eta(t')\rangle = \delta(t-t').
 \end{equation}
Above, $k_B$ is the Boltzmann constant, $T$ is the temperature of the heat bath and $\delta(z)$ is the Dirac delta-function. The equation of motion is completed by initial conditions for the position and velocity. We restrict ourself to
\begin{align}
  x(t=0) &= x_0 \label{eq:DHO_init1} \\
  v(t=0) &= \left.\frac{dx(t)}{dt}\right|_{t=0} = 0 \label{eq:DHO_init2},
\end{align}
i.e., the particle is at the initial time displaced by a distance $x_0$ from
its equilibrium position and then let go without imposing any initial velocity
(no external pushing or pulling).

Based on eq.~\eqref{eq:DHO} it is straightforward to derive an expression
for the expected position, $\langle x(t)\rangle$, at time $t$. By taking the
ensemble average of eq.~\eqref{eq:DHO} and then making the ansatz:
$\langle x(t)\rangle  = \exp(i\Omega t) $ we arrive at a second order
algebraic equation for $\Omega$ with two solutions:
\begin{equation}
\Omega_\pm = \frac{i}{2\tau} + \sqrt{\omega^2},
\end{equation}
where
\begin{align}
  \tau &= \frac{m}{\gamma},\\
  \omega^2 &= \omega_0^2 - \frac{1}{4\tau^2}
\end{align}
and
\begin{equation}
\omega_0 = \sqrt{\frac{\kappa}{m}}.
\end{equation}
Thus, for the case $\omega^2<0$ the solution for $\langle x(t)\rangle$ is an exponentially damped function. For the case $\omega^2 > 0$, the solution is a complex valued exponential which can be written in terms of real-valued exponentials multiplied by sinus and cosinus functions. Also incorporating the initial conditions used here, eqs.~\eqref{eq:DHO_init1} and \eqref{eq:DHO_init2}, we find the solution for the mean to be
\begin{equation}\label{eq:DHO_mean}
\langle x(t) \rangle = x_0 \left( \cos(\omega t) + \frac{\param_1}{\omega} \sin(\omega t) \right) \exp(-\param_1 t),
\end{equation}
with
\begin{equation}
\param_1 = \frac{1}{2\tau}.
\end{equation}
The case when $\omega = 0$ (i.e., $\omega_0 = 1/(2\tau)$) is referred to as critical damping. For this case we can obtain the solution from eq.~\eqref{eq:DHO_mean} by taking the limit of $\omega\to 0$ to find
\begin{equation}
  \langle x(t) \rangle = x_0 \left( 1 + \param_1 t \right) \exp(-\param_1 t).
\end{equation}
The case of critical damping is used in the simulations in the main text, where $\param_1$ is used as a fitting parameter.

Using the full stochastic eq.~\eqref{eq:DHO}, we can also derive an explicit expression for the covariance matrix $\exact{C}(t,\tilde{t}) = \langle [x(t) - \langle x(t)\rangle ] [x(\tilde{t})-\langle x(\tilde{t})\rangle ]\rangle $. For simplicity we limit ourselves to the case $\omega^2\ge 0$. We start by rewriting eq.~\eqref{eq:DHO} as a set of two coupled first order equations\cite{norrelykke2011}

\begin{equation}\label{eq:first_order_ODE1}
  \frac{d}{dt} \begin{pmatrix}x(t)\\v(t)\end{pmatrix}
  = -\bm{M} \begin{pmatrix}x(t)\\v(t)\end{pmatrix}
  + \begin{pmatrix}0\\ \frac{\sqrt{2D}}{\tau}\eta(t)\end{pmatrix},
\end{equation}
with $D = k_B T/\gamma$ being the particle diffusion constant and
\begin{equation}
  \bm{M} =
  \begin{pmatrix}
    0        & -1             \\
    \omega_0^2 & \frac{1}{\tau} \\
  \end{pmatrix},
\end{equation}
which has the formal solution
\begin{equation}
  \begin{pmatrix}x(t)\\v(t)\end{pmatrix}
  = \begin{pmatrix} \langle x(t)\rangle \\ \langle v(t)\rangle \end{pmatrix}
  + \frac{\sqrt{2D}}{\tau} \int_0^t
  \exp(-\bm{M}) (t-t') \begin{pmatrix}0\\ \eta(t')\end{pmatrix}  dt',
\end{equation}
where
\begin{equation}
  \begin{pmatrix}\langle x(t)\rangle \\ \langle v(t)\rangle \end{pmatrix}
  = \exp(-\bm{M}t) \begin{pmatrix}x_0\\v_0\end{pmatrix}
\end{equation}
is the solution to the mean of eq. \eqref{eq:first_order_ODE1}  (using $\langle \eta(t)\rangle = 0$).
The covariance matrix now becomes:
\begin{equation}
  \exact{C}(t,\tilde{t}) = \frac{2D}{\tau^2} \int_0^t dt' \int_0^{\tilde{t}} dt'' \left(\exp(-\bm{M}(t-t') ) \right)_{12} \left( \exp(-\bm{M}(\tilde{t}-t'')) \right)_{12} \langle \eta(t')\eta(t'')\rangle.
\end{equation}
Without loss of generality, we assume that $t < \tilde{t}$, and carry out the
integral over $t''$ above to find:
\begin{equation}\label{eq:C_DHO}
  \exact{C}(t,\tilde{t}) = \frac{2D}{\tau^2} \int_0^t dt' \left(\exp(-\bm{M}(t-t'))\right)_{12} \left(\exp(-\bm{M}(\tilde{t}-t'))\right)_{12}.
\end{equation}
Using for $\omega^2>0$ the explicit form for the matrix exponential above as provided by N{\o}rrelykke \textit{et~al.}\cite{norrelykke2011}
\begin{equation}
\exp(-\bm{M} t) = \exp(-\param_1 t) [\cos(\omega t)\bm{I} + \sin(\omega t) \bm{J}],
\end{equation}
with $\bm{I}$ the 2 by 2 identity matrix and
\begin{equation}
  \bm{J} =
  \begin{pmatrix}
    \frac{\param_1}{\omega}       & \frac{1}{\omega}           \\
    -\frac{\omega_0^2}{\omega} & -\frac{\param_1}{\omega} \\
  \end{pmatrix},
\end{equation}
eq. \eqref{eq:C_DHO} becomes:
\begin{equation}
  \exact{C}(t,\tilde{t}) = \frac{8D \param_1^2}{\omega^2 } \int_0^t dt' \exp(-\param_1(t-t')) \exp(-\param_1(\tilde{t}-t'))\sin [\omega (t-t')]\sin [\omega (\tilde{t}-t')].
\end{equation}
Carrying out the integral above we arrive at our final expression for the covariance matrix for DHO:
\begin{eqnarray}
  \exact{C}(t,\tilde{t}) & =&  \frac{2D\param_1}{\omega(\param_1^2+\omega^2)} \exp(-\param_1 |\tilde{t}-t| ) \left\{ \omega\cos [\omega (\tilde{t}-t)] + \param_1 \sin [\omega |\tilde{t}-t|]\right\} \nonumber \\
  && + \frac{2D\param_1^2}{\omega^2 } \exp(-\param_1 (\tilde{t}+t) ) \left(
    \frac{\param_1}{\param_1^2+\omega^2} \cos [\omega (\tilde{t}+t)]  -
    \frac{1}{\param_1} \cos [\omega (\tilde{t} - t)] -
    \frac{\omega}{\param_1^2+\omega^2} \sin [\omega (\tilde{t}+ t)]
  \right).\nonumber \\
\end{eqnarray}
In the limit $\omega \rightarrow 0$ (critical damping) we have
\begin{equation}
  \exact{C}(t,\tilde{t})  =  \frac{2D }{\param_1} \left( \exp(-\param_1 (\tilde{t}-t) ) \left( 1 + \param_1 (\tilde{t}-t)\right) - \exp(-\param_1 (\tilde{t}+t) ) \left( 1+\param_1 (\tilde{t}+t)  + 2\param_1^2 \tilde{t}t \right) \right).
\end{equation}
To arrive at this result we made a Taylor series expansion to second order in $\omega^2$ of the general expression.

We refrain from attempting to obtain an analytic expression for
the inverse covariance matrix for DHO, as it appears a daunting task beyond
the scope of the current study.

\subsection{Fractional Brownian motion}
\label{sec:fbm}
Our third example is the case of one-dimensional FBM,
which is a zero mean Gaussian process with autocorrelation function,~\cite{qian2003}
\begin{equation}
  \label{eq:fbm_autocorr}
v_{ij} =  \langle x(t_i) x(t_j) \rangle = c (t_i^{2H} + t_j^{2H} - |t_i-t_j|^{2H}),
\end{equation}
at discrete times $t_i = i\epsilon$ and where the parameter~$H$ denotes the
Hurst parameter~\cite{mandelbrot1968}.  For $H=1/2$, FBM becomes standard BM. Indeed, if we set $H=1/2$ in
eq.~\eqref{eq:fbm_autocorr} we find that $v_{ij} = c [(t_i + t_j) - |t_i-t_j|]
= 2c \min(t_i,t_j)$ which is identical to eq.~\eqref{eq:bias_11} if we choose
$c=D$. The inverse covariance matrix of eq.~\eqref{eq:fbm_autocorr} is
(currently) not known analytically.

From eq.~\eqref{eq:fbm_autocorr} we get the MSD, for $t_i=t_j$, as ($x(0)=0$)
\begin{equation}
  \label{eq:fbm_model}
  \langle x^2(t) \rangle = \param_1 t^{\param_2},
\end{equation}
where $\param_1 = 2c$ and $\param_2 = 2H$, i.e., the MSD
has, for~$H<1/2$, a sublinear (or superlinear, if $H>1/2$) dependence on
time,~$t$.

\subsection{Continuous time random walk (CTRW) }
\label{sec:ctrw}

Our last example uses CTRW in one dimension.
 Such a process
is defined through a waiting time density $\psi(\tau)$, and a jump length
probability density, $\zeta(\ell)$~\cite{metzler2000}. In our case we choose
\begin{equation}
\label{eq:ctrw_power_law}
\psi(\tau) = \frac{\alpha}{\tau^*} (1 + \tau/\tau^*)^{-1-\alpha}
\end{equation}
 with $0 < \alpha < 1$ so that we have infinite average
waiting time~$\langle \tau \rangle$.
The jump length probability density is  chosen to be a Gaussian:
\begin{equation}
  \label{eq:phi_ctrw}
  \zeta(\ell) = \frac{1}{\sqrt{2\pi a^2}} \exp \left( -\frac{\ell^2}{2a^2} \right)
\end{equation}
with a variance $a^2$. For such a process, the MSD follows (for long times)~\cite{metzler2000}: 
\begin{equation}
  \label{eq:ctrw_model}
  \langle x(t)^2 \rangle = \param_1 t^{\param_2}
\end{equation}
(with $x(0)=0$) where
\begin{equation}
\param_1 = \frac{2}{\Gamma(1+\alpha)\Gamma(1-\alpha)}\frac{a^2}{2(\tau^*)^{\alpha} },
\end{equation}
and
\begin{equation}
  \param_2 = \alpha.
\end{equation}

\section{Simulation procedures}
\label{sec:simulation}

In this section we provide details about the methods used to generate the data
for our prototypical example systems introduced in
section~\ref{sec:prototypical_systems}. Simulations
ran to a stop time~$t_{\rm stop}$. All simulation parameters are listed in Sec. \ref{sec:simulation_parameters}.

\subsection{Brownian motion (BM)}
\label{sec:simulation_bm}
BM in one dimension is simulated using random jump lengths drawn
from a normal distribution. In some detail, we start by taking the cumulative
sum of~$N$ random numbers from a Gaussian distribution with zero mean and
variance $a^2$, and square each element of the sum. Each step increments time
by~$\epsilon$. This is repeated~$M$ times and summed and averaged. In short,
the MSD was computed as:
\begin{equation}
  \sample{y}_i = \frac1M \sum_{m=1}^{M} \left[ \sum_{n=1}^i r_n^{(m)} \right]^2,
\end{equation}
where~$r_n^{(m)}$ is a random number drawn from a normal distribution,
associated with the length of the $n$th jump for trajectory~$m$. The
  diffusion constant for this type of process is $D=a^2/(2\epsilon)$.

\subsection{Damped harmonic oscillation (DHO)}
\label{sec:simulation_dho}

When simulating the harmonic oscillation in a heat bath, see eq.~\eqref{eq:DHO}, we follow the procedure described by  N{\o}rrelykke and Flyvbjerg\cite{norrelykke2011} (at critical damping, $\omega =0$).

\subsection{Fractional Brownian motion (FBM)}
\label{sec:simulation_fbm}
For FBM simulations we used an algorithm by Davies and
Harte~\cite{davies1987,chambers1995}.  When fitting the model in
eq.~\eqref{eq:fbm_model}, we include only time points $t \ge T_1$ since this model prediction for the MSD, as for CTRW (see section~\ref{sec:simulation_ctrw}),
is only valid for large simulation times.

\subsection{Continuous time random walk (CTRW)}
\label{sec:simulation_ctrw}
For generating the CTRW data we move a "particle" randomly with a step length drawn
from a Gaussian probability density, eq.~\eqref{eq:phi_ctrw}, at each time step and increment time with
a waiting time~$\tau$ from the power-law distribution in eq.~\eqref{eq:ctrw_power_law}. In more
detail: while the process time,
$t$, is smaller than the designated stop time we repeat the following procedure to generate one trajectory~$m$:
\begin{enumerate}
\item Draw a random waiting time, $\tau$, from the power-law in
  eq.~\eqref{eq:ctrw_power_law}.
\item Move the particle, by increasing the current displacement
  by a random number~$r$ drawn from a normal distribution.
\item Update the time~$t$ by~$\tau$.
\end{enumerate}
The procedure is repeated~$M$ times and averaged over, to yield the MSD.
 Since the prediction
in eq.~\eqref{eq:ctrw_model} is only valid for~$t \gg \tau^*$, for fitting
purposes, we include only time points $t \ge T_1$ in the $\chi^2$ expression,
eq. (\ref{eq:chi2_vectorform}), and in the associated
parameter covariance estimation formula, \eqGA{}.

\subsection{Simulation parameters}\label{sec:simulation_parameters}

Below we list the simulation parameters used in all simulations in the main text and for the Supplementary Figures. We also give values for the first sampling time, $T_1$, used in the fit procedure (some of the functional forms used for fitting are only valid for "large" times).

\begin{itemize}

\item \textbf{BM.} Time increment, $\epsilon=1$ (dimensionless). Step length variance, $a^2=1$ (dimensionless). Simulation stop time $t=10^4\epsilon$. First sampling time, $T_1=\epsilon $.

\item \textbf{DHO.} Spring constant $\kappa=1$ (dimensionless). Mass $m=1$ (dimensionless). Initial position, $x_0=1$ (dimensionless). Thermal energy, $k_B T = 10^{-2}$ (dimensionless). Simulation stop time, $t_{\rm stop}=20 \omega_0^{-1}$ (with $\omega_0 = \sqrt{\kappa/m}=1$). First sampling time, $T_1=\omega_0^{-1}$.

\item \textbf{FBM.} Hurst exponent, $H=1/4$, unless stated otherwise. Time increment, $\epsilon=1$ (dimensionless). Prefactor in covariance matrix, $c=1$ (dimensionless). Simulation stop time, $t_{\rm stop}=10^4\epsilon$. First sampling time, $T_1=200\epsilon$.

\item \textbf{CTRW.} Power-law exponent, $\alpha=0.5$. Step length variance, $a^2=1$ (dimensionless). Characteristic time scales $\tau^*=1$ (dimensionless). Simulation stop time, $t_{\rm stop}=10^8\tau^* $. First sampling time, $T_1=10^5\tau^* $.

\end{itemize}

\section{Bias effects in parameter estimation}
\label{sec:bias_effect}

In this section, we provide analytical expressions for the bias in parameter
(diffusion constant) estimation for BM. We find that for BM the
CCM method has a bias which increases strongly with the number
of sampling times,~$N$. In contrast, the WLS method provides a
(small) bias which is independent of~$N$ for large~$N$. To make notation
compact, we leave summations over repeated indices implicit (where no
confusion can occur) in this section.

\subsection{The origin of bias}
\label{sec:origin_of_bias}

In general the bias, i.e., the expected difference
between some observable based on sample estimates and the ``true'' value of
that observable, can be written as a series expansion in terms of~$1/M$,
where~$M$ is the number of trajectories~\cite{quenouille1956}. To understand why
this is so, in the present context, we recall that any sample mean or sample covariance,
$\sample{Q}_{ijk\ldots.}$ (where $i$, $j$, $k$ etc.\ labels sampling times), is an average
(normalized sum) over the $M$ trajectories. The multivariate central limit
theorem tells us that for large $M$ we can, for such averages, write
$\sample{Q}_{ijk\ldots} = \exact{Q}_{ijk\ldots.} + \gamma_{ijk\ldots}/\sqrt{M}$, where
$\gamma_{ijk\ldots}$ is a zero-mean ``noise''. Therefore any observable, $O$,
which is a function of one, or several, such sample estimates
(the optimal parameters $\estimated{\params}$ and their associated covariance matrix $\bm{\estimated{\Delta}}$,
see previous sections, are examples of such observables) will
(schematically) have a Taylor series expansion of the form:
\begin{equation}
  \label{eq:taylor-with-bias}
   O   =  \exact{O} +  \sum_{k=1}^\infty \frac{A_k}{\sqrt{M}M^{k-1}} +  \sum_{k=1}^\infty \frac{B_k}{M^k}
\end{equation}
for large $M$. The first term in the Taylor expansion is the sought quantity,
$\exact{O}$. Considering the remaining terms, we note that, by construction, we
have that $\langle A_1 \rangle =0$, and hence the first non-zero term of the
expectation value of the expression above is $\langle B_1\rangle /M \propto
1/M$. For the case that the observable, $O$, is a function of more than one
\emph{independent} sample estimates, then we have $\langle A_k \rangle = 0$ for
all~$k$. However, note that if~$O$ is a function of several sample estimates
which are \emph{dependent}, then in general $\langle A_k \rangle \neq 0$ for~$k\ge
2$. We can safely remove the first bias-term with a jackknife
procedure~\cite{miller1974}, see section~\ref{sec:jackknife}. Also
higher order bias terms can be removed formally. However, already at the
second order bias reduction level computational costs becomes considerable.

\subsection{Bias in parameter estimation of CCM for linear fit functions}
\label{sec:bias_general}

Consider equations~\eqref{eq:chi2_vectorform} and~\eqref{eq:bias_4}.
We write the sample estimator of the covariance matrix~eq.~\eqref{eq:C_sample}, and the
exact, $\exactm{Q}$, as related by
\begin{equation}
  \samplem{Q}_{ij} = \exactm{Q}_{ij} + \eta_{ij},
\end{equation}
where $\bm{\eta}$ represents their deviation.
We seek the ``noise'' in the inverse, $(\samplem{Q}^{-1})_{ij}$.
Using the normalization condition, and writing
\begin{equation}
  \label{eq:bias_21}
  (\samplem{Q}^{-1})_{ij} = ({\exactm{Q}}^{-1})_{ij} + \xi_{ij},
\end{equation}
we get
\newcommand{\xim}{\bm{\xi}}
\newcommand{\etam}{\bm{\eta}}
\begin{equation}
  \bm{I} = \samplem{Q} \ \samplem{Q}^{-1} = (\exactm{Q} + \etam)({\exactm{Q}}^{-1} +
  \xim) = \bm{I} + \etam{\exactm{Q}}^{-1} +\ \exactm{Q}\xim + \etam\xim.
\end{equation}
Thus, to first order $\etam{\exactm{Q}}^{-1} +\ \exactm{Q}\xim = 0$, and by definition $\etam
= \samplem{Q} - \exactm{Q}$:
\begin{equation}
  \label{eq:bias_22}
  \xim = {\exactm{Q}}^{-1} -\ {\exactm{Q}}^{-1}\samplem{Q} \ {\exactm{Q}}^{-1}.
\end{equation}

Using eq.~\eqref{eq:bias_21} in eq.~\eqref{eq:chi2_vectorform} and eq.~\eqref{eq:bias_4} yields
\begin{equation}
\begin{split}
  \estimated{\param} &= \frac{\samplem{y}^T ({\exactm{Q}}^{-1} + \xim) \bm{t}}
  {\bm{t}^T ({\exactm{Q}}^{-1} + \xim) \bm{t}}
  = \frac{\samplem{y}^T {\exactm{Q}}^{-1} \bm{t}}
  {\bm{t}^T {\exactm{Q}}^{-1}\bm{t}
    \left(1 + \frac{\bm{t}^T\xim\bm{t}}{\bm{t}{\exactm{Q}}^{-1}\bm{t}}
    \right)}
  + \frac{\samplem{y}^T \xim \bm{t}}
  {\bm{t}^T {\exactm{Q}}^{-1}\bm{t}
    \left(1 + \frac{\bm{t}^T\xim\bm{t}}{\bm{t}{\exactm{Q}}^{-1}\bm{t}}
    \right)}\\
  &\approx
  \frac{1}{\bm{t}^T{\exactm{Q}}^{-1}\bm{t}}
  \left(
    \samplem{y}^T{\exactm{Q}}^{-1}\bm{t} + \samplem{y}^T \xim \bm{t} -
    \frac{\samplem{y}^T{\exactm{Q}}^{-1}\bm{t}}{\bm{t}{\exactm{Q}}^{-1}\bm{t}}\bm{t}\xim\bm{t}
  \right),
\end{split}
\end{equation}
where we did a series expansion to first order in~$\xim$. Using
eq.~\eqref{eq:bias_22} we get
\let\xim\undefined
\let\etam\undefined
\begin{align}
  \nonumber
  \estimated{\param} &= \frac{\samplem{y}^T {\exactm{Q}}^{-1} \bm{t}} {\bm{t}^T
    {\exactm{Q}}^{-1} \bm{t}} +
  \frac{\samplem{y}^T ({\exactm{Q}}^{-1} - {\exactm{Q}}^{-1}\ \samplem{Q}\ {\exactm{Q}}^{-1}) \bm{t}} {\bm{t}^T
    {\exactm{Q}}^{-1} \bm{t}} -
  \frac{\samplem{y}^T {\exactm{Q}}^{-1} \bm{t}} {(\bm{t}^T
    {\exactm{Q}}^{-1} \bm{t})^2}
  \left(
    \bm{t}^T {\exactm{Q}}^{-1} \bm{t} - \bm{t}^T {\exactm{Q}}^{-1}\samplem{Q}\ {\exactm{Q}}^{-1} \bm{t}
  \right)\\
  \label{eq:bias_24}
  &= \frac{\samplem{y}^T {\exactm{Q}}^{-1} \bm{t}} {\bm{t}^T
    {\exactm{Q}}^{-1} \bm{t}}  \underbrace{ -
  \frac{\samplem{y}^T {\exactm{Q}}^{-1}\samplem{Q}\ {\exactm{Q}}^{-1} \bm{t}} {\bm{t}^T
    {\exactm{Q}}^{-1} \bm{t}} +
  \frac{\samplem{y}^T {\exactm{Q}}^{-1} \bm{t}} {(\bm{t}^T
    {\exactm{Q}}^{-1} \bm{t})^2} \bm{t}^T {\exactm{Q}}^{-1}\samplem{Q}\ {\exactm{Q}}^{-1} \bm{t}}_{\text{bias}=B}.
\end{align}
Note that the expectation value of the first term on the right hand side
evaluates to~$\exact{\param}$, hence the additional terms yield the bias,
whose expectation value, $\langle B \rangle$, we now seek. It is convenient to write
eq.~\eqref{eq:bias_24} on component form (repeated indices are summed over) with $B=B_1 + B_2$ where
\begin{subequations}
  \label{eq:bias_25and26}
  \begin{align}
    \label{eq:bias_25}
    B_1 &= - \frac{\sample{y}_k
      ({\exactm{Q}}^{-1})_{ki}\sample{Q}_{ij}({\exactm{Q}}^{-1})_{jl} t_l}
    {\bm{t}^T
    {\exactm{Q}}^{-1} \bm{t}} \\
    \label{eq:bias_26}
    B_2 &= \frac{\sample{y}_i ({\exactm{Q}}^{-1})_{ik} t_k t_j
      ({\exactm{Q}}^{-1})_{jm} \sample{Q}_{ml} ({\exactm{Q}}^{-1})_{ln} t_n} {(\bm{t}^T
    {\exactm{Q}}^{-1} \bm{t})^2}
  \end{align}
\end{subequations}
(the component form of the quantity appearing in the denominators above is $\bm{t}^T
    {\exactm{Q}}^{-1} \bm{t} = t_p ({\exactm{Q}}^{-1})_{pq} t_q$).
We thus see that the expected bias, $\langle B \rangle $, is determined by expectation value ($a,b,c\ldots$~label trajectories):
\begin{equation}
  \label{eq:bias_27}
  \langle \sample{y}_k \sample{Q}_{ij} \rangle = \frac{1}{M(M-1)}
  \langle
  \sum_{a=1}^M y_k^{(a)} \left[ \sum_{b=1}^M y_i^{(b)}y_j^{(b)} - \frac1M \sum_{b=1}^My_i^{(b)}\sum_{c=1}^My_j^{(c)} \right]
  \rangle.
\end{equation}

\subsection{Bias in parameter estimation of CCM for BM}
\label{sec:bias_corr}

Let us now consider the expected bias for CCM fitting for BM using the formal expression in section \ref{sec:bias_general}. We have:
\begin{equation}
  \langle y_k^{(a)} \rangle = \Big \langle \left[ x_k^{(a)} - x^{(a)}(0)\right]^2 \Big \rangle =
  {\exact{\sigma}}_k^2 = \exact{V}_{kk},
\end{equation}
where we in the last step used eq.~\eqref{eq:bias_11}. Also $\langle x_i^{(a)}
- x^{(a)}(0)\rangle = 0$, and since different realizations (trajectories) are
independent we have
\begin{equation}
  \label{eq:bias_29}
 \langle x_i^{(a)} x_j^{(b)} \rangle = \delta_{a,b}\exact{V}_{ij}.
\end{equation}
Higher order terms can be calculated using
Wick's theorem, eq.~\eqref{eq:bias_30} (for large~$i$, $x_i^{(a)}$ is a sum of many small
increments, from the central limit theorem it follows that~$x_i^{(a)}$ are
Gaussian). We have
\newcommand{\xia}{x_i^{(a)}}
\newcommand{\xjb}{x_j^{(b)}}
\begin{equation}
\begin{split}
  \langle y_i^{(a)} y_j^{(b)} \rangle &= \langle (\xia)^2 (\xjb)^2\rangle =
  \langle \xia \xia \xjb \xjb \rangle \\
  &= \langle \xia \xia \rangle \langle \xjb \xjb \rangle +
  \langle \xia \xjb \rangle \langle \xia \xjb \rangle +
  \langle \xia \xjb \rangle \langle \xia \xjb \rangle \\
  &= {\exact{\sigma}}_i^2 {\exact{\sigma}}_j^2
  + 2(\exact{V}_{ij})^2 \delta_{a,b}.
\end{split}
\end{equation}
\let\xia\undefined              
\let\xjb\undefined
Now, in the same way for higher order terms, we get
\newcommand{\xka}{x_k^{(a)}}
\newcommand{\xib}{x_i^{(b)}}
\newcommand{\xjc}{x_j^{(c)}}
\begin{equation}
\begin{split}
  \langle  y_k^{(a)} y_i^{(b)}  y_j^{(c)} \rangle &= \langle \xka \xka \xib
  \xib \xjc \xjc \rangle = [\text{tedious enumeration of all cases}]=\\
  \label{eq:bias_32}
  & \hspace{-1cm} = {\exact{\sigma}}_k^2 {\exact{\sigma}}_i^2 {\exact{\sigma}}_j^2 +
  2{\exact{\sigma}}_k^2(\exact{V}_{ij})^2\delta_{b,c} +
  2{\exact{\sigma}}_j^2(\exact{V}_{ki})^2\delta_{a,b} +
  2{\exact{\sigma}}_i^2(\exact{V}_{kj})^2\delta_{a,c} +
  8(\exact{V}_{ki})^2(\exact{V}_{kj})^2(\exact{V}_{ij})^2 \delta_{a,b}\delta_{b,c}\delta_{a,c},
\end{split}
\end{equation}
\let\xka\undefined
\let\xib\undefined
\let\xjc\undefined
(no sum over repeated indices). Eq.~\eqref{eq:bias_27} now becomes
\begin{equation}
  \label{eq:bias_33}
  \langle \sample{y}_k \sample{Q}_{ij} \rangle = \frac{1}{M(M-1)}
  \underbrace{\sum_{a=1}^M \sum_{b=1}^M \langle y_k^{(a)} y_i^{(b)} y_j^{(b)} \rangle}_{U_1}
  - \frac{1}{M^2(M-1)}\underbrace{\sum_{a,b,c} \langle y_k^{(a)} y_i^{(b)} y_j^{(c)} \rangle}_{U_2}.
\end{equation}
Using eq.~\eqref{eq:bias_32} we get:
\begin{subequations}
  \label{eq:bias_34and35}
  \begin{align}
    \nonumber
    U_1 &= \sum_{a,b}  \langle y_k^{(a)} y_i^{(b)} y_j^{(b)} \rangle \\
    \label{eq:bias_34}
    &= M^2 {\exact{\sigma}}_k^2 {\exact{\sigma}}_i^2 {\exact{\sigma}}_j^2 +
    2M^2{\exact{\sigma}}_k^2(\exact{V}_{ij})^2 + 2M
    {\exact{\sigma}}_j^2(\exact{V}_{ki})^2 + 2M {\exact{\sigma}}_i^2(\exact{V}_{kj})^2
    + 8M\exact{V}_{ki}\exact{V}_{ij}\exact{V}_{kj}\\
    \nonumber
    U_2 &= \sum_{a,b,c} \langle y_k^{(a)} y_i^{(b)} y_j^{(c)} \rangle \\
    \label{eq:bias_35}
    &= M^3 {\exact{\sigma}}_k^2 {\exact{\sigma}}_i^2 {\exact{\sigma}}_j^2 +
    2M^2\left[ {\exact{\sigma}}_k^2(\exact{V}_{ij})^2 +
      {\exact{\sigma}}_j^2(\exact{V}_{ki})^2 + {\exact{\sigma}}_i^2(\exact{V}_{kj})^2
    \right] + 8M\exact{V}_{ki}\exact{V}_{ij}\exact{V}_{kj}.
  \end{align}
\end{subequations}
Combining eq.~\eqref{eq:bias_34and35} with eq.~\eqref{eq:bias_33} results in:
\begin{equation}
  \label{eq:bias_36}
  \begin{split}
    \langle \sample{y}_k \sample{Q}_{ij} \rangle &= \frac{1}{M(M-1)} \left[
      (2M^2-2M){\exact{\sigma}}^2_k(\exact{V}_{ij})^2 +
      (8M-8)\exact{V}_{ki}\exact{V}_{ij}\exact{V}_{kj}
    \right]\\
    &= 2{\exact{\sigma}}_k^2 (\exact{V}_{ij})^2 +
    \frac8M\exact{V}_{ki}\exact{V}_{ij}\exact{V}_{kj}.
  \end{split}
\end{equation}
Using eq.~\eqref{eq:bias_36} in eq.~\eqref{eq:bias_25} we find
\begin{equation}
  \langle B_1 \rangle = - \frac{{\exact{\sigma}}_k^2 ({\exactm{Q}}^{-1})_{ki}\delta_{i,l}t_l
    + \frac8M \exact{V}_{ki}\exact{V}_{ij}\exact{V}_{kj}
    ({\exactm{Q}}^{-1})_{ki} ({\exactm{Q}}^{-1})_{jl} t_l}
  {\bm{t}^T{\exactm{Q}}^{-1}\bm{t}},
\end{equation}
where we used that $\exact{Q}_{ij}({\exactm{Q}}^{-1})_{jl}=\delta_{i,l}$.
Now consider $B_2$, eq.~\eqref{eq:bias_26}. We write eq.~\eqref{eq:bias_36}
according to (also see eq.~\eqref{eq:bias_13})
\begin{equation}
  \langle \sample{y}_i \sample{Q}_{ml} \rangle
  = {\exact{\sigma}}_i^2\exact{Q}_{ml} + \frac8M\exact{V}_{im}\exact{V}_{ml}\exact{V}_{li}.
\end{equation}
Eq.~\eqref{eq:bias_26} now becomes
\begin{equation}
  \begin{split}
    \langle B_2 \rangle &= \frac{({\exactm{Q}}^{-1})_{ik} t_k t_j
      ({\exactm{Q}}^{-1})_{jm} \left[ {\exact{\sigma}}_i^2 \exact{Q}_{ml} +
        \frac8M\exact{V}_{im}\exact{V}_{ml}\exact{V}_{li} \right]
      ({\exactm{Q}}^{-1})_{ln} t_n}
    {(\bm{t}^T{\exactm{Q}}^{-1}\bm{t})^2}\\
    &=
    \frac{{\exact{\sigma}}_i^2({\exactm{Q}}^{-1})_{ik}t_k}{\bm{t}^T{\exactm{Q}}^{-1}\bm{t}}
    + \frac8M \frac{({\exactm{Q}}^{-1})_{ik} t_k t_j ({\exactm{Q}}^{-1})_{jm}
      \exact{V}_{im}\exact{V}_{ml}\exact{V}_{li} ({\exactm{Q}}^{-1})_{ln} t_n}
    {(\bm{t}^T{\exactm{Q}}^{-1}\bm{t})^2}.
  \end{split}
\end{equation}
Combining $B_1$ and $B_2$ we arrive at an expression for the predicted first order
bias (eq.~\eqref{eq:bias_24}) for the suggested matrix, $\samplem{R}^{[CCM]}$; (notice
the cancellations of the first terms):
\begin{equation}
  \label{eq:bias_39}
  \langle B \rangle = \frac1M \frac{8}{\bm{t}^T{\exactm{Q}}^{-1}\bm{t}}
  \left(
    \frac{({\exactm{Q}}^{-1})_{ik} t_k t_j ({\exactm{Q}}^{-1})_{jm}
      \exact{V}_{im}\exact{V}_{ml}\exact{V}_{li} ({\exactm{Q}}^{-1})_{ln} t_n}
    {\bm{t}^T{\exactm{Q}}^{-1}\bm{t}}
    - \exact{V}_{ki}\exact{V}_{ij}\exact{V}_{jk}
    ({\exactm{Q}}^{-1})_{ki}({\exactm{Q}}^{-1})_{jl} t_l
  \right),
\end{equation}
which can be analytically evaluated. With this in mind we use eq.~\eqref{eq:bias_11},
with $t_i = i \epsilon$, and eq.~\eqref{eq:bias_16}, in
eq.~\eqref{eq:bias_39}. When evaluating the associated sums over repeated
indices in eq.~\eqref{eq:bias_39}, one uses:
\begin{equation}
  \min(i,j)=
  \begin{cases}
    i, & \text{if}\ i \le j \\
    j, & \text{if}\ i > j
  \end{cases}
\end{equation}
and then splits the sums accordingly. This splitting leads to sums on
the form
\begin{equation}
I(m,p) = \sum_k \frac{k^m}{(2k-1)^p},
\end{equation}
where $m$ and $p$ are positive integers. These sums are rewritten according to
\begin{equation}
  I(m,p) = \frac{1}{2^m}  \sum_k \frac{1}{(2k-1)^p} \left( (2k-1) +1 \right)^m =  \frac{1}{2^m}
  \sum_{q=1}^m \binom{m}{q} \sum_k (2k-1)^{q-p},
\end{equation}
where we used the binomial theorem. The full calculation is tedious but
straightforward. The final result is:
\begin{subequations}
  \label{eq:B_final}
  \begin{empheq}[box=\widefbox]{align}
    \langle B \rangle &= \frac{D}{M} G(N) \\
    G(N) &= -\frac{a}{d} + \frac{b}{d^2} \\
    a &= \frac{N}{2} + s_1 -\frac{s_2}{2}\\
    b &= \frac{1}{16} (3s_1 -s_3) \\
    d &= \frac{s_1}{8} \\
    s_n & = \sum_{k=1}^N \frac{1}{(2k-1)^n}.
  \end{empheq}
\end{subequations}

\subsubsection{Asymptotic expansion}
Let us now investigate eq.~\eqref{eq:B_final} for
large~$N$. To that end, we write~$s_n$, defined above, according to
\begin{equation}
\label{eq:s_1}
s_n = \sum_{k=1}^N \left( \frac{1}{(2k-1)^n} + \frac{1}{(2k)^n} -
  \frac{1}{(2k)^n} \right)  = \sum_{k=1}^{2N}k^{-n} - \frac{1}{2^n} \sum_{k=1}^N
  k^{-n}.
\end{equation}
In eq.~\eqref{eq:B_final}, there are three sums, $s_1$, $s_2$ and~$s_3$. Out
of these sums, $s_1$ decays most slowly with~$N$ and hence this sum is the
only one which needs to be kept for large~$N$. From eq.~(0.131) in
Gradshteyn \textit{et~al.}~\cite{gradshteyn2000} we have
\begin{equation}
\sum_{k=1}^N \frac{1}{k} = \gamma + \ln N + \frac{1}{2N} + \ordo(\frac{1}{N^2}),
\end{equation}
where $\gamma \approx 0.5772 $ is the Euler-Mascheroni constant. Combining the
result above with eq.~\eqref{eq:s_1} and eq.~\eqref{eq:B_final} we arrive at
the asymptotic expression
\begin{equation}
\label{eq:g_approx}
G(N) \approx - \frac{8N}{\ln N + \gamma + 2 \ln 2},
\end{equation}
where we used $\ln ab = \ln a + \ln b$.
For large $N$, eq.~\eqref{eq:g_approx} is a good approximation compared to the
exact bias, eq.~\eqref{eq:B_final}, see Supplementary Figure~\ref{fig:bias}.

\subsection{Bias in parameter estimation of WLS for BM}
\label{sec:bias_uncorr}
Let us now consider the second case, eq.~\eqref{eq:bias_6}, of
choosing~$\bm{R}$. According to eqs.~\eqref{eq:wls_linear} and~\eqref{eq:bias_6} we have the following:
\newcommand{\ecovnew}{\exactm{Q}_{\text{new}}}
\newcommand{\scovnew}{\samplem{Q}_{\text{new}}}
\begin{equation}
  \label{eq:bias_40}
  \estimated{\param} = \frac{\samplem{y}^T\scovnew^{-1}\bm{t}}{\bm{t}^T \scovnew^{-1}\bm{t}},
\end{equation}
where
\begin{align}
  \sample{Q}_{\text{new},ij} &= \sample{Q}_{ij}\delta_{i,j}\\
  \exact{Q}_{\text{new},ij} &= \exact{Q}_{ij}\delta_{i,j}\\
  ({\samplem{Q}}_{\text{new}}^{-1})_{ij} &= \delta_{i,j}/\sample{Q}_{ij}\\
  ({\exactm{Q}}_{\text{new}}^{-1})_{ij} &= \delta_{i,j}/\exact{Q}_{ij}.
\end{align}
The calculation starting from eq.~\eqref{eq:bias_22} to eq.~\eqref{eq:bias_24}
is identical to before, just replace $\samplem{Q}$ with $\scovnew$, and same
for exact results. Since our new matrices are diagonal,
eq.~\eqref{eq:bias_25and26} becomes
(we here reintroduce explicit sums for the sake of clarity)
\let\ecovnew\undefined
\let\scovnew\undefined
\begin{subequations}
  \label{eq:bias_45and46}
  \begin{align}
    \label{eq:bias_45}
    B_1 &=  -  \frac{\sum_k \sample{y}_k \frac{1}
      {(\exact{Q}_{kk})^2}\sample{Q}_{kk}t_k}{\sum_q t_q^2 / \exact{Q}_{qq}}\\
    \label{eq:bias_46}
    B_2 &= \frac{\sum_{j,k} \sample{y}_k \frac{1}{\exact{Q}_{kk}}t_k \cdot
      t_j^2\frac{1}{(\exact{Q}_{jj})^2}\sample{Q}_{jj}} {\left(\sum_q t_q^2 /
        \exact{Q}_{qq} \right)^2}.
  \end{align}
\end{subequations}
Also the calculation from eq.~\eqref{eq:bias_27} which leads up to
eq.~\eqref{eq:bias_36} is identical. From eq.~\eqref{eq:bias_45and46} we see that we need
\begin{subequations}
  \begin{align}
    \label{eq:bias_47}
    \langle \sample{y}_k \sample{Q}_{jj} \rangle
    =& 2 {\exact{\sigma}}_k^2 (\exact{V}_{jj})^2
    + \frac8M  (\exact{V}_{kj})^2  \exact{V}_{jj}\qquad   (j,k\ \text{fixed}),\\
    \label{eq:bias_48}
    \langle \sample{y}_k \sample{Q}_{kk} \rangle
    =& 2 {\exact{\sigma}}_k^6
    + \frac8M {\exact{\sigma}}_k^6\qquad    (k\ \text{fixed}).
  \end{align}
\end{subequations}
 Substituting eq.~\eqref{eq:bias_48} into eq.~\eqref{eq:bias_45}, and
using eq.~\eqref{eq:bias_13} $\exact{Q}_{kk}=2(\exact{V}_{kk})^2 =
2{\exact{\sigma}}_k^4$, and ${\exact{\sigma}}_k^2 = 2Dt_k$ we get (with sums
explicitly written)
\begin{equation}
\label{eq:bias_49}
\begin{split}
  \langle B_1 \rangle &= - \frac{\sum_k
    \frac{1}{(\exact{Q}_{kk})^2} \left( 2 {\exact{\sigma}}_k^6 + \frac8M
      {\exact{\sigma}}_k^6 \right) t_k} {\sum_q t_q^2 / \exact{Q}_{qq}}
  =  - \frac{(1+\frac4M)\sum_k 1/2D}{\sum_k 1/(2D)^2}\\
  &= -2D\left(1+ \frac4M\right).
\end{split}
\end{equation}
In much the same way, we insert eq.~\eqref{eq:bias_47} into
eq.~\eqref{eq:bias_46}
\begin{equation}
\label{eq:bias_50}
\begin{split}
  \langle B_2 \rangle &= \frac{\sum_{j,k} \left( 2
      {\exact{\sigma}}_k^2 {\exact{\sigma}}_j^4 + \frac8M {\exact{\sigma}}_j^2
      (\exact{V}_{kj})^2 \right) \frac{1}{2{\exact{\sigma}}_k^4}\frac{t_k
      t_j^2}{4{\exact{\sigma}}_j^8}}
  {\left( \sum_q t_q^2/2{\exact{\sigma}}_q^4 \right)^2} = \frac{ \sum_{j,k}
    \left( \frac14 \frac{1}{(2D)^3} + \frac1M \frac{1}{(2D)^5}
      \frac{(\exact{V}_{kj})^2}{t_k t_j} \right)}
  {1/64D^4 \left(\sum_k 1\right)^2} \\
  &= 2D + \frac{2}{MD}\frac{1}{N^2} \underbrace{\sum_j \sum_k
    \frac{(\exact{V}_{kj})^2}{t_k t_j}}_{I}.
\end{split}
\end{equation}
Consider the double sum, $I$, in eq.~\eqref{eq:bias_50}. We have time step
$t_j=\epsilon j$ and separate the sums into $j=k$ and $j\neq k$, which gives
$\exact{V}_{ij} = 2D\epsilon \min(i,j)$
\begin{equation}
\begin{split}
  I &= 4D^2 \sum_{k=1}^N\sum_{j=1}^N \frac{\left[\min(j,k)\right]^2}{jk} = 4D^2
  \left(
    \sum_{k=1}^N 1 + 2 \sum_{k=1}^N \sum_{j=1}^{k-1} \frac{\left[\min(j,k)\right]^2}{jk}
    \right)\\
    &= 4D^2\left(
      N + 2 \sum_{k=1}^N \frac1k \sum_{j=1}^{k-1}j
    \right) = 4D^2 \left( N + 2 \sum_{k=1}^N \frac1k \frac{k(k-1)}{2} \right)
    \\
    =& 4D^2\sum_{k=1}^N k = 2D^2 N(N+1),
\end{split}
\end{equation}
which inserted in eq.~\eqref{eq:bias_50} yields
\begin{equation}
  \langle B_2 \rangle = 2D + \frac{4D}{M}\left( \frac1N + 1 \right),
\end{equation}
from which we get the complete full bias together with eq.~\eqref{eq:bias_49}:
\begin{equation}
  \langle B \rangle = \langle B_1 \rangle + \langle B_2 \rangle =
  \frac{4D}{M}\left( \frac1N - 1 \right).
\end{equation}
Thus,
\begin{empheq}[box=\widefbox]{equation}
  \label{eq:bias_52}
  \estimated{\param} - \exact{\param} = - \frac{4D}{M}\left( 1 - \frac1N \right).
\end{empheq}
Note that the bias is independent of~$N$ for large~$N$.

\subsection{Lack of bias for BMALS}
\label{sec:bias_GA}

We now consider our third and final choice of $\bm{R}$-matrix for BM. Since $\langle \bar{y}_i \rangle = \exact{y}_i$ and  $\bm{R}$ is a true inverse covariance matrix (and hence no sample estimate, see eq. \eqref{eq:bias_8}) it follows immediately, by taking the expectation value of eq. \eqref{eq:wls_linear}, that the BMALS parameter estimate is unbiased.

\subsection{Lack of bias in parameter estimation of CCM for DHO}
\label{sec:bias_DHO}

For the DHO problem we choose as our observable the particle
position, i.e., we use $y_k^{(m)} = x_k^{(m)}$, where $m$ labels different
trajectories. For a good fit, the DHO parameter estimates
$\estimated{\params}$ are unbiased for CCM. To see this, consider the CCM minimization
criterion eq. \eqref{eq:Dchi-Dp} for DHO, which we write
\begin{align}\label{eq:DHO_mincrit}
 0 = \left.\frac{\partial \bm{f}}{\partial \param_a}\right|_{\params =
   \estimated{\params}} \samplem{Q}^{-1} (\bm{f}(\estimated{\params}) -\bm{y}^*)
- \left.\frac{\partial \bm{f}}{\partial \param_a}\right|_{\params =
   \estimated{\params}} \samplem{Q}^{-1} (\samplem{y}-\bm{y}^*).
\end{align}
As in previous subsections, we then expand the inverse sample covariance
matrix around its true value, i.e., we write $\samplem{Q}^{-1} =
{\exactm{Q}}^{-1} + \bm{\xi}$, where $\bm{\xi}$ is given in
eq. \eqref{eq:bias_22}. By expanding the right-hand side of
eq. \eqref{eq:DHO_mincrit} in $\bm{f}(\estimated{\params}) -\bm{y}^*$, $\samplem{y}-\bm{y}^*$ and $\bm{\xi}$, we arrive at
\begin{align}\label{eq:DHO_mincrit_2}
 0 = \underbrace{ \left. \frac{\partial \bm{f}}{\partial \param_a}\right|_{\params =
   \estimated{\params}} {\exactm{Q}}^{-1} (\bm{f}(\estimated{\params})
 -\bm{y}^*)}_{\bm{F}(\estimated{\params})}
 \end{align}
 \begin{align}
- \underbrace{\left. \frac{\partial \bm{f}}{\partial \param_a}\right|_{\params =
   \estimated{\params}} \left( ({\exactm{Q}}^{-1} - {\exactm{Q}}^{-1}\samplem{Q} \ {\exactm{Q}}^{-1}) (\bm{f}(\estimated{\params}) -\bm{y}^*) +  {\exactm{Q}}^{-1}
 (\samplem{y}-\bm{y}^*)
- ({\exactm{Q}}^{-1} - {\exactm{Q}}^{-1}\samplem{Q} {\exactm{Q}}^{-1}) (\samplem{y} -\bm{y}^*)\right)}_{\bm{G}(\estimated{\params})}.
\end{align}
Since $\bm{F}(\estimated{\params})$ involves only the true covariance matrix
and $\exactm{y}$, the solution to $\bm{F}(\estimated{\params})=0$ yields the true
parameter value, i.e., we have $F_a(\exact{\params})=0$. If the fit
is good,  then we obtain the solution to eq. \eqref{eq:DHO_mincrit_2} using
a Taylor expansion, i.e., we write $F_a(\estimated{\params})\approx
F_a(\exact{\params}) + \sum_b w_{ab} (\estimated{\param}_b - \exact{\param}_b)=\sum_b w_{ab} (\estimated{\param}_b - \exact{\param}_b) $,
where $ w_{ab} = \partial_b F_a(\estimated{\params})/\partial
\estimated{\param}_b|_{\estimated{\params}=\exact{\params}}$. Inserting this
into eq. \eqref{eq:DHO_mincrit_2} and solving for $\estimated{\param}_a$, we get
\begin{align}
\estimated{\param}_a  =\exact{\param}_a + \underbrace{\sum_b (\bm{w}^{-1})_{ab} G_b(\estimated{\params})}_{B_a}.
\end{align}
Thus, the bias in the estimated parameter, $\estimated{\param}_a$, is
determined by the expectation value of $B_a$.
 An application of Wick's theorem for Gaussian variables yields
\begin{align}\label{eq:DHO_nobias}
\langle \sample{y}_k \sample{Q}_{ij} \rangle = \langle \sample{y}_k\rangle
 \exact{Q}_{ij}.
\end{align}
  This result is a direct consequence of the fact that the
 positions at different times for the DHO process are distributed according to
 a multivariate Gaussian. This Gaussianity, in turn, follows from the fact
 that the harmonic oscillator position is a linear function of the imposed
 Gaussian noise, see eq. \eqref{eq:DHO}. Using eq. \eqref{eq:DHO_nobias} and the
 fact that $\langle \sample{y}_k\rangle = \exact{y}_k $ and $\langle
 \sample{Q}_{ij} \rangle =  \exact{Q}_{ij}$ we find that $\langle
 \bm{G}(\estimated{\params})\rangle = 0$ and thereby that indeed
$\langle B_a \rangle=0$, i.e., the CCM parameter estimate for DHO does not suffer from the bias problems discussed in the previous subsections.

\section{Approximate distribution for the estimated parameters}\label{sec:distribution_for_parameter_estimates}

In the main text we saw that if $M$ (the number of trajectories) is large enough the distribution for the estimated parameters is approximately Gaussian, see \figHistogram{}. To understand why this is so, we note that a set of random number, $\sample{y}_i$ ($i=1,\ldots N$), from the Gaussian distribution in eq. \eqref{eq:rho_wlsice} can be generated using
\begin{equation}
\sample{y}_i = \exact{y}_i + \frac{1}{\sqrt{M}} \eta_i
\end{equation}
where $\bm{\eta}$ is a zero mean Gaussian random number with ($M$-independent) covariance matrix $\bm{Q}$. Consider now a function $F(\samplem{y})$, and note that the estimated parameters, $\param_a$, are functions of this type. We then Taylor-expand:
\begin{equation}
F(\samplem{y})\approx F(\exactm{y}) +\frac{1}{\sqrt{M}} \bm{A}\cdot \bm{\eta} + O(\frac{1}{M}),
\end{equation}
where $\bm{A}$ is a matrix containing partial derivatives. Now assuming that the second term of the RHS above is non-zero, that the matrix $A$ is full rank, and that all terms higher than or equal to $1/M$ can be neglected, we have that the distribution for $F$ is another Gaussian. This follows from the fact that $\bm{A}\cdot \bm{\eta}$ is normally distributed if $\bm{\eta}$ are drawn from a multivariate Gaussian.~\cite{anderson2003}.

\section{Jackknife bias reduction}
\label{sec:jackknife}
Through data resampling, bias in data-fitting can often be
reduced. Let~$O$ be the parameter estimator, based on some data
set with $M$ trajectories. The associated true parameter is denoted by $\exact{O}$. Herein, we
choose $O$ as either the estimated parameters $\estimated{\params}$, obtained by minimizing eq. (\ref{eq:chi2_vectorform}),
or the associated covariance matrix $\bm{\estimated{\phi}}$, \eqPhiAb{}. As outlined
in section~\ref{sec:origin_of_bias}, one often expects such a finite data set
to yield a bias contribution of the form
\begin{equation}
  \label{eq:jack_bias}
  O = \exact{O} + \frac{a}{M} + \frac{b}{M^2} + \frac{c}{M^3} + \ordo\left(\frac{1}{M^4}\right).
\end{equation}
The bias terms can be reduced by increasing the data samples, $M$, or by using
the jackknife method~\cite{miller1974}. Let us split the sample into
$g$~groups, each of size~$h$, and define $O_{[-j]}$ as the parameter fitted
to a data sample with the $j$th group removed.

\subsection{First order jackknife bias reduction}
\label{sec:jack_firstorder}
The first order bias term can be removed through repeated fitting and
averaging over the sampled data set:
\begin{subequations}
  \label{eq:jack1}
  \begin{align}
    O^{(1)}  &= \frac1g \sum_{j=1}^g O_{[-j]}\\
    O^{(0,1)}_J &= gO - (g-1)O^{(1)}.
  \end{align}
\end{subequations}
By using eq.~\eqref{eq:jack_bias} which has bias terms proportional to $M=hg$ for
the full fitting, $O$, and $h(g-1)$ for the reduced sample estimator in
eq.~\eqref{eq:jack1}, we see that we are left with
\begin{equation}
  \label{eq:jack1_bias}
  \begin{split}
    O^{(0,1)}_J &= \exact{O}  - \frac{b}{h^2} \frac{1}{g(g-1)} -
      \frac{c}{h^3}\left( \frac{1}{(g-1)^2} - \frac{1}{g^2} \right) +
      \ordo(g^{-3})\\
    &\approx \exact{O} - \frac{b}{M^2} - 2\frac{c}{M^3},
  \end{split}
\end{equation}
lacking the first order bias term. Although the higher order terms remain,
their contribution is often lower than the first order term.

\subsection{Second order jackknife bias reduction}
\label{sec:jack_secondorder}
For further bias reduction we can apply a second order correction.
In a similar spirit to what is done in
the first order jackknife, we split the data into $g$ groups, and define
$O_{[-j,-j']}$ as the parameter estimator based on a data set with the
$j$th and $j'$th group removed, each of size~$h$. Following
Schucany~\textit{et~al.}~\cite{schucany1971} we get
\begin{subequations}
  \label{eq:jack2}
  \begin{align}
    \label{eq:jack2_2}
    O^{(2)}  &= \frac{2}{g(g-1)} \sum_{j<j'}^g O_{[-j,-j']}\\
    \label{eq:jack2_12}
    O^{(1,2)}_J &= (g-1)O^{(1)} - (g-2)O^{(2)}\\
    \label{eq:jack2_012}
    O^{(0,1,2)}_J &= \frac{g}{2}O^{(0,1)}_J - \frac{g-2}{2}O^{(1,2)}_J.
  \end{align}
\end{subequations}
If we combine our result with eq.~\eqref{eq:jack_bias}, we are only left with the third order term
and the ones that follows it,
\begin{equation}
  \label{eq:jack2_bias}
  \begin{split}
    O^{(0,1,2)}_J &= \exact{O} + \frac{c}{h^3}\frac{1}{g(g-1)(g-2)} +  \ordo(g^{-4})\\
    &\approx \exact{O} + \frac{c}{M^3}.
  \end{split}
\end{equation}

\subsection{Variance for jackknife-bias-reduced estimators}
\label{sec:jack-variance}

In this section, we use eq.~\eqref{eq:param_taylor} to show that
$\estimated{\param}_a - \exact{\param}_a$ is insensitive (to lowest orders in
$1/M$) to the jackknifing procedure. As a consequence, the covariance
estimation formula, \eqGA{}, remains valid also for jackknifed parameter estimations.

For later convenience, we define the derivative
in eq.~\eqref{eq:param_taylor} as
\begin{equation}
  A_{a,i} = \left.\frac{\partial \estimated{\param}_a}{\partial
      \sample{y}_i}\right|_{\samplem{y} = \exactm{y}},
\end{equation}
which we will use in the following.

\subsubsection{First order jackknife bias reduction}
\label{sec:jack_var_firstorder}
To first order the jackknife estimator is obtained by dividing the $M$
trajectories into $g$~groups of size~$h$. Define the observable
$\sample{O}_{[-j],i}$ as the estimate for observable $O$, in point $i$, with
group $j$ removed. In particular,
\begin{equation}
  \label{eq:variance_3}
  \sample{y}_{[-j],i} = \frac{1}{M-h}\sum_{m \neq m_j} y_i^{(m)} = \frac{1}{M-h}
  \left(
    \sum_{m=1}^M y_i^{(m)} - \sum_{m_j} y_i^{(m)}
  \right).
\end{equation}
The corresponding non-jackknifed estimator is
\begin{equation}
  \label{eq:variance_4}
  \sample{y}_i = \frac{1}{M}\sum_{m=1}^M y_i^{(m)}.
\end{equation}
The bias of the first order jackknife estimator of $\exact{\param}_a$ within the WLS-ICE method
(see section~\ref{sec:newmethod}) is
\begin{equation}
  \begin{split}
    \param^{(0,1)}_{J,a} - \exact{\param}_a =& g \estimated{\param}_a -
    (g-1)\left[ \frac{1}{g}\sum_{j=1}^g\param_{[-j],a} \right] -
    \exact{\param}_a = \frac{1}{h} \left[
      M\estimated{\param}_a - (M-h)\frac{1}{g}\sum_{j=1}^g \param_{[-j],a} \right]\\
    =& \frac{1}{h}\sum_i A_{a,i}\left( M(\sample{y}_i - \exact{y}_i) -
      (M-h) \frac1g \sum_{j=1}^g \left(\sample{y}_{[-j],i} - \exact{y}_i\right) \right)\\
    =& \frac{1}{h}\sum_i A_{a,i}\left( \sum_{m=1}^M (y_i^{(m)}-\exact{y}_i) -
      \frac{1}{g} \sum_{j=1}^g \left( \sum_{m=1}^M (y_i^{(m)}-\exact{y}_i) -
        \sum_{m_j} (y_i^{(m_j)} - \exact{y}_i)    \right)  \right)\\
    =& \sum_i A_{a,i}\left( \frac{1}{gh} \sum_{j=1}^g \sum_{m_j} (y_i^{(m_j)}
      - \exact{y}_i) \right) = \sum_i A_{a,i} \left( \frac{1}{M}
      \sum_{m=1}^M (y_i^{(m)} - \exact{y}_i) \right)\\
    =& \estimated{\param}_a - \exact{\param}_a,
  \end{split}
\end{equation}
where we used eq.~\eqref{eq:param_taylor} to get to the second and last (fifth) row, and
eq.~\eqref{eq:variance_3}-\eqref{eq:variance_4} for the third row. Thus
\begin{equation}
  \label{eq:variance_5}
  \param_{J,a}^{(0,1)} - \exact{\param}_a = \estimated{\param}_a - \exact{\param}_a.
\end{equation}
Hence, jackknifing a parameter estimate does not change the (co)variance:
\begin{equation}
  \label{eq:variance_6}
  (\param_{J,a}^{(0,1)} -
  \exact{\param}_a)(\param_{J,b}^{(0,1)} - \exact{\param}_b) = (\estimated{\param}_a - \exact{\param}_a)(\estimated{\param}_b - \exact{\param}_b).
\end{equation}

\subsubsection{Second order jackknife bias reduction}
\label{sec:jack_var_secondorder}
For the second order bias removal, the $M$ trajectories are again divided into
$g$ groups. We define, as before, $\sample{O}_{[-j,-j'],i}$ as the estimate
for observable $O$, in point $i$, with group $j$ and $j'$ removed.
In particular
\begin{equation}
  \label{eq:variance_8}
  \sample{y}_{[-j,-j'],i} = \frac{1}{M-2h}\sum_{m \neq m_j,m_{j'}} y_i^{(m)} =
  \frac{1}{M - 2h} \left( \sum_m^M y_i^{(m)} - \sum_{m_j} y_i^{(m_j)} - \sum_{m_{j'}} y_i^{(m_{j'})}  \right).
\end{equation}
The average over all groups for $\param_a$ is
\begin{equation}
  \label{eq:variance_9}
  \param_a^{(2)} = \frac{1}{g(g-1)} \sum_{j\neq j'} \param_{[-j,-j']}.
\end{equation}
The second order jackknife is now (as given by eq.~\eqref{eq:jack2_012})
\begin{equation}
  \label{eq:variance_10}
  \param_{J,a}^{(0,1,2)} = \frac{g}{2}\param_{J,a}^{(0,1)} - \frac{g-2}{2}\param_{J,a}^{(1,2)}.
\end{equation}
Using eq.~\eqref{eq:jack2_12} we note
\begin{equation}
  \begin{split}
    \param_{J,a}^{(1,2)} - \exact{\param}_a =& \frac{1}{h}
    \left((M-h)\left[\frac{1}{g}\sum_{j=1}^g \param_{[-j],a}\right] -
      (M-2h)\left[\frac{1}{g(g-1)}\sum_{j\neq j'} \param_{[-j,-j'],a}
      \right] \right) - \exact{\param}_a\\
    =& \frac{1}{h}\sum_i A_{a,i}\Biggl( \frac{1}{g} \sum_{j=1}^g \sum_{m=1}^M
    (y_i^{(m)} - \exact{y}_i) - \frac{1}{g}\sum_{j=1}^g\left( \sum_{m_j}
      y_i^{(m_j)} - \exact{y}_i \right)\\
    &\hspace{-1cm} - \Biggl[ \frac{1}{g(g-1)} \sum_{j,j'} \sum_{m=1}^M (y_i^{(m)} -
    \exact{y}_i) - \frac{1}{g(g-1)} \sum_{j,j'} \sum_{m_j}^M (y_i^{(m_j)} -
    \exact{y}_i) - \frac{1}{g(g-1)} \sum_{j,j'} \sum_{m_{j'}}^M
    (y_i^{(m_{j'})} - \exact{y}_i) \Biggr]
    \Biggr)\\
    & \hspace{-2cm}= \frac{1}{h} \sum_i A_{a,i}\left( - \frac{1}{g} \sum_{j=1}^g\sum_{m_j}
      (y_i^{(m_j)} - \exact{y}_i) + \frac{1}{g-1}
      \sum_{{j'}}\frac{1}{g}\sum_j\sum_{m_j} (y_i^{(m_j)} - \exact{y}_i) +
      \frac{1}{g-1} \sum_j\frac{1}{g}\sum_{j'} (y_i^{(m_{j'})} -
      \exact{y}_i) \right)\\
    &\hspace{-2cm}= \frac{1}{h} \sum_i A_{a,i}\left(-\frac1g + \frac1g + \frac1g \right)
    \sum_j \sum_{m_j} (y_i^{(m_j)} - \exact{y}_i).
  \end{split}
\end{equation}
Thus
\begin{equation}
  \label{eq:variance_13}
  \param_{J,a}^{(1,2)} - \exact{\param}_a = \sum_i A_{a,i} \frac1M \sum_j \sum_{m_j}
  (y_i^{(m_j)} - \exact{y}_i) = \estimated{\param}_a - \exact{\param}_a
\end{equation}
and
\begin{equation}
  \label{eq:variance_14}
  (\param_{J,a}^{(0,1,2)} - \exact{\param}_a) = \estimated{\param}_a - \exact{\param}_a.
\end{equation}
Thus the second order jackknife estimator has the same variance and covariance
as non-jackknifed estimators.

\section{Estimation of errors on estimated parameters, using jackknife and bootstrap procedures}\label{sec:resampling}

\subsection{Jackknife error estimation}

In the heuristic jackknife error estimation one makes use of the quantities $O_{[-j]}$, see section \ref{sec:jackknife}, and calculates\cite{efron1981,efron1994}
\begin{equation}
\sigma_J^2 = \frac{g-1}{g} \sum_{j=1}^g [O_{[-j]} - O^{(1)}]^2
\end{equation}
where $ O^{(1)}$ is given in eq. \eqref{eq:jack1}.
Then $\sigma_J$ serves as an estimate for the error on the estimated
parameter. Note that in contrast to jackknife {\em bias reduction} which is
mathematically justified (based on the expected fluctuations around
estimated mean values using the central limit theorem), there is in the
general case no corresponding simple justification for the jackknife error
estimation procedure for the present type of data.

\subsection{Bootstrap error estimation}

In the bootstrap error estimation, the scheme is:

\begin{itemize}
\item First, bootstrap \cite{efron1986,press2007,efron1994} our original $M$ trajectories, i.e., pick
  $M$ trajectories from the original data {\em with replacement} (the same
  trajectory may be picked several times). Denote by
  ($\tilde{y}_i^{(m)},t_i$) the associated observables  and compute the
  synthetic mean value of the chosen observable  $\bar{y}_i = M^{-1}\sum_m \tilde{y}_i^{(m)}$.

\item Make a weighted least squares (WLS) fit to the synthetic MSDs
with respect to the fitting parameters. This fitting yields
parameters $\tilde{\param}_i$.

\end{itemize}

By repeating the two steps above many times (here, 100 times) we get a set of fit
parameters  $\tilde{\param}_i$ ($i=1,2,\ldots,100$). From this set we simply
compute the standard deviation as an estimator of the error for the fit
parameters.\cite{efron1986,press2007}

\section{Coefficient of determination}
\label{sec:goodness_of_fit}
We determine the goodness of fit by using the $R^2$ coefficient of
determination, defined as
\begin{equation}
R^2 = 1- \frac{S_{\text{res}}}{S_{\text{tot}}}.
\end{equation}
The method is based on a sum of squares over the
$N$~sampling points of, in our case, the mean positions or the MSD, $\samplem{y}$; hence, measuring
the deviation from the sample mean in \emph{time},
\begin{align}
  \sample{Y} &= \frac1N \sum_{i=1}^N \sample{y}_i\\
  S_{\text{tot}}  &= \sum_{i=1}^N (\sample{y}_i - \sample{Y})^2\\
  S_{\text{res}}  &= \sum_{i=1}^N \left( f(t_i;\params) - \sample{y}_i \right)^2.
\end{align}
Heuristically, a model that fits data perfectly has an $R^2=1$, while if it does
not fit at all, $R^2 \ll 1$, see Supplementary Figure~\ref{fig:fitgoodness}.

\section{Settings in "Particle Tracker" plug-in}\label{sec:particleTracker}

For detecting and linking particles into trajectories from the Supplementary
movies S1, S5 and S6 from the study by Chenouard et al.\cite{chenouard2014} we used the ImageJ plug-in
"Particle Tracker" \cite{particletracker} (November 2016 version) with the following  settings:
\begin{itemize}
\item 3D-data: no
\item radius: 3
\item cutoff: 3
\item radius: 0.1
\item LinkRange: 1 (default: 2)
\item displacement: 10.00
\item Dynamics: Brownian
\end{itemize}
and the following advanced options:
\begin{itemize}
\item Object features: 1.000
\item dynamics: 1.000
\item optimizer: greedy
\end{itemize}
All the settings listed above are default values except our choice for "LinkRange".


\renewcommand*{\thefootnote}{\arabic{footnote}}
\setcounter{footnote}{0}


\begin{thebibliography}{10}
\expandafter\ifx\csname url\endcsname\relax
  \def\url#1{\texttt{#1}}\fi
\expandafter\ifx\csname urlprefix\endcsname\relax\def\urlprefix{URL }\fi
\providecommand{\bibinfo}[2]{#2}
\providecommand{\eprint}[2][]{\url{#2}}

\bibitem{saxton2008}
\bibinfo{author}{Saxton, M.~J.}
\newblock \bibinfo{title}{Single-particle tracking: connecting the dots}.
\newblock \emph{\bibinfo{journal}{Nature Methods}}
  \textbf{\bibinfo{volume}{5}}, \bibinfo{pages}{671--672}
  (\bibinfo{year}{2008}).

\bibitem{brockmann2006}
\bibinfo{author}{Brockmann, D.}, \bibinfo{author}{Hufnagel, L.} \&
  \bibinfo{author}{Geisel, T.}
\newblock \bibinfo{title}{The scaling laws of human travel}.
\newblock \emph{\bibinfo{journal}{Nature}} \textbf{\bibinfo{volume}{439}},
  \bibinfo{pages}{462--465} (\bibinfo{year}{2006}).

\bibitem{desouza2012}
\bibinfo{author}{de~Souza, N.}
\newblock \bibinfo{title}{Pulling on single molecules}.
\newblock \emph{\bibinfo{journal}{Nature methods}}
  \textbf{\bibinfo{volume}{9}}, \bibinfo{pages}{873--877}
  (\bibinfo{year}{2012}).

\bibitem{seifert2012}
\bibinfo{author}{Seifert, U.}
\newblock \bibinfo{title}{Stochastic thermodynamics, fluctuation theorems and
  molecular machines}.
\newblock \emph{\bibinfo{journal}{Reports on Progress in Physics}}
  \textbf{\bibinfo{volume}{75}}, \bibinfo{pages}{126001}
  (\bibinfo{year}{2012}).

\bibitem{jarzynski1997}
\bibinfo{author}{Jarzynski, C.}
\newblock \bibinfo{title}{Nonequilibrium equality for free energy differences}.
\newblock \emph{\bibinfo{journal}{Physical Review Letters}}
  \textbf{\bibinfo{volume}{78}}, \bibinfo{pages}{2690} (\bibinfo{year}{1997}).

\bibitem{kou2004}
\bibinfo{author}{Kou, S.} \& \bibinfo{author}{Xie, X.~S.}
\newblock \bibinfo{title}{Generalized langevin equation with fractional
  gaussian noise: subdiffusion within a single protein molecule}.
\newblock \emph{\bibinfo{journal}{Physical Review Letters}}
  \textbf{\bibinfo{volume}{93}}, \bibinfo{pages}{180603}
  (\bibinfo{year}{2004}).

\bibitem{szymanski2009}
\bibinfo{author}{Szymanski, J.} \& \bibinfo{author}{Weiss, M.}
\newblock \bibinfo{title}{Elucidating the origin of anomalous diffusion in
  crowded fluids}.
\newblock \emph{\bibinfo{journal}{Physical Review Letters}}
  \textbf{\bibinfo{volume}{103}}, \bibinfo{pages}{038102}
  (\bibinfo{year}{2009}).

\bibitem{rothe2012}
\bibinfo{author}{Rothe, H.~J.}
\newblock \emph{\bibinfo{title}{Lattice gauge theories: an introduction, 4th
  ed.}}, vol.~\bibinfo{volume}{74} (\bibinfo{publisher}{World Scientific},
  \bibinfo{year}{2012}).

\bibitem{press2007}
\bibinfo{author}{Press, W.~H.}, \bibinfo{author}{Teukolsky, S.~A.},
  \bibinfo{author}{Vetterling, W.~T.} \& \bibinfo{author}{Flannery, B.~P.}
\newblock \emph{\bibinfo{title}{Numerical Recipes 3rd Edition: The Art of
  Scientific Computing}} (\bibinfo{publisher}{Cambridge University Press},
  \bibinfo{address}{New York, NY, USA}, \bibinfo{year}{2007}),
  \bibinfo{edition}{3rd} edn.

\bibitem{bos2007}
\bibinfo{author}{Van~den Bos, A.}
\newblock \emph{\bibinfo{title}{Parameter estimation for scientists and
  engineers}} (\bibinfo{publisher}{John Wiley \& Sons}, \bibinfo{year}{2007}).

\bibitem{sivia2006}
\bibinfo{author}{Sivia, D.} \& \bibinfo{author}{Skilling, J.}
\newblock \emph{\bibinfo{title}{Data analysis: a Bayesian tutorial}}
  (\bibinfo{publisher}{OUP Oxford}, \bibinfo{year}{2006}).

\bibitem{gottlieb1988}
\bibinfo{author}{Gottlieb, S.}, \bibinfo{author}{Liu, W.},
  \bibinfo{author}{Renken, R.~L.}, \bibinfo{author}{Sugar, R.~L.} \&
  \bibinfo{author}{Toussaint, D.}
\newblock \bibinfo{title}{Hadron masses with two quark flavors}.
\newblock \emph{\bibinfo{journal}{Physical Review D}}
  \textbf{\bibinfo{volume}{38}}, \bibinfo{pages}{2245--2265}
  (\bibinfo{year}{1988}).

\bibitem{michael1994}
\bibinfo{author}{Michael, C.}
\newblock \bibinfo{title}{{Fitting correlated data}}.
\newblock \emph{\bibinfo{journal}{Physical Review D}}
  \textbf{\bibinfo{volume}{49}}, \bibinfo{pages}{2616--2619}
  (\bibinfo{year}{1994}).

\bibitem{seibert1994}
\bibinfo{author}{Seibert, D.}
\newblock \bibinfo{title}{Undesirable effects of covariance matrix techniques
  for error analysis}.
\newblock \emph{\bibinfo{journal}{Physical Review D}}
  \textbf{\bibinfo{volume}{49}}, \bibinfo{pages}{6240--6243}
  (\bibinfo{year}{1994}).

\bibitem{yoon2013}
\bibinfo{author}{Yoon, B.}, \bibinfo{author}{Jang, Y.-C.},
  \bibinfo{author}{Jung, C.} \& \bibinfo{author}{Lee, W.}
\newblock \bibinfo{title}{Covariance fitting of highly-correlated data in
  lattice {QCD}}.
\newblock \emph{\bibinfo{journal}{Journal of the Korean Physical Society}}
  \textbf{\bibinfo{volume}{63}}, \bibinfo{pages}{145--162}
  (\bibinfo{year}{2013}).

\bibitem{meroz2015}
\bibinfo{author}{Meroz, Y.} \& \bibinfo{author}{Sokolov, I.~M.}
\newblock \bibinfo{title}{A toolbox for determining subdiffusive mechanisms}.
\newblock \emph{\bibinfo{journal}{Physics Reports}}
  \textbf{\bibinfo{volume}{573}}, \bibinfo{pages}{1--29}
  (\bibinfo{year}{2015}).

\bibitem{hofling2013}
\bibinfo{author}{H{\"o}fling, F.} \& \bibinfo{author}{Franosch, T.}
\newblock \bibinfo{title}{Anomalous transport in the crowded world of
  biological cells}.
\newblock \emph{\bibinfo{journal}{Reports on Progress in Physics}}
  \textbf{\bibinfo{volume}{76}}, \bibinfo{pages}{046602}
  (\bibinfo{year}{2013}).

\bibitem{norregaard2017}
\bibinfo{author}{Norregaard, K.}, \bibinfo{author}{Metzler, R.},
  \bibinfo{author}{Ritter, C.~M.}, \bibinfo{author}{Berg-S{\o}rensen, K.} \&
  \bibinfo{author}{Oddershede, L.~B.}
\newblock \bibinfo{title}{Manipulation and motion of organelles and single
  molecules in living cells}.
\newblock \emph{\bibinfo{journal}{Chemical reviews}}
  \textbf{\bibinfo{volume}{117}}, \bibinfo{pages}{4342--4375}
  (\bibinfo{year}{2017}).

\bibitem{berglund2010}
\bibinfo{author}{Berglund, A.~J.}
\newblock \bibinfo{title}{Statistics of camera-based single-particle tracking}.
\newblock \emph{\bibinfo{journal}{Physical Review E}}
  \textbf{\bibinfo{volume}{82}}, \bibinfo{pages}{011917}
  (\bibinfo{year}{2010}).

\bibitem{michalet2012}
\bibinfo{author}{Michalet, X.} \& \bibinfo{author}{Berglund, A.~J.}
\newblock \bibinfo{title}{Optimal diffusion coefficient estimation in
  single-particle tracking}.
\newblock \emph{\bibinfo{journal}{Physical Review E}}
  \textbf{\bibinfo{volume}{85}}, \bibinfo{pages}{061916}
  (\bibinfo{year}{2012}).

\bibitem{vestergaard2014}
\bibinfo{author}{Vestergaard, C.~L.}, \bibinfo{author}{Blainey, P.~C.} \&
  \bibinfo{author}{Flyvbjerg, H.}
\newblock \bibinfo{title}{Optimal estimation of diffusion coefficients from
  single-particle trajectories}.
\newblock \emph{\bibinfo{journal}{Physical Review E}}
  \textbf{\bibinfo{volume}{89}}, \bibinfo{pages}{022726}
  (\bibinfo{year}{2014}).

\bibitem{persson2013}
\bibinfo{author}{Persson, F.}, \bibinfo{author}{Lind{\'e}n, M.},
  \bibinfo{author}{Unoson, C.} \& \bibinfo{author}{Elf, J.}
\newblock \bibinfo{title}{Extracting intracellular diffusive states and
  transition rates from single-molecule tracking data}.
\newblock \emph{\bibinfo{journal}{Nature Methods}}
  \textbf{\bibinfo{volume}{10}}, \bibinfo{pages}{265--269}
  (\bibinfo{year}{2013}).

\bibitem{monnier2015}
\bibinfo{author}{Monnier, N.} \emph{et~al.}
\newblock \bibinfo{title}{Inferring transient particle transport dynamics in
  live cells}.
\newblock \emph{\bibinfo{journal}{Nature Methods}}
  \textbf{\bibinfo{volume}{12}}, \bibinfo{pages}{838--840}
  (\bibinfo{year}{2015}).

\bibitem{el2015}
\bibinfo{author}{El~Beheiry, M.}, \bibinfo{author}{Dahan, M.} \&
  \bibinfo{author}{Masson, J.-B.}
\newblock \bibinfo{title}{Inferencemap: mapping of single-molecule dynamics
  with bayesian inference}.
\newblock \emph{\bibinfo{journal}{Nature Methods}}
  \textbf{\bibinfo{volume}{12}}, \bibinfo{pages}{594--595}
  (\bibinfo{year}{2015}).

\bibitem{robson2013}
\bibinfo{author}{Robson, A.}, \bibinfo{author}{Burrage, K.} \&
  \bibinfo{author}{Leake, M.~C.}
\newblock \bibinfo{title}{Inferring diffusion in single live cells at the
  single-molecule level}.
\newblock \emph{\bibinfo{journal}{Phil. Trans. R. Soc. B}}
  \textbf{\bibinfo{volume}{368}}, \bibinfo{pages}{20120029}
  (\bibinfo{year}{2013}).

\bibitem{krog2017}
\bibinfo{author}{Krog, J.} \& \bibinfo{author}{Lomholt, M.~A.}
\newblock \bibinfo{title}{Bayesian inference with information content model
  check for langevin equations}.
\newblock \emph{\bibinfo{journal}{Physical Review E}}
  \textbf{\bibinfo{volume}{96}}, \bibinfo{pages}{062106}
  (\bibinfo{year}{2017}).

\bibitem{gershenfeld1999}
\bibinfo{author}{Gershenfeld, N.~A.}
\newblock \emph{\bibinfo{title}{The nature of mathematical modeling}}
  (\bibinfo{publisher}{Cambridge university press}, \bibinfo{year}{1999}).

\bibitem{metzler2000}
\bibinfo{author}{Metzler, R.} \& \bibinfo{author}{Klafter, J.}
\newblock \bibinfo{title}{The random walk's guide to anomalous diffusion: a
  fractional dynamics approach}.
\newblock \emph{\bibinfo{journal}{Physics Reports}}
  \textbf{\bibinfo{volume}{339}}, \bibinfo{pages}{1--77}
  (\bibinfo{year}{2000}).

\bibitem{pigeon2017}
\bibinfo{author}{Pigeon, S.}, \bibinfo{author}{Fogelmark, K.},
  \bibinfo{author}{S\"oderberg, B.}, \bibinfo{author}{Mukhopadhyay, G.} \&
  \bibinfo{author}{Ambj\"ornsson, T.}
\newblock \bibinfo{title}{Tracer particle diffusion in a system with hardcore
  interacting particles}.
\newblock \emph{\bibinfo{journal}{Journal of Statistical Mechanics: Theory and
  Experiment}} \textbf{\bibinfo{volume}{2017}}, \bibinfo{pages}{123209}
  (\bibinfo{year}{2017}).

\bibitem{mehrer2009}
\bibinfo{author}{Mehrer, H.} \& \bibinfo{author}{Stolwijk, N.~A.}
\newblock \bibinfo{title}{Heroes and highlights in the history of diffusion}.
\newblock \emph{\bibinfo{journal}{Diffusion Fundamentals}}
  \textbf{\bibinfo{volume}{11}}, \bibinfo{pages}{1--32} (\bibinfo{year}{2009}).

\bibitem{bloch2013}
\bibinfo{author}{Bloch, S.~C.}
\newblock \emph{\bibinfo{title}{Introduction to Classical and Quantum Harmonic
  Oscillators}} (\bibinfo{publisher}{John Wiley \& Sons},
  \bibinfo{year}{2013}).

\bibitem{bouchaud1994}
\bibinfo{author}{Bouchaud, J.-P.} \& \bibinfo{author}{Sornette, D.}
\newblock \bibinfo{title}{The black-scholes option pricing problem in
  mathematical finance: generalization and extensions for a large class of
  stochastic processes}.
\newblock \emph{\bibinfo{journal}{Journal de Physique I}}
  \textbf{\bibinfo{volume}{4}}, \bibinfo{pages}{863--881}
  (\bibinfo{year}{1994}).

\bibitem{yuan2014}
\bibinfo{author}{Yuan, N.}, \bibinfo{author}{Fu, Z.} \& \bibinfo{author}{Liu,
  S.}
\newblock \bibinfo{title}{Extracting climate memory using fractional integrated
  statistical model: A new perspective on climate prediction}.
\newblock \emph{\bibinfo{journal}{Scientific Reports}}
  \textbf{\bibinfo{volume}{4}} (\bibinfo{year}{2014}).

\bibitem{barkai2012}
\bibinfo{author}{Barkai, E.}, \bibinfo{author}{Garini, Y.} \&
  \bibinfo{author}{Metzler, R.}
\newblock \bibinfo{title}{Strange kinetics of single molecules in living
  cells}.
\newblock \emph{\bibinfo{journal}{Physics Today}}
  \textbf{\bibinfo{volume}{65}}, \bibinfo{pages}{29} (\bibinfo{year}{2012}).

\bibitem{tsai1997}
\bibinfo{author}{Tsai, C.-C.}
\newblock \bibinfo{title}{Slip, stress drop and ground motion of earthquakes: A
  view from the perspective of fractional \uppercase{B}rownian motion}.
\newblock \emph{\bibinfo{journal}{Pure and Applied Geophysics}}
  \textbf{\bibinfo{volume}{149}}, \bibinfo{pages}{689--706}
  (\bibinfo{year}{1997}).

\bibitem{metzler2004}
\bibinfo{author}{Metzler, R.} \& \bibinfo{author}{Klafter, J.}
\newblock \bibinfo{title}{The restaurant at the end of the random walk: recent
  developments in the description of anomalous transport by fractional
  dynamics}.
\newblock \emph{\bibinfo{journal}{Journal of Physics A: Mathematical and
  General}} \textbf{\bibinfo{volume}{37}}, \bibinfo{pages}{R161}
  (\bibinfo{year}{2004}).

\bibitem{weigel2011}
\bibinfo{author}{Weigel, A.~V.}, \bibinfo{author}{Simon, B.},
  \bibinfo{author}{Tamkun, M.~M.} \& \bibinfo{author}{Krapf, D.}
\newblock \bibinfo{title}{Ergodic and nonergodic processes coexist in the
  plasma membrane as observed by single-molecule tracking}.
\newblock \emph{\bibinfo{journal}{Proceedings of the National Academy of
  Sciences}} \textbf{\bibinfo{volume}{108}}, \bibinfo{pages}{6438--6443}
  (\bibinfo{year}{2011}).

\bibitem{machta2013}
\bibinfo{author}{Machta, B.~B.}, \bibinfo{author}{Chachra, R.},
  \bibinfo{author}{Transtrum, M.~K.} \& \bibinfo{author}{Sethna, J.~P.}
\newblock \bibinfo{title}{Parameter space compression underlies emergent
  theories and predictive models}.
\newblock \emph{\bibinfo{journal}{Science}} \textbf{\bibinfo{volume}{342}},
  \bibinfo{pages}{604--607} (\bibinfo{year}{2013}).

\bibitem{kepten2013}
\bibinfo{author}{Kepten, E.}, \bibinfo{author}{Bronshtein, I.} \&
  \bibinfo{author}{Garini, Y.}
\newblock \bibinfo{title}{Improved estimation of anomalous diffusion exponents
  in single-particle tracking experiments}.
\newblock \emph{\bibinfo{journal}{Physical Review E}}
  \textbf{\bibinfo{volume}{87}}, \bibinfo{pages}{052713}
  (\bibinfo{year}{2013}).

\bibitem{metzler2014}
\bibinfo{author}{Metzler, R.}, \bibinfo{author}{Jeon, J.-H.},
  \bibinfo{author}{Cherstvy, A.~G.} \& \bibinfo{author}{Barkai, E.}
\newblock \bibinfo{title}{Anomalous diffusion models and their properties:
  non-stationarity, non-ergodicity, and ageing at the centenary of single
  particle tracking}.
\newblock \emph{\bibinfo{journal}{Physical Chemistry Chemical Physics}}
  \textbf{\bibinfo{volume}{16}}, \bibinfo{pages}{24128--24164}
  (\bibinfo{year}{2014}).

\bibitem{transtrum2010}
\bibinfo{author}{Transtrum, M.~K.}, \bibinfo{author}{Machta, B.~B.} \&
  \bibinfo{author}{Sethna, J.~P.}
\newblock \bibinfo{title}{Why are nonlinear fits to data so challenging?}
\newblock \emph{\bibinfo{journal}{Physical Review Letters}}
  \textbf{\bibinfo{volume}{104}}, \bibinfo{pages}{060201}
  (\bibinfo{year}{2010}).

\bibitem{flyvbjerg1989}
\bibinfo{author}{Flyvbjerg, H.} \& \bibinfo{author}{Petersen, H.~G.}
\newblock \bibinfo{title}{Error estimates on averages of correlated data}.
\newblock \emph{\bibinfo{journal}{The Journal of Chemical Physics}}
  \textbf{\bibinfo{volume}{91}}, \bibinfo{pages}{461--466}
  (\bibinfo{year}{1989}).

\bibitem{berg2008}
\bibinfo{author}{Berg, B.~A.} \& \bibinfo{author}{Billoire, A.}
\newblock \emph{\bibinfo{title}{Markov chain Monte Carlo simulations}}
  (\bibinfo{publisher}{Wiley Online Library}, \bibinfo{year}{2008}).

\bibitem{kampen1992}
\bibinfo{author}{Van~Kampen, N.~G.}
\newblock \emph{\bibinfo{title}{Stochastic processes in physics and
  chemistry}}, vol.~\bibinfo{volume}{1} (\bibinfo{publisher}{Elsevier},
  \bibinfo{year}{1992}).

\bibitem{gplv3}
\bibinfo{title}{\textsc{gnu} {G}eneral {P}ublic {L}icense}.
\newblock \urlprefix\url{http://www.gnu.org/licenses/gpl.html}.

\bibitem{miller1974}
\bibinfo{author}{Miller, R.~G.}
\newblock \bibinfo{title}{The jackknife --- a review}.
\newblock \emph{\bibinfo{journal}{Biometrika}} \textbf{\bibinfo{volume}{61}},
  \bibinfo{pages}{1--15} (\bibinfo{year}{1974}).

\bibitem{efron1994}
\bibinfo{author}{Efron, B.} \& \bibinfo{author}{Tibshirani, R.~J.}
\newblock \emph{\bibinfo{title}{An introduction to the bootstrap}}
  (\bibinfo{publisher}{CRC press}, \bibinfo{year}{1994}).

\bibitem{chenouard2014}
\bibinfo{author}{Chenouard, N.} \emph{et~al.}
\newblock \bibinfo{title}{Objective comparison of particle tracking methods}.
\newblock \emph{\bibinfo{journal}{Nature Methods}}
  \textbf{\bibinfo{volume}{11}}, \bibinfo{pages}{281} (\bibinfo{year}{2014}).

\bibitem{sbalzarini2005}
\bibinfo{author}{Sbalzarini, I.~F.} \& \bibinfo{author}{Koumoutsakos, P.}
\newblock \bibinfo{title}{Feature point tracking and trajectory analysis for
  video imaging in cell biology}.
\newblock \emph{\bibinfo{journal}{Journal of Structural Biology}}
  \textbf{\bibinfo{volume}{151}}, \bibinfo{pages}{182--195}
  (\bibinfo{year}{2005}).

\bibitem{particletracker}
\bibinfo{author}{Sbalzarini, I.~F.} \& \bibinfo{author}{Koumoutsakos, P.}
\newblock \bibinfo{title}{Particletracker} (\bibinfo{year}{2016}).
\newblock \urlprefix\url{{http://imagej.net/Particle_Tracker}}.
\newblock \bibinfo{note}{Version November 2016}.

\bibitem{kariya2004}
\bibinfo{author}{Kariya, T.} \& \bibinfo{author}{Kurata, H.}
\newblock \emph{\bibinfo{title}{Generalized least squares}}
  (\bibinfo{publisher}{John Wiley \& Sons}, \bibinfo{year}{2004}).

\bibitem{savin2005}
\bibinfo{author}{Savin, T.} \& \bibinfo{author}{Doyle, P.~S.}
\newblock \bibinfo{title}{Static and dynamic errors in particle tracking
  microrheology}.
\newblock \emph{\bibinfo{journal}{Biophysical Journal}}
  \textbf{\bibinfo{volume}{88}}, \bibinfo{pages}{623--638}
  (\bibinfo{year}{2005}).

\bibitem{martin2002}
\bibinfo{author}{Martin, D.~S.}, \bibinfo{author}{Forstner, M.~B.} \&
  \bibinfo{author}{K{\"a}s, J.~A.}
\newblock \bibinfo{title}{Apparent subdiffusion inherent to single particle
  tracking}.
\newblock \emph{\bibinfo{journal}{Biophysical Journal}}
  \textbf{\bibinfo{volume}{83}}, \bibinfo{pages}{2109--2117}
  (\bibinfo{year}{2002}).

\bibitem{calderon2016}
\bibinfo{author}{Calderon, C.~P.}
\newblock \bibinfo{title}{Motion blur filtering: A statistical approach for
  extracting confinement forces and diffusivity from a single blurred
  trajectory}.
\newblock \emph{\bibinfo{journal}{Physical Review E}}
  \textbf{\bibinfo{volume}{93}}, \bibinfo{pages}{053303}
  (\bibinfo{year}{2016}).

\bibitem{chaichian2001}
\bibinfo{author}{Chaichian, M.} \& \bibinfo{author}{Demichev, A.}
\newblock \bibinfo{title}{Path integrals in physics, vol. 1: Stochastic
  processes and quantum mechanics}.
\newblock \emph{\bibinfo{journal}{IOP, Bristol, UK}}  (\bibinfo{year}{2001}).

\bibitem{norrelykke2011}
\bibinfo{author}{N{\o}rrelykke, S.~F.} \& \bibinfo{author}{Flyvbjerg, H.}
\newblock \bibinfo{title}{Harmonic oscillator in heat bath: Exact simulation of
  time-lapse-recorded data and exact analytical benchmark statistics}.
\newblock \emph{\bibinfo{journal}{Physical Review E}}
  \textbf{\bibinfo{volume}{83}}, \bibinfo{pages}{041103}
  (\bibinfo{year}{2011}).

\bibitem{qian2003}
\bibinfo{author}{Qian, H.}
\newblock \bibinfo{title}{Fractional {B}rownian motion and fractional
  {G}aussian noise}.
\newblock In \emph{\bibinfo{booktitle}{Processes with Long-Range
  Correlations}}, \bibinfo{pages}{22--33} (\bibinfo{publisher}{Springer},
  \bibinfo{year}{2003}).

\bibitem{mandelbrot1968}
\bibinfo{author}{Mandelbrot, B.~B.} \& \bibinfo{author}{Van~Ness, J.~W.}
\newblock \bibinfo{title}{Fractional {B}rownian motions, fractional noises and
  applications}.
\newblock \emph{\bibinfo{journal}{SIAM Review}} \textbf{\bibinfo{volume}{10}},
  \bibinfo{pages}{422--437} (\bibinfo{year}{1968}).

\bibitem{davies1987}
\bibinfo{author}{Davies, R.~B.} \& \bibinfo{author}{Harte, D.}
\newblock \bibinfo{title}{Tests for \uppercase{H}urst effect}.
\newblock \emph{\bibinfo{journal}{Biometrika}} \textbf{\bibinfo{volume}{74}},
  \bibinfo{pages}{95--101} (\bibinfo{year}{1987}).

\bibitem{chambers1995}
\bibinfo{author}{Chambers, M.}
\newblock \bibinfo{title}{The simulation of random vector time series with
  given spectrum}.
\newblock \emph{\bibinfo{journal}{Mathematical and Computer Modelling}}
  \textbf{\bibinfo{volume}{22}}, \bibinfo{pages}{1--6} (\bibinfo{year}{1995}).

\bibitem{quenouille1956}
\bibinfo{author}{Quenouille, M.~H.}
\newblock \bibinfo{title}{Notes on bias in estimation}.
\newblock \emph{\bibinfo{journal}{Biometrika}} \textbf{\bibinfo{volume}{43}},
  \bibinfo{pages}{353--360} (\bibinfo{year}{1956}).

\bibitem{gradshteyn2000}
\bibinfo{author}{Gradshteyn, I.} \& \bibinfo{author}{Ryzhik, I.}
\newblock \bibinfo{title}{Table of integrals, series and products (corrected
  and enlarged edition prepared by \uppercase{A}. \uppercase{J}effrey and
  \uppercase{D}. \uppercase{Z}willinger)}.
\newblock \emph{\bibinfo{journal}{Academic Press, New York}}
  (\bibinfo{year}{2000}).

\bibitem{anderson2003}
\bibinfo{author}{Anderson, T.~W.}
\newblock \emph{\bibinfo{title}{An introduction to multivariate statistical
  analysis, 3rd ed.}} (\bibinfo{publisher}{Wiley New York},
  \bibinfo{year}{2003}).

\bibitem{schucany1971}
\bibinfo{author}{Schucany, W.}, \bibinfo{author}{Gray, H.} \&
  \bibinfo{author}{Owen, D.}
\newblock \bibinfo{title}{On bias reduction in estimation}.
\newblock \emph{\bibinfo{journal}{Journal of the American Statistical
  Association}} \textbf{\bibinfo{volume}{66}}, \bibinfo{pages}{524--533}
  (\bibinfo{year}{1971}).

\bibitem{efron1981}
\bibinfo{author}{Efron, B.} \& \bibinfo{author}{Stein, C.}
\newblock \bibinfo{title}{The jackknife estimate of variance}.
\newblock \emph{\bibinfo{journal}{The Annals of Statistics}}
  \bibinfo{pages}{586--596} (\bibinfo{year}{1981}).

\bibitem{efron1986}
\bibinfo{author}{Efron, B.} \& \bibinfo{author}{Tibshirani, R.}
\newblock \bibinfo{title}{Bootstrap methods for standard errors, confidence
  intervals, and other measures of statistical accuracy}.
\newblock \emph{\bibinfo{journal}{Statistical science}} \bibinfo{pages}{54--75}
  (\bibinfo{year}{1986}).

\end{thebibliography}
\end{document}